\documentclass[UTF8,a4paper]{article}
\usepackage{amsmath}
\usepackage{bm}
\usepackage{graphicx}
\usepackage{subfigure}
\usepackage{ mathrsfs }
\usepackage{epstopdf}
\usepackage{ marvosym }
\usepackage{ tipa }
\usepackage{geometry}
\usepackage{ wasysym }
\usepackage{ulem}
\usepackage{indentfirst}
\usepackage{amsfonts,amssymb,dsfont}
\usepackage{accents}

\usepackage{setspace}
\usepackage{threeparttable}
\usepackage{amsmath,mathtools,amsthm}
\usepackage{array}
\usepackage{extarrows}
\usepackage{upgreek}

\makeatletter
\renewcommand*\env@matrix[1][\arraystretch]{%
	\edef\arraystretch{#1}%
	\hskip -\arraycolsep
	\let\@ifnextchar\new@ifnextchar
	\array{*\c@MaxMatrixCols c}}
\makeatother

\usepackage[pdfstartview=XYZ,
bookmarks=true,
colorlinks=true,
linkcolor=blue,
urlcolor=blue,
citecolor=blue,
pdftex,
bookmarks=true,
linktocpage=true, % makes the page number as hyperlink in table of content
hyperindex=true
]{hyperref}
\usepackage{orcidlink}

\title{\bf Composite polaron formed on surface of two-dimensional lattice system in weak coupling regime}
\author{Chen-Huan Wu 
	\orcidlink{0000-0003-1020-5977} 
	\thanks{chenhuanwu1@gmail.com}
	\\College of Physics and Electronic Engineering, Northwest Normal University, Lanzhou 730070, China}

\begin{document}
	
	\maketitle

	\begin{abstract}
We investigate the properties of composite polaron containing the effects of electron-phonon coupling and 
interaction between impurity and electron-hole pair.
A model of a two-dimensional electron gas occupying the surface of two-dimensional Dirac honeycomb lattice
is constructed.
We focus on the weak coupling regime throughout the paper.
Our results are meaningful to the study of pairing mechanism as
well as the phonon-mediated high-temperature superconductivity.
\\
\\
% $PACS\ number(s)$: 71.10.Hf, 71.10.Li, 71.36.+c\\
%$ Keywords$: Fermi polaron;
%Medium T-matrix;
%Self-energy;
%Spectral function;
%Pair propagator;
%Ladder approximation\\

\end{abstract}
\begin{small}

\section{Introduction}

The mobile impurity is widely studied in ultracold atomic systems like the Fermi gases\cite{Scazza F,Haussmann R} 
or Bose-Einstein condensates\cite{Li W}, degenerate Bose-Fermi mixtures\cite{Fratini E,Wang D W}, 
  %{Variational polaron method for Bose-Bose mixtures}
and the solid state systems, like the semiconductor\cite{Sidler M}, (semi) Dirac systems\cite{polaron1,polaron2}
metals\cite{Wong W P D,Shvonski A}, and the transition metal dichalcogenides\cite{Dong X Y}.
  %{Miyata K, Meggiolaro D, Trinh M T, et al. Large polarons in lead halide perovskites[J]. Science advances, 2017, 3(8): e1701217.}
For both the Mott insulator state\cite{Fukuhara T} and superfluid state\cite{Qin F,Caldas H,Wang D W,Kinnunen J J}, 
the mobile spin impurity as well as the formed polaron (by dressing the particle-hole pairs) are also discussed.
In dealing with the many-body problem of the polaron system,
the leading order $1/N$ expansion base on the random-phase approximation (RPA) is similar to the self-consistent medium $T$-matrix
which with a dynamically screened interaction (the GW approximation),
and the $1/N$ order requires a self-energy term contributed by the particle-hole loop,
to describes the pairing fluctuation especially in strong coupling regime like the unitary Fermi gas.
At $N\rightarrow\infty$ limit, the self-energy vanishes and the lifetime of the Fermi quasiparticle $\rightarrow\infty$
according to ${\rm Re}\Sigma({\rm Im}\Sigma)\sim \frac{1}{N}$ and ${\rm Im}\Sigma(\omega)=-\frac{1}{2\tau}{\rm sgn}[\omega]$.
  %{Quantum critical transport in the unitary Fermi gas}
  %{Vertex corrections to the dc conductivity in anisotropic multiband systems}
Here the self-energy is the Coulomb-interaction-induced which including the exchange part and the correlation part,
   %{Anisotropy and Interplane Interactions in the Dielectric Response of Graphite}
   %{Plasmons and the spectral function of graphene DOI: 10.1103/PhysRevB.77.081411}
the correlation part here is due to the quantum fluctuations of the Fermi sea,
and it leads to the dielectric response.
The particle-hole pairs (coupled by the Coulomb potential) 
  %{Plasmons and the spectral function of graphene}
as the low-energy excitations are promoted in two-dimensional system\cite{Koschorreck M},
In the presence of RPA with the collective electronic modes, 
  %{Plasmon-polaron of the topological metallic surface states}
the quasiparticles interact with the particle-hole excitations which forms the polaron when the quasiparticle is a mobile impurity,
such a scenario can be realized by using a two-dimensional electron gas coupled with the two-dimensional Dirac system or the 
surface state of three-dimensional Dirac system,
the Dirac material here can also be replaced by the topological crystals like the Chern insulator,
then the impurity also exhibit some topological features due to the coupling with the environment
which is in a topologically nontrivial state\cite{Qin F,Camacho-Guardian A}.
Such scenario can be realized even in the absence of gases but just relys on the exciton-electron interaction within a microcavity diagram
\cite{Sidler M,Kasprzak J}.
As we known,
  %{Plasmon-pole approximation for many-body effects in extrinsic graphene}
the plasmon excitation $\omega_{p}\sim \sqrt{q}$ in long-wavelength limit where $q$ is the scattering momentum,
and it obtains more spectral weight when enters into the particle-hole continuum region,
  %{Plasmons and the spectral function of graphene}
due to the Laudau damping which leads to a weaker interaction (and thus with larger residue $\mathcal{Z}$\cite{polaron2}).
The coupling between the plasmon and the polaron would be more strong in the massless two-dimensional Dirac system 
compared to the massive one\cite{Polini M,Wang Z}.
The appearing of avoided crossing in the polaron band structure as a result of the strong coupling
is also an interesting phenomenon.
Furthermore, in the presence of strong coupling, the avoided crossing will disappears gradually
with the increase of temperature (away from the superfluid phase again),
and then the particle branch merges with the hole branch to form a new (adiabatic) polaron dispersion\cite{Haussmann R},
and this new dispersion is shifted by the binding energy $(-E_{b})$ or the polaron shift in atomic limit ($t\rightarrow 0$)
which is
$(-\frac{g^{2}}{2m_{e-h}\omega^{2}_{e-h}})$.
  %{Spectral functions and rf response of ultracold fermionic atoms}
Besides, in this limit, the polaron conductivity behaves as $\sigma\sim \frac{W}{T}e^{\Omega/T}$
due to the smallness of bandwidth $W$.
  %{dynamical mean-field theory of transport of small polarons}

The adiabaticity and non-adiabaticity are closely related to the physical properties of the quantum sytem\cite{Altman E},
including the instability and the phase transitions.
In this paper, we consider the composite polaron which is a fermi polaron dressed by the surrounding acoustic phonons,
in the surface of a two-dimensional (2D) 
honeycomb lattice.
Unlike the optical phonon modes, the acoustic phonons have a lower energy (which vanishes in the BZ center) 
and thus provide an effective scattering channel\cite{Hwang E H}.
Also, the energy transition into the acoustic phonon bath is the dominant cooling way
for the moving electron (impurity)\cite{Bistritzer R}.
This energy transition is related to the acoustic phonon dispersion (mainly the linear one).
   %{electronic cooling in graphene}
   %{acoustic phonon scattering limited carrier mobility in two-dimensional extrinsic graphene}
While the effect of surface optical phonons can be
exploited by deposit the material on a polar substrate.
  %{plasmon-phonon coupling in a valley-spin-polarized two-dimensional electron system: a theoretical study on monolayer silicene}
In adiabatic limit, where the Fermi velocity of electron is much larger than the sound velocity,
the electron-phonon scattering is elastic (this is valid even for the neutral limit) and interband transitions vanish,
thus the electron will not loss energy.
But for another case, when the non-relativistic
electron moves faster than the sound wave but still interacts with the acoustic phonons, i.e., in the nonadiabatic regime,
the electron may losses energy by the Cherenkov radiabation.

The existence of Dirac cone in 2D lattice can be seen as a perturbation to the free electron (impurity)
as discussed in this paper.
The resulting mobile polaron exhibits composite dispersion
modified by the interaction between impurity and the electron-hole pair as well as the electron-phonon coupling.
We find the modification of dispersion induced by weak electron-phonon coupling is linear with the impurity momentum in the stable 
(adiabatic) regime.
While for strong electron-phonon coupling,
the avoided crossing appears in the polaron band structure as a result of nonadiabaticity.
We know that the band gap in the middle point of Brillouin zone edge will be reduced gradually with the rise of doping level.
  %{Doped Graphene as Tunable Electron-Phonon Coupling Material}
At the nonadiabatic regime,
the band gap in $M$-point can even disappears,
which means that, the excitation (acoustic plasmon mode\cite{Shvonski A,Zhang S,Jia X}) can
 enters into the adjacent Brillouin zone and be free from the influence of umklapp scattering
even at higher temperature.
The lattice periodicity as well as the Brillouin zone symmetry is missing.
In this case, the superfluid can be observed by the crossing Bloch bands at low enough temperature,
which can not be stoped by the periodic perturbing potential originates from the lattice\cite{Shvonski A,Chen Z}.
  %{Breakdown of the Migdal-Eliashberg theory in the strong-coupling adiabatic regime}
Note that the avoided crossing can also appears in the ultracold fermi atomic systems in strong-interacting limit\cite{Haussmann R}.
The finite temperature effect on the self-energy induced by electron-phonon coupling is also been discussed.
  %{Optical conductivity of black phosphorus with a tunable electronic structure}
In this paper, we set a small fermi energy throughout this paper to avoid the problem of Pauli blocking and support the contact interaction.
   %{fermi polaron-polaritons in charge-tunable atomically thin semiconductor}
Also, in nonadiabatic case without the perturbations,
we treat the impurity dispersion (kinetic energy) as the conventional non-relativistic electron dispersion which is parabolic.
  %{quantum transport in Weyl semimetal thin films in the presence of spin-orbit coupled impurities}

For the impurity motion-dependent phenomenons, like the coherence or decoherence,
which in some degrees related to the Fermi liquid or non-Fermi lqiuid behavior\cite{ms18},
are important in studying the formation of the polaron.
The decoherence process may happen in the
 extreme case which with a very light
impurity immersed into a bath of heavy particles.
While in the opposite limit: for a heavy impurity in 1D system, 
the decoherence can also happen at zero temperature limit due to the orthogonality catastrophe\cite{Rosch A,Kantian A,Meden V}.
  %{Heavy Particle in a d-Dimensional Fermionic Bath: A Strong Coupling Approach}
  %{Competing regimes of motion of 1D mobile impurities}
  %{Excitation spectra and rf response near the polaron-to-molecule transition from the functional renormalization group}
  %{Orthogonality catastrophe in a one-dimensional system of correlated electrons}
In adiabatic case,
the existence of the absorption and emission of phonons also leads to decoherence effect
and reduces the coherent band motion.
  %{Light-enhanced electron-phonon coupling from nonlinear electron-phonon coupling}
  %{Polarons and bipolarons in strongly interacting electron-phonon systems}

To extracting more information about the single and multi-particle excitations
and quasiparticles,
we apply the method of range-separated hybrid functionals (RSH),
where the spacial effect can be estimated in terms of the RSH parameters
(for short range and long-range exchange), 
and a range-separation parameter.
Distinct from the generalized gradient approximation (GGA),
the method of RSH exhibits better the role of spatial effects,
e.g., when the polaron-molecule (dimer ground state) transition happen
and
for example when a molecule is being stretched or is subject to
an applied electric field where charge transfer takes place.

For electron-phonon coupling considered in this paper,
we assume the acoustic phonons with large enough sound velocity to ensure the phonon energy ($\hbar \Omega_{ph}$)
larger than the band gap, and thus single-phonon linewidth (quantum) is nonzero due to the existence of electron-phonon coupling.
For in-plane acoustic phonons near Brillouin zone center, they can cause the intraband
carrier transition by the deformation potential scattering,
  %{Carrier recombination and generation rates for intravalley and intervalley phonon scattering in graphene}
  %{Intrinsic electrical transport properties of monolayer silicene and MoS from first principles}
by cannot cause interband carrier transition due to the nonconservation of energy\cite{Rana F,Li X,Borysenko K M}.
However, for large enough sound velocity and away from the zone center, the interband transition rate is finite\cite{Hwang E H} and 
with a finite scattering angle in the relaxation time approximation.
This scattering angle in ladder diagrams will introduces a $(1-{\rm cos}\theta)$ term and related to the chiral characters,
   %{Vertex corrections to the dc conductivity in anisotropic multiband systems}
which we will discuss indetailin other place\cite{polaron1}.
To deal with the impurity-phonon interaction, we apply
the Fourier transformed Su-Schrieffer-Heeger (SSH) model\cite{Perebeinos V,Marchand D J J,Alexandrov A}
  %{Electron-Phonon Interaction and Transport in Semiconducting Carbon Nanotubes}
  %{Semiclassical transport and phonon scattering of electrons in semiconducting carbon nanotubes}
  %{Light Bipolarons Stabilized by Peierls Electron-Phonon Coupling}
  %{Sharp Transition for Single Polarons in the One-Dimensional Su-Schrieffer-Heeger Model}
  %{Bipolarons bound by repulsive phonon-mediated interactions}
  %{Fr?hlich polaron and bipolaron: recent developments}
where we only consider the carriers transition from valence band to conduction band,
and thus the electron-hole recombination does not exists.
  %{Carrier recombination and generation rates for intravalley and intervalley phonon scattering in graphene}
While in the presence of strong field or the polar substrate,
an electric field may be produced by the phonons with high energy,
and then the polaronic effect may also comes from the coupling between the electron-hole pair with the ions,
as recently realized in ionic solid LiF\cite{Sio W H} where the method of optimally
tuned and range-separated hybrid functionals (OT-RSH) performs well\cite{Bhandari}.
Away from the dilute electron density limit\cite{Nocera,Parish}, in a system that dominated by the coupling strength,
a polaron-molecule transition can be realized at a large binding energy, which corresponds to small momentum cutoff during the calculation of two-body coupling function.
As an example,
in N-type doped transition metal oxides, the localized electrons form the electron-trapped
surface polarons can modify the charge transfer and strengthen the polarization.
Such charge transfer can weaken or break the covalent bond through the formation of active sites,
like the reduced metal ion or the synergistic effect between electron-deficient and electron-rich sites\cite{Jiang}.
During this process, the electron transition cleave the existing covalent bond
and dissociate the molecule.
Such actived covalent bond in the moleculas can be experimentally observed through the far-infrared near-field spectrum\cite{Kalosakas}, where the vibrational polaronic dynamics can be directly measured.
This can also be realized by applying high pressure to molecular solid can change it into  extended structures during which process the
repulsive interatomic interactions are weakened\cite{Dias}
and the long-range order will be replaced by the short-range one.
In this case, the suppression of long-range order can be estimated by analying the deviation between the real charge density distribution with that calculated under range-separated potential.
Such phenomenon can be observed in serveral materials containing the transition metals,
like the amorphous WTe$_{2}$ in
semiconducting phase\cite{Peng} and the TiO$_{2}$\cite{Jiang,Kick},
where the hot(thermally-activated) carriers hopping as well as the electron-phonon coupling contribute to the formation of polaron.

%we note that, although the phonons are exist, they are assumed with low energy (i.e., the
%acoustic phonons with small phonon wave vector $q'$),
%and thus the intervalley scattering does not taken into account.
  %{Intrinsic electrical transport properties of monolayer silicene and MoS 2 from first principles}
  %{Phonon linewidths and electron-phonon coupling in graphite and nanotubes}
  %{First-principles method for electron-phonon coupling and electron mobility: Applications to two-dimensional materials}
  %{First-principles analysis of electron-phonon interactions in graphene}
%The coupling 

\section{$T$-matrix approach and the related coupling between non-relativistic particles}

Different to the usual case of a Bose-Fermi mixture in ultracold atomic system,
we can assuming a two-dimensional weakly-interacting Bose gas coupled to the (two-dimensional) crystal,
for average-distance between bosons in the order of van der Waals length $l_{vdW}$
which is comparable or even larger than the $s$-wave scattering length\cite{Rath S P},
  %{Field-theoretical study of the Bose polaron}
that's in contrast to the case near Feshbach resonances (with large $a$)
 where the weakly-bound three-body states (trimer) would be formed\cite{Blume D,Wang Y}.
The van der Waals interaction here can be controlled by
 choosing materials with different orbital characters.
  %{Emergence of interfacial polarons from electron-phonon coupling in graphene/h-BN van der Waals heterostructures}
To obtain the analytical solution of the scattering problem,
  %{Fast, adaptive, high-order accurate discretization of the Lippmann--Schwinger equation in two dimensions}
the Bethe-Salpeter equation (BSE) is often reduced to the Lippmann-Schwinger equation (LSE)
by ignoring the center-of-mass momenta, i.e., the vacuum $T$-matrix.
  %{pairing instabilities in quasi-two-dimensional fermi gases}
  %{bipolarons in bose-einstein condensate}
The zero center-of-mass momentum can be realized experimentally by increasing the strength of spin-orbit coupling,
  %{Molecule and Polaron in a Highly Polarized Two-Dimensional Fermi Gas with Spin-Orbit Coupling}
like the molecule state in BEC limit, note that
while for polaron state, the impurity with zero momentum does not corresponds to the zero center-of-mass case,
the impurity and hole with momentum $k_{F}$ and $-k_{F}$, respectively, corresponds to it\cite{Chen J G}.
The BSE
is more difficult to solved due to its singularity\cite{Levine M,Karmanov V A,Camacho-Guardian A2}
especially in the presence of coupled center-of-mass and relative motions by the Fermi sea,
  %{pairing instabilities in quasi-two-dimensional fermi gases}
it can be described by the non-self-consistent many-body $T$-matrix\cite{polaron1,polaron2}
  %{Bipolarons in a Bose-Einstein condensate}
  %{Mass imbalance effect in resonant Bose-Fermi mixtures}
\begin{equation} 
\begin{aligned}
T(p+q,&\square;p+q-k';\omega)\\
=&g^{b}(p+q,\square;p+q-k')\\         
      &\sum_{k,\Omega}g^{b}(p+q,\square;k)(-G^{\phi}_{0}(p+q-k;\omega+\Omega)G^{\psi}_{0}(\square+k;\Omega))T(p+q-k,\square+k;p+q-k-k';\omega)\\
=&g^{b}(p+q,\square;p+q-k')        
      +g^{b}(p+q,\square;k)\Pi(p+q;\omega)T(p+q-k,\square+k;p+q-k-k';\omega),
\end{aligned}
\end{equation}
where $g^{b}$ are the bare impurity-majority interactions,
specially, $g^{b}(p+q,\square;k)$ is the interaction induced by the polarization operator
(consist of the two bare Green's functions).
  %{pairing instabilities in quasi-two-dimensional fermi gases}
$k,\ k'$ are the relative momentum.
$G^{\psi}_{0}$ and $G^{\phi}_{0}$ are the bare Fermionic and Bosonic Green's function, as
given by $G^{\psi}_{0}(\nu,k)=[\nu+i0^{+}-\frac{\hbar^{2}k^{2}}{2m_{\psi}}+\mu_{\uparrow}]^{-1}$ and
$G^{\phi}_{0}(\omega+\Omega-\nu,p+q-k)=[\omega+\Omega-i\nu+2i0^{+}-\frac{\hbar^{2}(p+q-k)^{2}}{2m_{\phi}}+\mu_{\downarrow}]^{-1}$, respectively.
%is the bare impurity propagator (not the scalar-field one).
$\Pi(p+q)$ is the polarization operator.
  %{Pairing instabilities in quasi-two-dimensional Fermi gases}

By ignoring the harmonic oscillation quantum numbers,
  %{Single-particle properties and pseudogap effects in the BCS-BEC crossover regime of an ultracold Fermi gas above Tc}
  %{Multichannel scattering and Feshbach resonances Effective theory, phenomenology, and many-body effects}
  %{Pairing instabilities in quasi-two-dimensional Fermi gases}
  %{Quasiparticle Properties of a Mobile Impurity in a Bose-Einstein Condensate}
  %{Quasiparticle Properties of a Mobile Impurity in a Bose-Einstein Condensate}
and solve the LSE up to first order,
we can obtain
\begin{equation} 
\begin{aligned}
T(\omega)=\frac{g(\omega)}{1-g(\omega)\Pi(\omega)},
\end{aligned}
\end{equation}
and thus
\begin{equation} 
\begin{aligned}
T^{-1}(\omega)=(g(\omega))^{-1}-\Pi(\omega).
\end{aligned}
\end{equation}
Straightly, when we solve the LSE up to second order (we omit the common parameter $\omega$ in the following),
we have
\begin{equation} 
\begin{aligned}
T^{-1}=(g+g^{2}\Pi)^{-1}-(\frac{1}{g\Pi^{2}}+\frac{1}{\Pi})^{-1}.
\end{aligned}
\end{equation}
This $T$-matrix here is certainly the single channel one due to the infinitely large hyperfine splitting energy,
while
for finite range case,
i.e., both the intraspeices and interspecies (hyperfine levels) coupling are considered,
the multichannel $T$-matrix (consider the center-of-mass frame) becomes\cite{Massignan P,Bruun G M}
  %{Polarons and dressed molecules near narrow Feshbach resonances}
  %{Multichannel scattering and Feshbach resonances Effective theory, phenomenology, and many-body effects}
  %{Pairing instabilities in quasi-two-dimensional Fermi gases}
\begin{equation} 
\begin{aligned}
T_{12}=\frac{g_{1}}{1-g_{1}\Pi_{1}}+(\frac{g_{12}}
{1-g_{1}\Pi_{1}})^{2}\frac{1}{\Pi^{-1}_{2}-g_{2}-\frac{(g_{12})^{2}\Pi_{1}}{1-g_{1}\Pi_{1}}},
\end{aligned}
\end{equation}
where the subscript $1$ denotes the open channel and $2$ denotes the close channel.

At first, we discuss the zero energy scattering matrix in the atomic limit in perturbation theory (without consider the effect of hopping).
  %{Critical slowdown of non-equilibrium polaron dynamics}
By solving the LSE for low-energy scattering,
the regularized intraspecies and bare interspecies coupling parameter (or called the binary contact interaction) can also be obtained\cite{Rath S P,polaron1}
  %{Thermodynamics of spin-orbit-coupled Bose-Einstein condensates}
  %{Mass imbalance effect in resonant Bose-Fermi mixtures}
  %{Stability condition of a strongly interacting boson-fermion mixture across an interspecies Feshbach resonance support}
\begin{equation} 
\begin{aligned}
g_{\phi\phi}=&\frac{4\pi \hbar^{2}a_{\psi\psi}}{m_{\phi}}=[\frac{1}{g_{\phi\phi}^{b}}+\int\frac{d^{2}k}{(2\pi)^{2}}\frac{g(k)}{\hbar^{2}k^{2}}]^{-1},\\
g^{b}_{\psi\phi}(\Lambda)=&[\frac{m_{r}}{2\pi \hbar^{2}a}+\int\frac{d^{2}k}{(2\pi)^{2}}\frac{1}{\varepsilon_{k\uparrow}+\varepsilon_{k\downarrow}}]^{-1},\\
=&[\frac{m_{r}}{2\pi \hbar^{2}a}+\int\frac{d^{2}k}{(2\pi)^{2}}\frac{2m_{r}}{\hbar^{2}k^{2}}]^{-1},\\
\end{aligned}
\end{equation}
  %{The nonlinear Dirac equation in Bose–Einstein condensates: Foundation and symmetries}
  %{multichannel scattering and feshbach resonances effective theory, phenomenology, and many-body effect}
  %{field-theoretical study of the bose polaron}
where $\phi$ denotes the bosonic field, $g_{\phi\phi}^{b}$ is the bare coupling parameter,
here $g_{\phi\phi}<0$ for attractive interaction.
$g(k)$ is the interaction range-dependent function,
when it's momentum-independent (for contact interaction or the short-range limit),
the integral within the above equation does not converges,
then the momentum cutoff is required which is of the order of inversed interaction range, e.g., $\Lambda\sim 1/l_{vdW}$.
  %{field-theoretical study of the bose polaron}
  %{van der Waals energy and pressure in dissipative media: fluctuational electrodynamics and mode summation}
The cutoff $\Lambda$ goes to infinite when $g_{\psi\psi}^{b}$ goes to zero and in this case the average potential energy is zero.
  %{normal state of highly polarized fermi gases: full many-body treatment}
This usually corresponds to the short-range limit with very large hyperfine splitting energy 
between the two lowest energy hyperfine states\cite{Bruun G M,Combescot R} in open channel and closed channel respectively for 
a statistically mixed atomic system (like the fermi-fermi or fermi-bose mixtures).
Such van der Waals interaction in the interface of a heterostructure system 
also provides a possible formation of the polaron\cite{Chen C}
in the atomic layer stacking system\cite{Narayanaswamy A} which favors more the van der Waals interaction.
Besides the van der Waals force,
the attractive potential (of impurity) mediated by the crystal phonon forms the Cooper parts that produced the supetconductivity phenomenon.
  %{induced p-wave pairing in bose-fermi mixture}

The fermi polaron formed in solid state
is stable in weak coupling regime,
  %($s$-wave interaction)
  %according to our previous studies\cite{polaron1,polaron2} which focusing on the formation of polaron near the spin-polarized Dirac cone,
  %{Spin-Polarized Dirac-Cone-Like Surface State with  Character at W(110)}
and
can be treated uncoupled with the environment,
that is similar to the bare pairing in closed channel\cite{Bruun G M}
which has higher energy than the open one (dressed),
but in gaseous systems where the interaction can be turned via a magnetic field-induced Feshbach resonance,
the single channel description with zero-range 
%contact 
potential fails in weak interaction regime\cite{Massignan P}.
  %{polarons and dressed molecules near narrow feshbach resonances}
  %{multichannel scattering and feshbach resonances effective theory, phenomenology, and many-body effect}
The properties of the polaron is closely related to the dispersion of the majority component in the bath, 
for the most common kind of the polaron in solid system---the electron-phonon coupling induced Bose polaron, 
since the dispersion of longitudinal acoustic phonon is $\sim v_{s}q'$ ($q'$ is the phonon momentum) with sound velocity
$v_{s}$ (see, e.g., Ref.\cite{Ge X J}).
%it is more resembles to the polaron formed in Bose gases than in Fermi gases due to the lower mass of Bosons,
%  %{critical slowdown of non-equilibrium polaron dynamics}
%as can be seen from the Bogoliubov dispersion in a BEC at nearly zero temperature.

The validity of one-particle-hole ansatz\cite{Scazza F}
in dealing with the polaron problem in the presence of many-body effect has been verified.
  %{Normal State of Highly Polarized Fermi Gases: Full Many-Body Treatment}
  %{Polaron-to-molecule transition in a strongly imbalanced Fermi gas}
However, according to Ref.\cite{Combescot R},
this validity requires the nearly perfect destructive interference
of the contributions from other particle-hole excitations (brings the higher order terms) apart from the formed polaron,
  %{Polaron-to-molecule transition in a strongly imbalanced Fermi gas}
  %{diagrammatic monte carlo study of quasi-two-dimensional fermi polarons}
in the polaron ground state as a superpositions of the coherent many-body states.
However, such a coherence requires the quantum system isolated from the envionment,
like a isolated impurity interact with the thermal bath or the solid medium (see Ref.\cite{Diestler D J} and the references therein).
One another excellent experimental platform for the preparation of coherence is the cold atomic gases.
  %{Critical slowdown of non-equilibrium polaron dynamics}
Although the long-range coherence will destoryed by the temperature fluctuation\cite{Hadzibabic Z}, the short-range one can be observed even at room temperature\cite{Visuri A M}.
While in the extreme case which with a very light impurity immersed into a bath of heavy particles,
the decoherence process may happen\cite{Visuri A M}.
  %{decoherence of an impurity in a one-dimensional fermionic bath with mass imbalance}
Besides, 
for mobile impurity in a superfluid state (created by a degenerate Bose gas) 
moves slower than the speed of sound in condenstate which is about 1.33$\times 10^{-4}$ m/s
(the speed of sound in condensate is $\sim 10^{-3}$ m/s\cite{Andrews M R}),
or for the mobile impurity in a BEC which has been accelerated above the speed of sound\cite{Nielsen K K}
where the decoherence occur,
the Bose polaron can be formed.
In the latter case where the formation of polaron is usually related to the quenching dynamics
at zero-temperature limit, 
  %{Critical slowdown of non-equilibrium polaron dynamics}
since the interference patterns of the coherent states are destroyed and the superposition
turns into the statistical mixture, it can be represented by the statistical functions and the
overlap between the impurity with the majority component.
  %{Frequency-dependent polarizability, plasmons, and screening in the two-dimensional pseudospin-1 dice lattice}
In condensate, the speed of sound is proportional to the intraspecies coupling $g_{\phi\phi}$,
and thus inversely proportional to the healing length which reads $\ell=1/\sqrt{8\pi a_{\phi\phi}n}$ ($n$ is the cobdensate density).
  %{feynman path-integral treatment of the bec-impurity polaron}
  %{Critical slowdown of non-equilibrium polaron dynamics}
  %{Neutral impurities in a Bose-Einstein condensate for simulation of the Fr?hlich-polaron}
  %{the nonlinear dirac equation in bose-einstein condensates: foundation and symmetries}
The healing length here is comparable to the size of the screening cloud around the impurity (can be atom, electron, or ion).
   %{decoherence of an impurity in a one-dimensional fermionic bath with mass imbalance}
In the extreme case of strong intraspecies interaction with the leading instability (e.g., when close to the Feshbach resonance),
due to the vanishing healing length,
the dispersion of the Bosons (the phonon) in the presence of polaron and the effect of potential
could be still similar to the above-mentioned one $\sim sk$ ($s$ here is the speed of sound 
$s=\hbar/\sqrt{2m_{\phi}\ell}$ which becomes very large now;
   %{feynman path-integral treatment of the bec-impurity polaron}
   %{ground-state properties of interacting bose polarons}
like the Debye model in metal).
   %{Electron-phonon interactions from first principles}
For the optical lattice as a two-dimensional artificial solid state system with continuous symmetry,
the healing length is very small due to the harmonic transverse confinement,
and thus leads to a larger speed of sound than that in the quantum gases.
  %{The nonlinear Dirac equation in Bose–Einstein condensates: Foundation and symmetries}
And specially, for two-dimensional system,
the intraspecies atomic interaction decreases with the increasing oscillator length\cite{Merkl M}.
  %{Chiral Confnement in Quasirelativistic Bose-Einstein Condensates}
Compared to the graphene-like two-dimensional Dirac materials,
the ultracold atomic system in a hexagonal (or triangular) optical lattice provides a
more controllable and clean (without the impurity or defeat
unless added by the additional laser or magnetic field) 
  %{Transport of spin-orbit coupled Bose–Einstein condensates in lattice with defects}
platform to probe the polaron behavior as well as the self-trapping of the mobile impurity in equilibrium or unequilibrium cases\cite{Haddad L H,Kim N Y}.
  %{The nonlinear Dirac equation in Bose–Einstein condensates: Foundation and symmetries}

\section{Polaron model involving boson-impurity and electron-electron interactions}

For composite polaron state which
contains both the interactions between impurity and the induced particle-hole excitation, and that between impurity and phonon
(Peierls electron-phonon interaction\cite{Marchand D J J}),
  %{Light Bipolarons Stabilized by Peierls Electron-Phonon Coupling}
  %{Phonon-mediated repulsion, sharp transitions and (quasi)self-trapping in the extended Peierls-Hubbard model}
we can write the many-body Hamiltonian as 
%{Observation of Fermi Polarons in a Tunable Fermi Liquid of Ultracold Atoms}
  %{Decoherence of an impurity in a one-dimensional fermionic bath with mass imbalance}
  %{Feynman path-integral treatment of the BEC-impurity polaron}
  %{Phonons in a strongly coupled electron-phonon system}
  %{Dropping an impurity into a Chern insulator A polaron view on topological matter}
  %{Frohlich polaron and bipolaron recent developments}
  %{dynamical mean-field theory of transport of small polarons}
  %{Spectral functions and rf response of ultracold fermionic atoms}
  %{Fermi polaron-polaritons in charge-tunable atomically thin semiconductors}
  %{Photoemission spectrum and effect of inhomogeneous pairing fluctuations in the BCS-BEC crossover regime of an ultracold Fermi gas}
\begin{equation} 
\begin{aligned}
	\label{871}
H=&\sum_{k}\varepsilon_{k\uparrow}c_{k\uparrow}^{\dag}c_{k\uparrow}+\sum_{p}\varepsilon_{p\downarrow}c_{p\downarrow}^{\dag}c_{p\downarrow}
+\frac{1}{N}\sum_{k,p,q}g_{q}c_{p-q\downarrow}^{\dag}c_{k+q\uparrow}^{\dag}c_{k\uparrow}c_{p\downarrow}\\
&+\sum_{q'}\Omega_{ph}^{0}(q')b_{q'}^{\dag}b_{q'}
+\frac{1}{\sqrt{N}}\sum_{p,k,q',\sigma}|g_{k+q',k}|
c^{\dag}_{p+q',\sigma}c_{p,\sigma}(b^{\dag}_{-q'}+b_{q'}),
\end{aligned}
\end{equation}
where $N=S/s_{0}$ is the total number of the unit cell where $S$ is the total area and $s_{0}$ is the area each unit cell.
$g_{q}^{-1}=-\sum_{k}[E_{b}+\varepsilon_{k\uparrow}+\varepsilon_{k\downarrow}+W]^{-1}$.
Here we have transform the position representation into momentum basis through the unitary transformation,
  %{Dropping an impurity into a Chern insulator A polaron view on topological matter} 
and use the single Einstein mode of the phonon without the second-quantization.
  %{Acoustic phonon quantization in buried waveguides and resonators}
  %{Light Bipolarons Stabilized by Peierls Electron-Phonon Coupling}
 The first term shows the contribution of the bath particles,
which can, in the absence of spin flipping, includes the spin degree of freedom 
by simply times a factor of 2.
Then the ground state wave function reads (here consider the one particle-hole pair and one phonon limit)
\begin{equation} 
\begin{aligned}
|\psi\rangle_{FP}=&(\psi_{0}c^{\dag}_{p\downarrow}+\sum_{k>k_{F},q<k_{F}}\psi_{kq}c_{p+q-k\downarrow}^{\dag}c_{k\uparrow}^{\dag}c_{q\uparrow})
|0\rangle_{\uparrow},\\
|\psi\rangle_{CP}=&(\phi'_{0}c^{\dag}_{p\downarrow}+\sum_{k>k_{F},q<k_{F}}\phi'_{kq}c_{p+q-k\downarrow}^{\dag}c_{k\uparrow}^{\dag}c_{q\uparrow}
+
\sum_{|q'|\le q_{D}}\phi_{q'}
b_{-q'}^{\dag}c_{p-(-q')\downarrow}^{\dag}\\
&+
\sum_{k>k_{F},q<k_{F},|q'|\le q_{D}}\phi_{kqq'}
c_{p+q-k-(-q')\downarrow}^{\dag}c_{k\uparrow}^{\dag}c_{q\uparrow}b_{-q'}^{\dag})|0\rangle
,
\end{aligned}
\end{equation}
where $|\psi\rangle_{FP}$ and $|\psi\rangle_{CP}$ denote the ground state wave function of the fermi polaron and composite polaron.
Here $q_{D}$ denotes the Debye radiu.
  %{Many-body effects in the normal-state polaron system}
For the wave function $|\psi\rangle_{CP}$,
we ignore the coupling between the phonons and the electron-hole pair (with total momentum $k-q$)
since we assume the binding energy within the electron-hole pair (especially for the light-excited one) 
is much larger than that between impurity and the phonons,
that is
in consideration of the large exciton binding energy which could be as large as 0.3 eV in the
$\pi$-conjugated system\cite{Knupfer M} or
0.5 eV in the semiconductor\cite{Muntwiler M},
   %{fermi polaron-polaritons in charge-tunable atomically thin semiconductors}
   %{Size of Electron-Hole Pairs in p-Conjugated Systems}
  %{From two-dimensional electron gas to localized charge Dynamics of polaron formation in organic semiconductors}
while that of the electron-phonon polaron is only 0.3 meV in a continuum model\cite{Verissimo-Alves M},
  %{Electron-Phonon Interaction and Transport in Semiconducting Carbon Nanotubes}
or 70 meV when consider the optical phonon\cite{Perebeinos V}.
   %{Polaron-phonon interaction in a finite-size lattice A perturbative approach}
For some other materials like the MoS$_{2}$,
several types of excitons can be found\cite{Wu F},
and the binding energy is dependent of the separation between atomic sites and the dielectric constant.
  %{Exciton band structure of monolayer MoS2}
We note that such binding energy can be reduced by the screening from fermi liquid or the band gap effect.
For the composite polaron,
by minimizing $\langle \psi|E-H|\psi\rangle$,
we have (considering the case that emits a phonon)
  %{Fermi polaron-polaritons in charge-tunable atomically thin semiconductors}
  %{Repulsive Fermi Polarons in a Resonant Mixture of Ultracold 6Li Atoms}
\begin{equation} 
\begin{aligned}
\langle \psi_{CP}|E_{\downarrow}-H|\psi\rangle_{CP}=&
E_{\downarrow}(|\phi'_{0}|^{2}+\sum_{kq}|\phi_{kq}|^{2}+\sum_{q'}|\phi_{q'}|^{2}+\sum_{kqq'}|\phi_{kqq'}|^{2})\\
&  -[|\phi'_{0}|^{2}\varepsilon_{p\downarrow}+\sum_{kq}(\varepsilon_{p+q-k\downarrow}+\varepsilon_{k\uparrow}-\varepsilon_{q\downarrow})|\phi_{kq}|^{2}\\
&    +\sum_{q'}(\varepsilon_{p-(-q')\downarrow}+\varepsilon_{-q'})|\phi_{q'}|^{2}
    +\sum_{kqq'}(\varepsilon_{p+q-k-(-q')}+\varepsilon_{k\uparrow}-\varepsilon_{q\uparrow}+\Omega_{ph}^{0}(q'))|\phi_{kqq'}|^{2}\\
&    +|\phi'_{0}|^{2}\sum_{q}g_{q}
    +\sum_{kq}(\phi^{'*}_{0}\phi_{kq}g_{k-q}+c.c.)
    +\sum_{q'}(\phi^{'*}_{0}\phi_{q'}g_{-q'}+c.c.)\\
&    +\sum_{kqq'}(\phi^{'*}_{0}\phi_{kqq'}g_{k-q+(-q')}+c.c.)\\
&    +\sum_{kqq'}(\phi^{'*}_{kq}\phi_{q'}g_{k-q+(-q')}+c.c.)
    +\sum_{kqq'}(\phi^{'*}_{kq}\phi_{kqq'}g_{q'}+c.c.)
    +\sum_{kqq'}(\phi^{'*}_{q'}\phi_{kqq'}g_{k-q}+c.c.)
].
\end{aligned}
\end{equation}
Here we ignore the change of the momentum of electron-hole pair.
Then for $\frac{\partial \langle\psi_{CP}|H|\psi_{CP}\rangle}{\partial \phi}=0$,
we write
\begin{equation} 
\begin{aligned}
\varepsilon_{p\downarrow}\phi_{0}+\phi_{0}\sum_{q}g_{q}+\sum_{kq}\phi_{kq}g_{k-q}+\sum_{q'}\phi_{q'}g_{-q'}+\sum_{kqq'}\phi_{kqq'}g_{k-q+(-q')}=E\phi_{0},\\
(\varepsilon_{p+q-k\downarrow}+\varepsilon_{k\uparrow}-\varepsilon_{q\downarrow})\phi_{kq}+\phi_{0}g_{k-q}+\sum_{kqq'}\phi_{q'}g_{k-q+(-q')}+\sum_{kqq'}\phi_{kqq'}g_{-q'}=E\phi_{kq},\\
(\varepsilon_{p-(-q')\downarrow}+\varepsilon_{-q'})\phi_{q'}+\phi_{0}g_{-q'}+\sum_{kqq'}\phi_{kq}g_{k-q+(-q')}+\sum_{kqq'}\phi_{kqq'}g_{k-q}=E\phi_{q'},\\
(\varepsilon_{p+q-k-(-q')}+\varepsilon_{k\uparrow}-\varepsilon_{q\uparrow}+\Omega_{ph}^{0}(q'))\phi_{kqq'}
+\phi_{0}g_{k-q+(-q')}+\sum_{kqq'}\phi_{kq}g_{-q'}+\sum_{kqq'}\phi_{q'}g_{k-q}
=E\phi_{kqq'},
\end{aligned}
\end{equation}
The variational paramters are obtained by the minimization as
\begin{equation} 
\begin{aligned}
\phi_{0}=&\frac{\frac{g}{N}\sum_{q}\chi_{q}}{E_{\downarrow}-\varepsilon_{p\downarrow}},\\
\phi_{kq}=&\frac{\frac{g}{N}\chi_{q}}
           {E_{\downarrow}-(\varepsilon_{p+k-q\downarrow}+\varepsilon_{k\uparrow}-\varepsilon_{q\uparrow})},\\
\phi_{q'}=&\frac{\frac{g}{N}\sum_{q}\chi_{q}}
           {E_{\downarrow}-(\varepsilon_{p-(-q')\downarrow}+\Omega_{ph}^{0}(-q'))},\\
\phi_{kqq'}=&\frac{\frac{g}{N}\chi_{q}}
           {E_{\downarrow}-(\varepsilon_{p+k-q-(-q')\downarrow}+\varepsilon_{k\uparrow}-\varepsilon_{q\uparrow}+\Omega_{ph}^{0}(-q'))},\\
\chi_{q}=&\psi_{0}+\sum_{k}\psi_{kq}+\sum_{q'}\psi_{q'}+\sum_{kq'}\psi_{kqq'},\\
\end{aligned}
\end{equation}
where the self-consistent energy term $(E-\varepsilon_{p\downarrow})$ reads
\begin{equation} 
\begin{aligned}
E-\varepsilon_{p\downarrow}=&\frac{1}{N}\sum_{q<k_{F}}\left[g^{-1}-\frac{1}{N}\sum_{k=k_{F}}^{\Lambda}\frac{1}{E+i0
-(\varepsilon_{p+k-q-(-q')\downarrow}+\varepsilon_{k\uparrow}-\varepsilon_{q\uparrow}+\Omega_{ph}^{0}(-q'))}
\right]^{-1},
\end{aligned}
\end{equation}
  %{Bose polarons near quantum criticality}
  %{Fermi polaron-polaritons in charge-tunable atomically thin semiconductors}
  %{Highly polarized Fermi gases in two dimensions}
which equals the binding energy when $p=0$.
Note that in effective mass approximation we 
also have ${\rm Re}[\Sigma_{e}(\omega,p)+\Pi_{e}(p)]=E-\varepsilon_{p\downarrow}$.
%{Observation of Fermi Polarons in a Tunable Fermi Liquid of Ultracold Atoms}
  %{Observation of repulsive Fermi polarons in a resonant mixture of ultracold 6 Li atoms supp}
  %{Repulsive Fermi Polarons in a Resonant Mixture of Ultracold 6Li Atoms}
  %{Field-theoretical study of the Bose polaron}
  %{Bose polarons near quantum criticality}
  %{Molecule and Polaron in a Highly Polarized Two-Dimensional Fermi Gas with Spin-Orbit Coupling}
For small electron chemical potential, we define
\begin{equation} 
\begin{aligned}
g^{-1}(\Lambda)=&-\frac{1}{N}\sum_{k,q'}^{\Lambda}\frac{1}{E_{b}+\varepsilon_{k\uparrow}+\varepsilon_{k\downarrow}+\Omega_{ph}^{0}(-q')+W}.
\end{aligned}
\end{equation}
Unlike the renormalized coupling parameter in Eq.(6),
we can see that this coupling parameter depends on both the binding-energy $E_{b}$ and momentum (ultraviolet) cutoff $\Lambda$.
  %{Fermi polaron-polaritons in charge-tunable atomically thin semiconductors}
  %{Dimensionality-induced BCS-BEC crossover in layered superconductors}
  %{Spectral functions and rf response of ultracold fermionic atoms}
  %{Field-theoretical study of the Bose polaron}
and the logarithmically diverges of $g^{-1}(\Lambda)=-\infty$ can be seen
($g(\Lambda)\rightarrow 0^{-}$) when $\Lambda\rightarrow\infty$ (where the quantum correction vanishes).
   %{Highly polarized Fermi gases in two dimensions}
The normalization condition is ${}_{CP}\langle \psi|\psi\rangle_{CP}=
|\phi_{0}|^{2}+\sum_{k>k_{F},q<k_{F}}|\phi_{kq}|^{2}+\sum_{q'\le q_{D}}|\phi_{q'}|^{2}+
\sum_{k>k_{F},q<k_{F},|q'|\le q_{D}}|\phi_{kqq'}|^{2}=1$.
Note that such a composite polaron cannot found in a superfluid bath since in this case the propagating impurity will not excites the electron-hole
excitation.
%For the first wave function (Fermi polaron),
%we have
%\begin{equation} 
%\begin{aligned}
%\psi_{kq}=\frac{g_{q}}{N}\frac{\psi_{0}+\sum_{k}\psi_{kq}}{E-(\varepsilon_{p+q-k\downarrow}+\varepsilon_{k\uparrow}-\varepsilon_{q\uparrow})},
%\end{aligned}
%\end{equation}
The binding energy $E_{b}$ here is positive for attractive potential, i.e., the fermi polaron here
with negative $g_{q}$,
thus the energy of such a bound state must be negative $-E_{b}<0$,
  %{Fermi polaron-polaritons in charge-tunable atomically thin semiconductors}
  %{Highly polarized Fermi gases in two dimensions}
  %{Dimensionality-induced BCS-BEC crossover in layered superconductors}
  %{Polarons and dressed molecules near narrow Feshbach resonances}
  %{Electron-Phonon Interaction and Transport in Semiconducting Carbon Nanotubes}
which corresponds to the negative self-energy of the attractive fermi polaron.

For an analysis, we approximately suppose that there only exist two possibilities: isolate impurity or
interaction with both the electron-hole pair and the phonons.
Then the variational wave function reads (here consider the one particle-hole pair and multi-phonon case)
\begin{equation} 
\begin{aligned}
|\psi\rangle_{CP}=&(\phi'_{0}c^{\dag}_{p\downarrow}
+\sqrt{N_{ph}}
\sum_{k>k_{F},q<k_{F},|q'|\le q_{D}}\phi_{kqq'}
c_{p+q-k-(-q')\downarrow}^{\dag}c_{k\uparrow}^{\dag}c_{q\uparrow}b_{-q'}^{\dag})|0\rangle
,
\end{aligned}
\end{equation}
where the factor $\sqrt{N_{ph}}$ resulted by the $b_{-q'}$ acting on the unperturbed state $|0\rangle$.
  %{Bose polarons near quantum criticality}
Unlike the anharmonic phonon modes induced by the high intensity laser\cite{Zijlstra E S,Cinquanta E,Carneiro L M}
  %{Phonon Anharmonicities in Graphite and Graphene}
%{Quasiparticle Properties of a Mobile Impurity in a Bose-Einstein Condensate}
%{Landau Effective Interaction between Quasiparticles in a Bose-Einstein Condensate}
  %{Laser-induced phonon-phonon interactions in bismuth}
  %{Ultrafast THz Probe of Photoinduced Polarons in Lead-Halide Perovskites}
where the interphonons couping can not be ignored,
we ignore the interphonons couping here as the modes are created by the self-induced lattice polarization.
  %{Enhanced Superconducting State in FeSe=SrTiO3 by a Dynamic Interfacial Polaron Mechanism}
The residue $Z$ will not affected largely by this approximation since the penetration length (coherence length) 
of the 2D system here (e.g., the tight-binding model) or the optical lattice which with deep trapping potential,
are much shorter\cite{Kopnin N B} than the ideal BEC\cite{Christensen R S}.
  %{BCS Superconductivity of Dirac Electrons in Graphene Layers}
  %{Quasiparticle Properties of a Mobile Impurity in a Bose-Einstein Condensate}
The variational parameters are obtained by the minimization as
\begin{equation} 
\begin{aligned}
\phi_{0}=&\frac{\frac{gN_{ph}}{N}\sum_{q}\chi_{q}}{E_{\downarrow}-\varepsilon_{p\downarrow}}
=\frac{gn_{ph}\sum_{q}\chi_{q}}{E_{\downarrow}-\varepsilon_{p\downarrow}},\\
\phi_{kqq'}=&\frac{\frac{g}{N}\chi_{q}}
           {E_{\downarrow}-(\varepsilon_{p+k-q-(-q')\downarrow}+\varepsilon_{k\uparrow}-\varepsilon_{q\uparrow}+\Omega_{ph}^{0}(-q'))},\\
\chi_{q}=&\psi_{0}+\sum_{kq'}\psi_{kqq'},\\
\end{aligned}
\end{equation}
where $n_{ph}$ denotes the phonon density.
The normalization condition is ${}_{CP}\langle \psi|\psi\rangle_{CP}=
|\phi_{0}|^{2}+N_{ph}
\sum_{k>k_{F},q<k_{F},|q'|\le q_{D}}|\phi_{kqq'}|^{2}=1$.
Here the momentum distribution of polaron is shifted from the $\delta_{p}$-function
by the electron-phonon coupling term $|\phi_{kqq'}|^{2}$.
    %{Bose polarons near quantum criticality}
Through the above procedure,
we can obtain
  %{Bose polarons near quantum criticality}
  %{Highly polarized Fermi gases in two dimensions}
  %{Fermi polaron-polaritons in charge-tunable atomically thin semiconductors}
\begin{equation} 
\begin{aligned}
\frac{\phi_{kqq'}}{\phi_{0}}=&\frac{g\frac{1}{N}\chi_{q}(E-\varepsilon_{p\downarrow})}{g\frac{N_{ph}}{N}\sum_{q}\chi_{q}
(E-(\varepsilon_{p+k-q-(-q')\downarrow}+\varepsilon_{k\uparrow}-\varepsilon_{q\uparrow}+\Omega_{ph}^{0}(-q')))}\\
=&\frac{1}{N_{ph}N}\frac{E-\varepsilon_{p\downarrow}}
{E-(\varepsilon_{p+k-q-(-q')\downarrow}+\varepsilon_{k\uparrow}-\varepsilon_{q\uparrow}+\Omega_{ph}^{0}(-q')))},
\end{aligned}
\end{equation}
where the energy is
\begin{equation} 
\begin{aligned}
E-\varepsilon_{p\downarrow}=&\frac{1}{N}\sum_{q<k_{F}}\left[g^{-1}-\frac{1}{N}\sum_{k=k_{F}}^{\Lambda}\frac{1}{E+i0
-(\varepsilon_{p+k-q-(-q')\downarrow}+\varepsilon_{k\uparrow}-\varepsilon_{q\uparrow}+N_{ph}\Omega_{ph}^{0}(-q'))}
\right]^{-1},\\
=&\frac{1}{N}\sum_{q<k_{F}}\left[
-\frac{1}{N}\sum_{k,q'}^{\Lambda}\frac{1}{E_{b}+\varepsilon_{k\uparrow}+\varepsilon_{k\downarrow}+N_{ph}\Omega_{ph}^{0}(-q')+W}\right.\\
&\left.-\frac{1}{N}\sum_{k=k_{F},q'}^{\Lambda}\frac{1}{E+i0
-(\varepsilon_{p+k-q-(-q')\downarrow}+\varepsilon_{k\uparrow}-\varepsilon_{q\uparrow}+N_{ph}\Omega_{ph}^{0}(-q'))}
\right]^{-1}.
\end{aligned}
\end{equation}
  %{Bose polarons near quantum criticality}
  %{Fermi polaron-polaritons in charge-tunable atomically thin semiconductors}
  %{Highly polarized Fermi gases in two dimensions}
Note that for the majority component,
the eigenenergy should contains the chemical potential,
e.g., in above equation we have $\varepsilon_{k\uparrow}=\frac{k^{2}}{2m_{\uparrow}}-\mu_{\uparrow}$
and $\varepsilon_{q\uparrow}=\frac{q^{2}}{2m_{\uparrow}}+\mu_{\uparrow}$,
i.e., the pair energy is measured from $2\mu_{\uparrow}$.
  %{Bose Condensation in an Attractive Fermion Gas: From Weak to Strong Coupling Superconductivity}

\section{Electron-phonon coupling}

\subsection{Polaronic effect induced by electron-phonon coupling (boson-impurity interaction)}

The transition of the quasifree polaron (large size) to the self-trapped polaron (small size)
could be
to realized for solid state system in strong coupling region.
  %{Feynman path-integral treatment of the BEC-impurity polaron}
Similar to the Fr{\"o}hlich Hamiltonian\cite{Mahan G D},
  %{new theoretical approaches to bose polarons}
  %{Feynman path-integral treatment of the BEC-impurity polaron}
in solid state, the electron-phonon coupling can be described by the approximation up to first order of the atomic displacements.
  %(like the phonon displacement in momentum space ${\bf q}'$),
  %{Electron-phonon interactions from first principles}
  %{electron-phonon coupling in two-dimensional silicene and germanene}
%and since in most cases, the electron-phonon coupling strength is much smaller than the electron (impurity) hopping 
%(or tunneling energy),
  %{decoherence of an impurity in a one-dimensional fermionic bath with mass imbalance}
%i.e., $\varepsilon_{e-ph}/t\ll 1$ where $t$ denotes the hopping strength,
We at first discuss the weak coupling adiabatic case, where the 
Migdal-Eliashberg theory is applicable,
  %{Polaron-phonon interaction in a finite-size lattice: A perturbative approach}
  %{ Polarons and bipolarons in strongly interacting electron-phonon systems}
 then the (single species) self-energy of the phonon (no matter transverse or longitudinal modes)
  %{From electron to small polaron: an exact cluster solution}
  %{electron-phonon coupling in two-dimensional silicene and germanene}
  %{Electron-phonon interactions from first principles}
  %{First-principles analysis of electron-phonon interactions in graphene}
  %{Universal infrared absorbance of two-dimensional honeycomb group-IV crystals}
  %{Anomaly of Optical Phonons in Bilayer Graphene}
  %{Many-body effects in the normal-state polaron system}
  %{vertex corrections to the dc conductivity in anisotropic multiband systems}
  %{Electron-phonon interaction using Wannier functions}
can be obtained by the summation of the electronic states (excited by the electron-phonon interaction) with different crystal momenta during the transition
% (including the virtual processes)
\begin{equation} 
\begin{aligned}
\Sigma_{ph}(q',\Omega_{ph})=\int\frac{d^{2}k}{(2\pi)^{2}}|g_{p,p+q'}|^{2}\frac{N_{F}(\varepsilon_{k+q'})-N_{F}(\varepsilon_{k})}
{\varepsilon_{k+q'}-\varepsilon_{q'}-\Omega_{ph}-i0},
\end{aligned}
\end{equation}
where $g_{p,p+q'}$ is the well known electron-phonon coupling matrix element\cite{Li X,Matthes L,Borysenko K M,Yan J A,Yan J A2}
  %{Electrons and Phonons - J. M. Ziman - Oxford University Press}
which is inversely proportional to the atomic mass and the bare phonon frequency $\Omega^{0}_{ph}(q')$, 
just similar to the descriptions in collective Holstein approximation.
     %{polarons and bipolarons in strongly interacting electron-phonon system}
  %{Instrinsic electrical transport properties of monolayer silicene and mos2 from first principle}
  %{First-principles analysis of electron-phonon interactions in graphene}
  %{universal infrared absorbance of two-dimensional honeycomb group-IV crystals}
  %{First-principles analysis of electron-phonon interactions in graphene}
  %{electron-phonon coupling in two-dimensional silicene and germanene}
  %{electron-phonon interactions for optical-phonon modes in few-layer graphene: first-principles calculations}
We have in perturbational treatment
\begin{equation} 
\begin{aligned}
g_{p,p+q'}=\sqrt{\frac{\hbar}{2M\Omega_{ph}^{0}}}\langle k+q'|\frac{\delta U}{\delta u_{q}}|k\rangle,
\end{aligned}
\end{equation}
where $M$ is the atomic mass of the lattice and $\delta U$ is the variation of the self-consistent Kohn-Sham potential.
$\delta u_{q}$ denotes the tight-binding amplitude.
Here the change of quantum nunbers are not shown which make the self-energy nonzero even in the limit $q'\rightarrow 0$.
  %{anomaly of optical phonons in bilayer graphene}
We can see that this self-energy is just the polarization loop (particle-hole bubble)
with the crossing phonon vertex correction.
Note that in Migdal approximation, the effect of phonon vertex correction is small and even negligible\cite{Alexandrov A S}
  %{breakdown of the Migdal-Eliashberg theory in the strong-coupling adiabatic regime}
due to the weak electron-phonon coupling strength (compared to the electron tunneling),
$\varepsilon_{e-ph}(\sim |g_{p,p+q'}|^{2})\ll 3t\Omega_{ph}$ ($3t$ here is the approximated 
%$|\varepsilon_{e-ph}|(\sim (\frac{3t}{a})^{2}|g_{p,p+q'}|^{2})\ll 3t\Omega_{ph}$ ($3t$ here is the approximated 
%{dynamical mean-field theory of transport of small polarons}
%{http}
%{Periodically driven small polarons P. E. Kornilovitch}
   %{Sharp Transition for Single Polarons in the One-Dimensional Su-Schrieffer-Heeger Model}
   %{Bipolarons bound by repulsive phonon-mediated interactions}
bare half bandwidth of a two-dimensional honeycomb lattice system, which is 8.4 eV in the intrinsic graphene;
for large polaron, the half bandwidth can also be written as $\frac{3}{2ma^{2}}$\cite{Devreese J T}),
%{Many-body effects in the normal-state polaron system}
%{Phonons in a strongly coupled electron-phonon system}
%{Fr?hlich polaron and bipolaron recent developments}
  %{Dynamical mean-field theory of transport of small polarons}
  %{Periodically driven small polarons}
  %{Inhomogeneous dynamical mean-field theory of the small polaron problem}
in which case the effective mass of electron ($m^{*}=m(1+\frac{\varepsilon_{e-ph}}{3t\Omega^{0}_{ph}})$) is closes to the rest one.
 %(while in the strong coupling adiabatic case, $\frac{\varepsilon_{e-ph}}{ t}\geq \frac{t}{\Omega^{0}_{ph}(q')}\gg 1$,
 %the effective mass is much larger than the rest one).
  %{Many-body effects in the normal-state polaron system}
Note that the above phonon self-energy induced by the propagating electron
is valid only within the Migdal-Eliashberg approximation, 
i.e., the weakly coupled adiabatic limit $\frac{\varepsilon_{e-ph}}{3t\Omega^{0}_{ph}}\ll 1,\ \frac{t}{\Omega^{0}_{ph}(q')}\ll 1$.
(In the insulator phase with large electron-phonon couping, the Migdal-Eliashberg approximation breaks down)
  %{Breakdown of the Migdal-Eliashberg theory in the strong-coupling adiabatic regime}
  %{From electron to small polaron: an exact cluster solution}
The real part of above phonon self-energy provides the phonon frequency shift while its imaginary part provides
the phonon linewidth (or the population decay\cite{Kohstall C}) since it does not contains the phonon-related vertex correction.
  %{electron-phonon coupling in two-dimensional silicene and germanene}
  %{vertex corrections to the dc conductivity in anisotropic multiband systems}
  %{Anomaly of Optical Phonons in Bilayer Graphene}
  %{Electron-phonon interactions for optical-phonon modes in few-layer graphene: First-principles calculations}
The phonon linewidth is proportional to the electron-phonon coupling matrix element
and inversely proportional to the fermi velocity of electron,
i.e., the self-energy will induces a higher linewidth in nonadiabatic case,
  %{electron-phonon coupling in two-dimensional silicene and germanene}
in other word, the linewidth increases with the increase of doping level.
  %{Doped graphene as tunable electron? phonon coupling material}
Note that here the contributions to linewidth from the anharmonic term (inter-phonon interaction)
and the interaction between phonons and the electron-hole pairs are neglected.
  %{phonon linewidths and electron-phonon coupling in graphite and nanotubes}

As diagrammatically shown in the Fig.1, the boson self-energy loop (phonon)
obtained by many-body diagrammatic method
is comprised of two fermion propagators (the impurity) and two boson propagators (with external momentum and external frequency) where we consider the current-current correlation here.
The phonon vertex correction reads
\begin{equation} 
\begin{aligned}
\sigma_{j}\Gamma_{ph}=\sigma_{j}-\sum_{q',\Omega_{ph}}
TD_{ph}(q',\Omega_{ph})|g_{k,k+q}|^{2}G_{0}(k+q',\omega+\Omega_{ph})\sigma'_{j}\Gamma'_{ph}G_{0}(k+q+q',\omega+\Omega+\Omega_{ph}),
\end{aligned}
\end{equation}
where $\omega$ is the fermion frequency and $\Omega/\Omega_{ph}$ is the bosonic frequency.
Here the Pauli matrices $\sigma$ can also be replaced by the Dirac $\gamma$ matrices\cite{Gusynin V P,Pyatkovskiy P K},
which then requires a new form of Dirac quasiparticle propagator.
In the weak-coupling region,
$D_{ph}^{0}(q',\Omega_{ph})=\frac{2\Omega^{0}_{ph}(q')}{\Omega_{ph}^{2}-\Omega^{0}_{ph}(q')}$
is the bare phonon propagator,
which should be replaced by the interacting one when the electron-phonon coupling is strong
(and thus with a larger phonon self-energy):
\begin{equation} 
\begin{aligned}
D_{ph}(q',\Omega_{ph})
=\frac{2\widetilde{\Omega}^{0}_{ph}(q')}{\Omega_{ph}^{2}-\widetilde{\Omega}^{0}_{ph}(q')-2\widetilde{\Omega}^{0}_{ph}(q')\Sigma_{ph}(q',\Omega_{ph})}
\end{aligned}
\end{equation}
where $\widetilde{\Omega}^{0}_{ph}=\Omega^{0}_{ph}(q')(1-2\frac{\varepsilon_{e-ph}}{t})$ is the renormalized phonon frequency.
%\cite{}.
  %{Breakdown of the Migdal-Eliashberg theory in the strong-coupling adiabatic regime}
  %{phonons in a strongly coupled electron-phonon system}
  %{vertex corrections to the dc conductivity in anisotropic multiband systems}

For gapless band structure, 
the states could be adiabatically pass through the massless Dirac point no matter in Dirac material or the artificial honeycomb optical lattice
\cite{Chen Z},
  %{bose-einstein condensate in a honeycomb optical lattice: fingerprint of superfluidity at the Dirac point}
  %{Electron-phonon interactions for optical-phonon modes in few-layer graphene: First-principles calculations}
while for the case that the phonon energies are smaller than the band gap, the 
interband contributions from the excited electrons vanish,
  %{https://www.phas.ubc.ca/~berciu/TEACHING/PHYS502/NOTES/elph.pdf}
  %{Electron-phonon interactions - UBC Physics & Astronomy}
and the adiabaticity is also broken due to the crossing of the lowest bands
or the large separation (larger than the phonon energy) of bands.
The electron-phonon coupling matrix element can be rewritten as (for optical phonon scattering)
%{first principles analysis of electron-phonon interactions in graphene}
   %{Electron-phonon interactions for optical-phonon modes in few-layer graphene: First-principles calculations}
   %{Mahan G D. Electron–optical phonon interaction in carbon nanotubes[J]. Physical Review B, 2003, 68(12): 125409.}
\begin{equation} 
\begin{aligned}
%g_{p,p+q'}=&g_{p,p+q'}^{0}[u^{*}_{k+q',s}u'_{k,s'}\frac{1}{\varepsilon_{k}}
%           ({\bf t}_{k}\cdot {\pmb \phi}_{q'}-{\bf t}_{k+q'}\cdot{\pmb \phi}'_{q'})
%+u^{*'}_{k+q',s}u_{k,s'}\frac{1}{\varepsilon_{k}}
%           (-{\bf t}_{-k}\cdot {\pmb \phi}'_{q'}+{\bf t}_{-k-q'}\cdot{\pmb \phi}_{q'})],
g_{p,p+q'}=&g_{p,p+q'}^{0}[u^{*}_{p+q',s}u'_{p,s'}\frac{1}{\varepsilon_{p}}
           ({\bf t}_{p}\cdot {\pmb \phi}_{q'}-{\bf t}_{p+q'}\cdot{\pmb \phi}'_{q'})
+u^{*'}_{p+q',s}u_{p,s'}\frac{1}{\varepsilon_{p}}
           (-{\bf t}_{-p}\cdot {\pmb \phi}'_{q'}+{\bf t}_{-p-q'}\cdot{\pmb \phi}_{q'})],
\end{aligned}
\end{equation}
with
\begin{equation} 
\begin{aligned}
g_{p,p+q'}^{0}=
\frac{\varepsilon_{e-ph}}{\Omega_{ph}^{0}}\sqrt{\frac{s_{0}}{M}}
\approx &\frac{6\hbar v_{F}}{a^{2}\sqrt{2\frac{2M}{s_{0}}\Omega_{ph}^{0}S}}
=\frac{3\sqrt{3}t}{a\sqrt{4MN\Omega_{ph}^{0}}},
\end{aligned}
\end{equation}
where $N=S/s_{0}$ is the number of unit cell,
 and $2M$ is the atomic mass per unit cell.
   %{electronic cooling in graphene}
For optical phonon, the $\Omega_{ph}^{0}$ in the denominator is the optical phonon
frequency in the BZ center ($\Gamma$-point) which is
independent of the dispersion.
In graphene, the $g_{p,p+q'}^{0}$ is estimated as $5.3$ eV/$\AA$\cite{Perebeinos V}.
This expression is obviously a tight-binding treament which is in the basis of tight-binding amplitudes $u$,
  %{Electron-phonon interactions for optical-phonon modes in few-layer graphene: First-principles calculations}
  %{d+id'  Chiral Superconductivity in Bilayer Silicene}
where ${\bf t}_{k}$ is the hopping vector of the electron and ${\bf \phi}$ is the phonon eigenvector 
(vibrational modes) whose direction is related to the vibrational direction.
  %{Dynamic disorder, phonon lifetimes, and the assignment of modes to the vibrational spectra of methylammonium lead halide perovskites}
  %{Vibrational spectra and phonon dispersion analysis of a single-walled zigzag carbon nanotube: A first principles study}
$u$ and $u'$ correspond to the two basis vectors.
$s$ and $s'$ are the band indices.
We can see that when the nearest-neighbor hopping direction is orthogonal to vibrational direction, the electron-phonon interaction vanishes.
When the polaron-phonon interaction is ignored (like the closed channel),
the hopping has a simple expression 
\begin{equation} 
\begin{aligned}
t_{k}&=\int d{\bf r}w^{*}({\bf r}-{\bf R})H_{0}w({\bf r}-{\bf R})\\
&=\frac{1}{2N}\sum_{{\bf R},{\pmb \delta}}\varepsilon_{k}{\pmb \delta}e^{i{\bf k}\cdot{\bf R}}
\end{aligned}
\end{equation}
  %{Wannier states and Bose–Hubbard parameters for 2D optical lattices}
  %{Phonons of graphene and graphitic materials derived from the empirical potential LCBOPII}
where $2N$ is the site number here since we assume the primitive unit cell which contains two sites,
  %{from electron to small polaron: an exact cluster solution}
${\pmb \delta}$ is the unit vector of the nearest-neighbor hopping,
${\bf R}$ is the lattice vector of the primitive unit cell,
$H_{0}$ is the noninteracting Hamiltonian,
$w({\bf r}-{\bf R})$ is the Wannier function which containing the Bloch eigenstates (for linear Schrodinger equation)\cite{Blakie P B,Wu B2}.
  %{the nonlinear dirac equation in bose-einstein condensates: foundation and symmetries}
  %{Phonons of graphene and graphitic materials derived from the empirical potential LCBOPII}
When the polaron-phonon interaction is taken into account,
% (like the open channel),
the self-energy of an impurity electron induced by the electron-phonon interaction is
  %{Electron-phonon interactions from first principles}
  %{Electron-Phonon Interaction and Transport in Semiconducting Carbon Nanotubes}
  %{Electron-phonon interaction using Wannier functions}
\begin{equation} 
\begin{aligned}
\Pi_{e}(p)=&\int\frac{d^{2}q'}{(2\pi)^{2}}|g_{k,k+q'}|^{2}
D_{ph}(q',\Omega_{ph})G_{0}(p+q',\omega+\Omega_{ph})\\
=&\int\frac{d^{2}q'}{(2\pi)^{2}}|g_{p,p+q'}|^{2}
(\frac{1+N_{B}(\Omega^{0}_{ph}(q'))}{\varepsilon_{p-q'}-\varepsilon_{p}+\Omega^{0}_{ph}(q')+i0}
 +\frac{N_{B}(\Omega^{0}_{ph}(q'))}{\varepsilon_{p+q'}-\varepsilon_{p}-\Omega^{0}_{ph}(q')+i0}),
\end{aligned}
\end{equation}
where the analytical continuation $i\Omega\rightarrow \Omega+i\eta^{+}$ is used\cite{Kim S,Pyatkovskiy P K}.
  %{Field-theoretical study of the Bose polaron}
  %{Dynamical polarization, screening, and plasmons in gapped graphene}
  %{Vertex corrections to the dc conductivity in anisotropic multiband systems}
  %{quasiparticles properties of a mobile impurity in a bose-einstein condensate}
Note that the signs of imaginary part of self-energy and the spectral function (positive or negative) are
dependent on the sign of infinite small quantity $\eta$.
The hopping in real space needs to containing the effects of the emission or absorption of phonons,
and becomes
$t'_{ij}=t_{ij}{\rm exp}[\sum_{q'}\frac{|g_{p,p+q'}|}{\sqrt{N}}(c_{i}^{\dag}c_{i}d_{q'}-c_{j}^{\dag}c_{j}d_{q'})-h.c.]$,
which contains the effect of electron-phonon Fr{\"o}hlich interaction.
  %{phonons in a strongly coupled electron-phonon system}
as described by the electron scattering (inelastic) rate,
which can be obtained by the imaginary part of electron self-energy
%{Electron transport and hot phonons in carbon nanotubes}
%{Phonon linewidths and electron-phonon coupling in graphite and nanotubes}
%{Electron-phonon interaction using Wannier functions}
%{anomalous dirac plasmons in 1D topological electrides}
  %{Vertex corrections to the dc conductivity in anisotropic multiband systems}
  %{Instrinsic electrical transport properties of monolayer silicene and mos2 from first principle} 
  %{Electron-phonon interactions for optical-phonon modes in few-layer graphene: First-principles calculations}
  %{First-principles analysis of electron-phonon interactions in graphene}
  %{electron-phonon interaction and transport in semiconducting carbon nanotubes}
  %{Electron-phonon interaction using Wannier functions}
\begin{equation} 
\begin{aligned}
\frac{1}{\tau_{e}}=\frac{2\pi}{\hbar}\sum_{q'}|g_{p,p+q'}|^{2}[(N_{B}(\Omega^{0}_{ph}(q'))+1)\delta(\varepsilon_{k-q'}-\varepsilon_{k}+\Omega^{0}_{ph}(q'))
+N_{B}(\Omega^{0}_{ph}(q'))\delta(\varepsilon_{k+q'}-\varepsilon_{k}-\Omega^{0}_{ph}(q'))],
\end{aligned}
\end{equation}
the two terms within the square bracket correspond to the emission and absorption of a phonon respectively,
as persistently happen during the coherent propagation of the phonon (dressed by the cloud of phonons).
Apparently, at zero-temperature, the scattering rate (relaxation)
  %{Ultrafast THz Probe of Photoinduced Polarons in Lead-Halide Perovskites}
  %{Electron-Phonon Interaction and Transport in Semiconducting Carbon Nanotubes}
  %{Semiclassical transport and phonon scattering of electrons in semiconducting carbon nanotubes}
is related to the selection rule 
$\varepsilon_{k-q'}-\varepsilon_{k}=-\Omega^{0}_{ph}(q')$,
  %{Electron-phonon interaction using Wannier functions}
and it is coupled to the self-induced lattice polarization here.
  %{Enhanced Superconducting State in FeSe=SrTiO3 by a Dynamic Interfacial Polaron Mechanism}

For acoustic phonon,
the matrix element $|g_{p,p+q'}|^{2}$ in above equation should be replaced by 
\begin{equation} 
\begin{aligned}
\frac{s_{0}}{M}\frac{D^{2}q'^{2}}{2\Omega_{ph}^{0}}|\langle k+q'|k\rangle|^{2}=
\frac{s_{0}}{M}\frac{D^{2}q'^{2}}{2\Omega_{ph}^{0}}\frac{1+ss'{\rm cos}\theta}{2},
\end{aligned}
\end{equation}
where $D$ is the screened (quasistatic) deformation potential, $s_{0}$ is the area of the unit cell,
and $\theta$ denotes the angle between $k+q'$ and $k$.
  %{electronic cooling in graphene}
Note that here we discuss the case that the quasiparticle which dressed with the phonons is the fermi polaron: 
an electron dressed with electron-hole parts (when the electron-electron interaction (electron self energy) is taken into account).
While for the BEC in hexagonal optical lattice system,
the above tight-binding treatment fails to describes the modification of the band structure in Dirac cone by the interatomic interactions\cite{Chen Z},
however, we still find a common point.
For both the Dirac metarial and honeycomb optical lattice, 
  %{Bose-Einstein condensate in a honeycomb optical lattice: Fingerprint of superfluidity at the Dirac point}
  %{nonlinear evolution of quantum states in the adiabatic regime}
  %{Electron-phonon interactions for optical-phonon modes in few-layer graphene: First-principles calculations}
the adiabatic evolution of the electron eigenstates is broken when the band crossing happens,
which gives rise to the nonlinearity\cite{Liu J}.
This similarity requires the extreme low temperature in which case there are not thermally excited phonons.

For BEC in linear quantum regime, 
the adiabatic evolution maintains as long as the variance of the perturbation, 
which can simply described by the nonlinear Dirac equation with a momentum (or continuous wave function) dependent short range interaction term, 
is slower than the frequency (level spacing).
  %{the nonlinear dirac equation in bose-einstein condensates: foundation and symmetries}
  %{nonlinear landau-Zener tunneling}
For Dirac solid material, the electron-phonon coupling strength $|g_{p,p+q'}|^{2}$ changes linearly along the inreducible
 Brillouin zone\cite{Yan J A2} 
when the Dirac linear dispersion is presented or with band gap which is smaller than phonon energy $\Omega^{0}_{ph}$.
 %For the case that band gap larger than phonon frequency, the square of electron-phonon coupling may still exist but close to zero, 
 %and it is still linear (without a drop near the Dirac point).
This implies that $|g_{p,p+q'}|$ is independent of the electron momentum in the adiabatic case,
but depends only on the exchange momentum $q'$.
  %{Light Bipolarons Stabilized by Peierls Electron-Phonon Coupling}
  %{Sharp Transition for Single Polarons in the One-Dimensional Su-Schrieffer-Heeger Model}
That is consistent with the strong coupling adiabatic in Holstein (lattice) model
where the hopping $t_{k}$ is much larger than the phonon frequency, and thus the electron 
%(eigenstates)
 transition (or tunneling) between different levels (or eigenstates)
is suppressed, which leads to the exponentially reduced (collapsed) polaronic bandwidth.
  %{polarons and bipolarons in strongly interacting electron-phonon systems}
  %{nonlinear landau-zener tunneling}
  %{Theory of nonlinear Landau-Zener tunneling}
  %{Bose-Einstein condensate in a honeycomb optical lattice: Fingerprint of superfluidity at the Dirac point}
  %{Electron-phonon interactions for optical-phonon modes in few-layer graphene: First-principles calculations}
  %{Nonlinear evolution of quantum states in the adiabatic regime}
We note that, for the tunneling of BEC,
the level crossing between the eigenstates from valence band 
(but not from conduction band and valence band) can happen\cite{Chen Z,Liu J,Liu J2,Wu B} due to the additional eigenstates 
   %{Nonlinear evolution of quantum states in the adiabatic regime}
in nonlinear frame as described (at zero-temperature limit) by the Gross-Pitaevskii equation within mean-field approximation.
That is not unique to the BEC in an optical lattice, 
the multilayer Dirac materials also have such kind of level crossing (like the bilayer graphene or bilayer silicene\cite{Liu F}).

\subsection{Decoherence effect induced by electron-phonon coupling}

Now let us temporarily return to the problem of decoherence,
where the pairing instability can be described by the statistic mixture (Fermion or Bosonic distribution function) as shown in the pair propagator and the (non-self-consistent) $T$-matrix.
For an atomic system, the occurence of decoherence (accompanied by an exponential decay of the quasiparticle lifetime $\tau=\frac{-{\rm sgn}[\Omega]}{2{\rm Im}\Sigma(\Omega)}$)
  %{vertex corrections to the dc conductivity in anisotropic multiband system}
  %{Critical slowdown of non-equilibrium polaron dynamics}
requires that the velocity of the impurity is faster than the speed of sound in condensate,
and thus the momentum becomes larger than the Landau critical momentum.
That usually happen in the extreme case of mass-imbalance between impurity and the majority particles.
   %{decoherence of an impurity in a one-dimensional fermionic bath with mass imbalance}
While in the opposite case,
the superposition (coherence) mentioned in Ref.\cite{Combescot R}
is possible since they consider the case of zero impurity momentum which can be realized in a mixture where
 the two spin species have equal masses, such configuration has also been considered in Refs.\cite{Yi W,Vlietinck J}.
%{Diagrammatic Monte Carlo study of the Fermi polaron in two dimensions}
  %{Normal state of highly polarized Fermi gases: full many-body treatment}
  %{molecule and polaron in a highly polarized two-dimensional fermi gas with spin-orbit coupling}
  %{Critical slowdown of non-equilibrium polaron dynamics}
For the transition of electron states near Dirac cone, even the smallest perturbation could breaks the adiabaticity,
and 
the nonlinear dynamics induced by the on-site interaction will breaks the superposition of states as well as the coherence and the
noticeable interference patterns.
  %{Bose-Einstein condensates in time-dependent light potentials: Adiabatic and nonadiabatic behavior of nonlinear wave equations}
  %{from electron to small polaron: an exact cluster solution}
  %{theory of nonlinear Landau-Zener tunneling}
  %{Superfluidity of Bose-Einstein condensate in an optical lattice Landau–Zener tunnelling and dynamical instability}
 %{Decoherence of an impurity in a one-dimensional fermionic bath with mass imbalance}
Thus we conclude that, near Dirac cone, the transition of electrons is adiabatic both for the linear band structure or with a small band gap
(e.g., smaller than the bare phonon energy).
  %{polarons and bipolarons in strongly interacting electron-phonon systems}
While in the extreme adiabatic and strong coupling case where the electron hopping amplitude 
is much larger than the phonon frequency 
(or much larger than the Bloch energy of bosons $E_{B}$, which is intraspeices interaction-dependent for a bosonic gas\cite{Band Y B}),
  %{polarons and bipolarons in strongly interacting electron-phonon systems}
  %{electron-phonon interactions for optical-phonon modes in few-layer graphene: first-principles calculations}
  %{Breakdown of the Migdal-Eliashberg theory in the strong-coupling adiabatic regime}
since in this case the velocity of impurity (the electron) in real space is much larger than the bath 
component, as indicated by $t\gg \Omega^{0}_{ph}$ and infinite bath effective mass,
the coherence is decreased and the decoherence is enhanced,
as exhibited by the reduction of the coherent quantum motion.
  %{Quantum dynamics of a mobile spin impurity}
  %{Polarons and bipolarons in strongly interacting electron-phonon systems}
And it is indeed a diffusive motion (incoherent) which also contributes to the conductivity unless when a magnetic field is applied.
  %{Decoherence of an impurity in a one-dimensional fermionic bath with mass imbalance}
  %{Integer and half-integer quantum Hall effect in silicene: Infuence of an external electric feld and impurities}
In this case, the impurity with large velocity will excites, e.g., the electron-hole excitations or phonons, 
or even emits the Cherenkov radiation,
  %{Critical slowdown of non-equilibrium polaron dynamics}
and thus the superfluid can not emergents as the impurity losses energy during this process,
as observed recently of the self-trapped small polaron in haematite,
where the photoexcited carriers exciting the optical phonons during the process of charge-transfer\cite{Carneiro L M}.
  %{Ultrafast THz Probe of Photoinduced Polarons in Lead-Halide Perovskites}
Then the polaronic band (or electron band) collapse 
and leads to a charged ordered insulating state 
  %{Breakdown of the Migdal-Eliashberg theory in the strong-coupling adiabatic regime}
as long as the zero-point quantum fluctuation is weaker than the lattice distortion.
  %{small adiabatic polarons and bipolarons}
  %{Dynamical mean-field theory of transport of small polarons}
However, the self-trapping of polaron may be destroyed by the thermal fluctuation at high temperature
as recently studied in Ref.\cite{Sous J}.
  %{Light Bipolarons Stabilized by Peierls Electron-Phonon Coupling}
Note that the charged ordered insulating state mentioned above is essentially different from the Mott insulator state where the 
on-site interaction strength is very large and the impurity propagation induces the coherent quantum motion.
  %{Quantum dynamics of a mobile spin impurity}
Such extreme adiabatic limit corresponds to the local phonon modes and maximal decoherence 
with the impurity density matrix $\rho_{p}$ loses its off-diagonal elements: $c_{p+q'}^{\dag}c_{p}$ and $c_{p+q'}c_{p}^{\dag}$.
   %{Decoherence of an impurity in a one-dimensional fermionic bath with mass imbalance}
Experimently, the hopping here can be manipulated by the magnetic gradient modulation in atomic system\cite{Jotzu G} or by the
strain in a solid state system\cite{Shubnyi V O}.
  %{Decoherence of an impurity in a one-dimensional fermionic bath with mass imbalance}
  %{Effect of resonance impurities on Drude peak broadening in uniaxially strained graphene}

The perturbation from quantum fluctuation in a lattice system is more significant 
  %{comment t U expansion for the Hubbard model}
in 2D than that in the 3D,
as reflected by the high-order exapnsion about the ratio of $t/U$. 
For most cases, like for the ultracold atomic system\cite{Visuri A M} and the Holstein compounds\cite{Wellein G,Alexandrov A S}, 
the polaronic band collapse happen as long as $\varepsilon_{e-ph}/t\gtrsim 1$,
   %{Phonons in a strongly coupled electron-phonon system}
and for the former one,
the linear feature of the lattice near Dirac cone requires the hopping strength 
(or the coupling between modes) larger than the
nonlinear eigenvalue which forms the Bloch bands
$t>Ng_{\psi\psi}\mathcal{Z}_{\psi}\rightarrow 0$,
%{exciton-polariton condensates near the dirac point in a triangular lattice}
%{Superfluidity of Bose–Einstein condensate in an optical lattice Landau–Zener tunnelling and dynamical instability}
%{Bose-Einstein condensates in time-dependent light potentials: Adiabatic and nonadiabatic behavior of nonlinear wave equations}
  %{Disorder effect on spin excitation in double-exchange systems}
where $N$ is the particle number, $g_{\psi\psi}$ is the electron-electron interaction (characterized by the $s$-wave interaction),
and $\mathcal{Z}_{\psi}$ is the probability of a electron to be interacting.
  %{Breakdown of Fermi liquid theory in topological multi-Weyl semimetals}
On the other hand, when the band crossing happen, 
like the overlap between conduction band and valence band or the loop structure 
which emerges in the lowest Bloch band as obtained by the eigenvalues of the tight-binding Hamiltonian 
(or nonlinear Dirac equation) and the nonlinear Schrodinger equation, respectively, 
  %{decoherence of an impurity in a one-dimensional fermionic bath with mass imbalance}
  %{d+id'  Chiral Superconductivity in Bilayer Silicene}
the superconductivity\cite{Liu F} or superfluidity\cite{Chen Z,Berman O L,Min H} can be observed,
as the strong short-range interaction (the deformation potential-independent one) 
can helps the plane wave to overcomes the perturbations (like the periodic potential of optical lattice or the lattice defect).
  %{Effect of resonance impurities on Drude peak broadening in uniaxially strained graphene}
  %{Intrinsic electrical transport properties of monolayer silicene and MoS 2 from first principles}
As an example, the bipolarons formed in real space, which requires strong enough impurity-impurity interaction\cite{Camacho-Guardian A2},
are with highly nonadiabatic charectes in real space, which results in a different superconducting\cite{Alexandrov A S2}
mechanism compared to the BCS one.

We note that, the electron-phonon coupling term $\frac{\varepsilon_{e-ph}}{3t\Omega_{ph}^{0}(q')}$ obtained above is base on the 
combination of the
 polaron energy shift in single-site frame, i.e., assume the hopping is $t=0$ like in the atomic limit,
   %{Periodically driven small polarons  P. E. Kornilovitch}
  %{dynamical mean-field theory of transport of small polarons}
  %{Light Bipolarons Stabilized by Peierls Electron-Phonon Coupling}
and the finite bandwidth.
Usually, this description of coupling is valid in the tight-binding model as well as the orthorhombic YBCO compounds.
  %{dynamical mean-field theory of transport of small polarons}
The effect of hopping becomes dominative when the size of polaron is smaller than the lattice spacing,
  %{dynamical mean-field theory of transport of small polarons}
as described by the tight-binding Hamiltonian,
and
the value of hopping determines the dispersion of the mode and the motion (velocity) of the impurity.
For slow motion, the decay rate of the polaron (as a quasiparticle)
decreases and vanishes at long-time limit.
then leads to the negative polaron self-energy,
  %{Behavior of the impurity atom in a weakly-interacting Bose gas}
  %{Repulsive Fermi polarons in a resonant mixture of ultracold 6Li atoms}
i.e., the attractive polaron\cite{Panochko G,Scazza F,polaron2}.

\section{Self-energy effects, effective mass and residue of impurity 
due to the interaction with acoustic phonon and particle-hole pair}

The induced effective mass can be written as
 $\Delta m^{*}=-[\frac{\partial ^{2}E(p)}{\partial p^{2}}]^{-1}=[\frac{\partial ^{2}|E(p)|}{\partial p^{2}}]^{-1}$.
For the case that the dispersion of polaron is nearly quadratic, the
following approximated expression is valid\cite{Rath S P}
%base on the approximated expression
%{Quasiparticle Properties of a Mobile Impurity in a Bose-Einstein Condensate}
  %{Field-theoretical study of the Bose polaron}
  %\bibitem{Vlietinck J2}Vlietinck J, Ryckebusch J, Van Houcke K. Quasiparticle properties of an impurity in a Fermi gas[J]. Physical Review B, 2013, 87(11): 115133.
  %{Quasiparticle Properties of a Mobile Impurity in a Bose-Einstein Condensate}
  %{Repulsive Fermi Polarons in a Resonant Mixture of Ultracold 6Li Atoms}}
\begin{equation} 
\begin{aligned}
\Delta m^{*}=&\frac{p}{\partial_{p}E(p)}=\frac{1}{\frac{1}{m_{\downarrow}}+\frac{1}{p}\partial_{p}
{\rm Re}[\Sigma(E(p),p)]\big|_{E(p)}}\\
=&\frac{1/Z}{\frac{1}{m_{\downarrow}}+\frac{1}{p}\partial_{p}
{\rm Re}[\Sigma_{e}(\omega,p)+\Pi_{e}(p)]\big|_{E(p)}},
  %{Field-theoretical study of the Bose polaron}
  %{  %{Repulsive Fermi Polarons in a Resonant Mixture of Ultracold 6Li Atoms}}
  %{Repulsive Fermi Polarons in a Resonant Mixture of Ultracold 6Li Atoms}}
\end{aligned}
\end{equation}
where the second line of above expression corresponds to the linear expansion around the pole (quasienergy)
of the impurity Green's function.
  %{Field-theoretical study of the Bose polaron--\hbar^{-1}}
  %{Electron-phonon interactions from first principles}
  %{Repulsive Fermi Polarons in a Resonant Mixture of Ultracold 6Li Atoms}

The quasiparticle residue (spectral weight) $Z$ is given by the GW calculation result
%{Electron-phonon interactions from first principles}
  %{Molecule and Polaron in a Highly Polarized Two-Dimensional Fermi Gas with Spin-Orbit Coupling}
  %{Bose polarons near quantum criticality}
\begin{equation} 
\begin{aligned}
Z^{-1}\equiv &1-\partial_{\omega}{\rm Re}[\Sigma_{e}(\omega,p)+\Pi_{e}(p)]\big|_{\omega=E(p)}\\
=&|\phi_{0}|^{-2}=1+\sum_{k>k_{F},q<k_{F},|q'|\le q_{D}}(\frac{\phi_{kqq'}}{\phi_{0}})^{2}\\
\approx &1+\frac{1}{N_{ph}}(Z^{-1}_{{\rm min}}-\frac{E-\varepsilon_{p\downarrow}}{2N_{ph}\widetilde{g}}-1).
\end{aligned}
\end{equation}
Note that here $E(p)=\varepsilon_{p\downarrow}+{\rm Re}\Sigma(\omega,p)$ is the pole of the impurity Green's function
$G(\omega,p)=1/(\omega+i0-\varepsilon_{p\downarrow}-{\rm Re}\Sigma(\omega,p))$,
which can be rewritten as the dressed impurity propagator: $G(\omega,p)
  %{Marginal Fermi liquid in twisted bilayer graphene}
  %{Repulsive Fermi Polarons in a Resonant Mixture of Ultracold 6Li Atoms}
  %{Field-theoretical study of the Bose polaron}
=Z/(\omega+i\gamma-\varepsilon_{p\downarrow}-{\rm Re}\Sigma(E(p),p))$,
where $\gamma=-Z{\rm Im}\Sigma=-Z(\frac{-1}{2\tau}){\rm sgn}(\omega)$.
  %{Repulsive Fermi Polarons in a Resonant Mixture of Ultracold 6Li Atoms}
  %{Field-theoretical study of the Bose polaron}
  %{Fully self-consistent GW self-energy of the electron gas}
In the thrid line of above equation,
we use the approximated formula,
where $\widetilde{g}$ is the renormalized coupling parameter,
$\widetilde{g}^{-1}=g^{-1}
+\frac{1}{N}\sum_{k,q'}^{\Lambda}\frac{1}{\varepsilon_{k\uparrow}+\varepsilon_{k\downarrow}+\Omega_{ph}^{0}(-q')+W}$,
which is comparable with the mean-field polaron energy.
  %{Quasiparticle Properties of a Mobile Impurity in a Bose-Einstein Condensate}
  %{Bose polarons near quantum criticality}
In single phonon limit, it reduced to
$Z^{-1}\approx Z_{{\rm min}}^{-1}-\frac{E-\varepsilon_{\downarrow}}{2\widetilde{g}}$,
which means $Z^{-1}$ decreases with the raising self-consistent polaron energy $E-\varepsilon_{\downarrow}$.
Similar result is obtained for organic semiconductor\cite{Vukmirovi? N}.
  %{Electron-Phonon Coupling in Crystalline Organic Semiconductors Microscopic Evidence for Nonpolaronic Charge Carriers}
The $Z_{min}$ here is the minimal value of the quasiparticle residue of the 2D solid system,
which is generally larger than that of the ultracold fermi degenerate gases (0.39)\cite{Punk M} or the bose gases (0.67)\cite{ZZ Yan},
  %{Polaron-to-molecule transition in a strongly imbalanced Fermi gas}
for example,
the electron residue measured in semiconductors is 0.74\cite{Vukmirovi? N}.
That is partly due to the effect of linear dispersion in the presence of Dirac or Weyl cones (undoped semimetal)
with the broken down Fermi-liquid picture,
where the ratio of
  %the fermi circle become very small and
(long-ranged) Coulomb potential to the kinetic energy, $e^{2}/\hbar v_{F}\epsilon$ (the effective fine structure constant\cite{Burkov A A,Jafari S A},
where $\epsilon$ here is the environment dielectric constant),
becomes very large.
The residue can even becomes zero in the multi-Weyl semimetal\cite{Wang J R} with the dressed Coulomb potential.
While for the self-trapping of polaron in the Holstein model, the residue could as low as 0.1-0.2\cite{Vukmirovi? N}.

For electron self-energy induced by the electron-phonon coupling obtained above,
obviously, the Eq.() is for the case of single electron,
while for the many-electron case,
the electron self-energy should be replaced by\cite{Giustino F,Giustino F2}
   %{Electron-phonon interaction using Wannier functions}
\begin{equation} 
\begin{aligned}
\Pi_{e}(p)
=&\int\frac{d^{2}q'}{(2\pi)^{2}}|g_{p,p+q'}|^{2}
(\frac{1+N_{B}(\Omega^{0}_{ph}(q'))-N_{F}(p+q')}{\varepsilon_{p-q'}-\varepsilon_{p}+\Omega^{0}_{ph}(q')+i\eta}
 +\frac{N_{B}(\Omega^{0}_{ph}(q'))+N_{F}(p+q')}{\varepsilon_{p+q'}-\varepsilon_{p}-\Omega^{0}_{ph}(q')+i\eta}).
\end{aligned}
\end{equation}
%Specially,
%for the case that the density of impurity electron equals to that of phonon,
%i.e., $n_{e}=n_{ph}$,
%we have $\dot{n}_{e}=-n_{e}^{2}/\tau_{polaron}$,
In Fan-Migdal approximation,
the function $N_{F}$ can be treated as a step function, which vanishes for unoccupied Kohn-Sham state\cite{Giustino F2},
and the function $N_{B}$ exists at finite temperature.
While for adiabatic case with zero temperature,
the zero-point renormalization can be obtained as
$\int\frac{d^{2}q'}{(2\pi)^{2}}|g_{p,p+q'}|^{2}
\frac{1}{\varepsilon_{p-q'}-\varepsilon_{p}}$,
when the Debye-Waller term is neglected.

In long-wavelength limit,
the electron-phonon coupling matrix element 
$\frac{D^{2}\hbar q^{2}s_{0}}{2M\Omega_{ph}}(1-\frac{q'^{2}}{p^{2}})\frac{1+ss'{\rm cos}\theta}{2}$ 
  %{Acoustic phonon scattering limited carrier mobility in two-dimensional extrinsic graphene}
  %{electronic cooling in graphene}
can be reduced as $\frac{D^{2}\hbar q^{2}s_{0}}{2M\Omega_{ph}}$.
For the two-dimensional materials which with linear transverse acoustic (TA) 
and longitudinal acoustic (LA) phonon dispersion in small-$q'$ region, we can write
$\Omega_{ph}^{0}(q')=v_{s}q'$ where $v_{s}=v_{L}(v_{T})$ is the sound velocity of the LA mode (TA mode).
For simulation, we set $v_{L}=1.6v_{TA}$,
which is applicable for graphene, silicene, blue phosphorene, and MoS$_{2}$
according to the DFT based calculation results\cite{Ge X J,Li X}.
While
for out-of-plane (flexural) acoustic phonon (ZA) which is closely related to the thermal conductivity of the crystal lattice\cite{Ong Z Y},
  %{Effect of substrate modes on thermal transport in supported graphene}
the disperion can be approximately written as $q'^{\alpha}$ with $\alpha\ge 2$.
In the following, we at first set $\alpha=2$.

Fig.2 shows the self-energy and effective masses of the polaron induced by the electron-phonon coupling.
The real-part of the self-energy $\Pi_{e}(p)$ corresponds to the energy shift of the impurity,
which gives rise to the renormalization effect to the intrinsic impurity dispersion.
From the real part of the self-energy,
we can see that there exist kinks (for TA and LA modes) or discontinuity (for ZA mode)
in their first derivative,
and the coordinates for these kinks or discontinuity are dependent on the sound velocity of the corresponding mode,
e.g., we can see these special points are local on $p=1$ and $p=1.6$ (since we set the impurity mass as $m=1$).
The imaginary part of the self-energy (also related to the impurity scattering rate and the coherence length)
  %{Electron-Phonon Interaction and Transport in Semiconducting Carbon Nanotubes}
  %{Vertex corrections to the dc conductivity in anisotropic multiband systems}
also shows the sound velocity-dependence:
The polaronic instability does not emerges until the velocity of impurity is larger than the speed of sound.
We will find that the induced effective masses also obey this rule,
i.e., the induced effective masses (calculated by using the exact expression)
becomes nonzero only when the impurity moves faster than the sound velocity
(in other words, when the momentum of impurity is larger than the Landau critical momentum $v_{s}p$),
and we know that the decoherence effect only emerges in this case.
   %{Critical slowdown of non-equilibrium polaron dynamics}
The induced effective masses exhibit a power-law dependence on the momentum $p$ (with a exponent larger than one).
In the last panel, we compare the results of the induced effective masses calculated by the exact expression (blue and red lines)
 and approximated expression (green and black lines)
for TA and LA phonons.
We can see that the approximated expression of the induced effective mass loss its accuracy in such case,
which reveals great differece in polaronic dynamics in solid state system (crystal lattice)
compared to the one in the BEC and atomic lattice system\cite{Rath S P}.
  %{Field-theoretical study of the Bose polaron}

Here we focus only on the low-energy case, while in the presence of high field,
the impurity scattering rate contributed by the optical phonons.
  %{Electron-Phonon Interaction and Transport in Semiconducting Carbon Nanotubes}
Also, the strong binding due to the optical phonon-electron coupling will gives rise to large polaronic energy shift\cite{Perebeinos V}
as well as the band gap renormalization 
(the band gap is reduced since the binding energy here is positive and the induced bound state energy is negative).
Certainly,
that requires much higher temperature to realizes the optical excitation,
for eaxmple, at 600 K or even 1600 K as reported in Ref.\cite{Carneiro L M},
in which case the acoustic
phonons would not be expected to directly participate the polaron formation.
Another important different is that at high temperatures 
the distribution of electrons (in multi-impurity case) and phonons obey the Maxwell-Boltzmann law.
This is also verified by the DFT result\cite{Zijlstra E S} that the high intensity laser-induced 
phonon-phonon interaction cannot be well described by the
usual polarization
 formula\cite{Allen P B} which contains two Dirac-Fermi distribution functions for electron and hole respectively.
  %{Phonon Anharmonicities in Graphite and Graphene}
  %{Electron-phonon interactions for optical-phonon modes in few-layer graphene First-principles calculations}
  %{P.B. Allen, Phys. Rev. B 6, 2577 (1972); P.B. Allen and R. Silberglitt, ibid. 9, 4733 (1974).}
  %{Anomaly of Optical Phonons in Bilayer Graphene}
In Fig.3, we show the spectral function GW calculation.
  %{Electron-phonon interactions from first principles}
We can see that, in the stable region ($p>0.6$) where ${\rm Im}\Pi_{e}(p)\rightarrow 0$ (see Fig.2),
the rebuilt polaron band (by the electron-ponon coupling) is linear with the momentum,
while in the unstable region, we can clearly see that the polaronic mode is damped,
then the damping dynamical could be described by the Laudau theory when the Fermi liquid picture is preserved.
  %{Breakdown of Fermi liquid theory in topological multi-Weyl semimetals}
  %{Breakdown of the Landau-Fermi liquid in two dimensions due to umklapp scattering}
  %{Repulsive Fermi Polarons in a Resonant Mixture of Ultracold 6Li Atoms}
For linear phonon mode,
the induced polaronic mode does not damped for static polaron ($p=0$),
as can be seen from the bright spots which form the high order dispersion.
Note that the dispersion showed here is not the final polaron dispersion,
but just exhibits the effect of self-energy (electron-phonon coupling) to the original impurity dispersion
(which is simply quadratic for the free electron occupying the surface of 2D system we discussing).
We can also see that the slope of linear dispersion for $q'^{2}$-phonon is smaller than the linear acoustic phonon.

Similarly, for $\Omega_{ph}^{0}\sim q^{'4}$ ($\alpha=4$),
we obtain
\begin{equation} 
\begin{aligned}
\Pi_{e}(p)=&\frac{D^{2}s_{0}}{4\pi M}
\frac{1}{2 i \eta}(-((2 \hbar^{2} p\ {\rm arctan}[\frac{(2 m q + \hbar^2 (-p + q))}{
    \sqrt{B}}])/  \sqrt{B}) - 2 {\rm ln}\ q \\
&+ 
 {\rm ln}(\hbar^2 q (-2 p + q) + 2 m (q^2 - i z)))\big|_{q'},
\end{aligned}
\end{equation}
where $B=-\hbar^{4}p^{2}-2\hbar^{2}mi\eta-4m^{2}i\eta$.
For finite (low) temperature where the distribution of phonon still obeys the Bose-Einstein one,
we obtain
\begin{equation} 
\begin{aligned}
\Pi_{e}(p)=&\frac{D^{2}s_{0}}{4\pi M}
		   \frac{1}{(2 b^2 i \eta \sqrt{A})}\\
	&(2 (-\hbar^2 p T + \beta m (T + 2 i \eta)) {\rm arctan}[\frac{(\beta m + \hbar^2 (-p + q))}
     {\sqrt{A}}] \\
	 &+ 
   T ((-2 \beta m + 2 \hbar^2 p) {\rm arctan}[\frac{(-\beta m + \hbar^2 (p + q))}
{        \sqrt{A}}] \\
&+ 
      \sqrt{A} (-4 {\rm ln}(q) + 
         {\rm ln}(2 \beta m q + \hbar^2 q (-2 p + q) - 2 i m \eta) + 
         {\rm ln}(-2 \beta m q + \hbar^2 q (2 p + q) - 
           2 i m \eta))))\big|_{q'},
\end{aligned}
\end{equation}
Note that although we focus on the long-wave limit of $q'$,
the detail integrate range of $q'$ depends on the setting of the temperature.
Our results are in agree with Ref.\cite{Perebeinos V}
that in low-energy limit the self-energy is heavily affected by the change of the temperature,
thus we select the not-too-small value of $q'$ for better understanding.
That can also be explained by the failure of RPA when the Fermi energy is smaller than or just comparable with the phonon frequency
(temperature-dependent),
  %{Temperature dependent phonon frequency shift and structural stability of free-standing graphene: a spectral energy density analysis}
  %{The effect of temperature on the phonon dispersion relation in graphene}
 %{Failure of the random-phase-approximation correlation energy}
and the above expression of electron self-energy becomes unphysical when $\Omega_{ph}^{0}(q')=0$ in the denominator.
 %{http}
 %that's also easy to be solved just by lifting the energy of zero-momentum polaron,
 %and then the Kohn anomaly should appears at $q'=2k_{F}$.
From Fig.2 and Fig.4,
we can see that the self-energy of impurity electron dressed by phonon could is not totally negative,
which suggrest that the interaction between the electron and phonons (acoustic or the breathing-mode one)
  %{Bipolarons bound by repulsive phonon-mediated interactions}
  %{Electron-Phonon Interaction and Transport in Semiconducting Carbon Nanotubes}
could be attractive or repulsive.
That is different with the normal attractive Fermi polaron\cite{Combescot R}.

While for the Fermi polaron formed by the excited electron-hole pair,
the self-energy of the impurity obtained by the method of
non-self-consistent $T$-matrix can be written as (at zero-temperature limit)
\begin{equation} 
\begin{aligned}
\Sigma_{e}(p,\omega)&=\int\frac{d^{2}q}{(2\pi)^{2}}T(p+q,\omega+\nu_{q})\\
&=\int\frac{d^{2}q}{(2\pi)^{2}}\left[g_{q}^{-1}-L(p+q,\omega+\nu_{q})\right]^{-1}\\
&=\int\frac{d^{2}q}{(2\pi)^{2}}\left[g_{q}^{-1}-\int^{\Lambda}_{k_{F}}\frac{d^{2}k}{(2\pi)^{2}}\frac{1}{\omega+i\eta-\varepsilon_{p+q-k\downarrow}
-\varepsilon_{k\uparrow}+\varepsilon_{q\uparrow}}\right]^{-1},
\end{aligned}
\end{equation}
where $\omega$ and $\nu_{q}$ are the impurity and hole frequency, respectively.
Within the expression of bare coupling $g^{-1}$ (see Eq.()),
we use the polaron shift in a single site 
  %{dynamical mean-field theory of transport of small polarons}
(i.e., in zero tunneling limit and $p=0$) instead of the binding energy of the bounded state
   %{Periodically driven small polarons}
which is 
$E_{b}={\rm min}\langle\psi_{CP}|E-\varepsilon_{p=0\downarrow}|\psi\rangle_{CP}=\hbar^{2}/2m_{r}a_{\phi\psi}^{2}$ 
($m_{r}$ is the reduced mass and $a_{\phi\psi}$ is the interspecies scattering length),
  %{Highly polarized Fermi gases in two dimensions}
  %{Normal State of Highly Polarized Fermi Gases Full Many-Body Treatment}
  %{Variational study of polarons in Bose-Einstein condensates}
  %{Attractive and repulsive Fermi polarons in two dimensions}
i.e., 
\begin{equation} 
\begin{aligned}
\frac{1}{g}=&-\frac{1}{N}\sum^{\Lambda}_{k}\frac{1}{\frac{g^{2}}{2m_{e-h}\omega^{2}_{e-h}}+\varepsilon_{k\uparrow}+\varepsilon_{k\downarrow}+W},
\end{aligned}
\end{equation}
  %{Periodically driven small polarons}
where $g$ is the interaction vertex between the impurity and the electron-hole pair,
   %{Sharp Transition for Single Polarons in the One-Dimensional Su-Schrieffer-Heeger Model}
$m_{e-h}$ and $\omega_{e-h}$ are the mass and frequency of the eletron-hole pair.
Here the summation over electron momentum $k$ can be replaced by in continuum limit through the relation 
$\frac{1}{N}\Sigma=\int\frac{d{\bf q}}{\Omega}$ ($\Omega$ is the BZ area).
Obviously, the self-energy of this Fermi polaron depends on both the impurity momentum and frequency,
i.e., it is related to the response to the effective electric field.
  %{Many-body effects in the normal-state polaron system}

Fig.4 shows the self-energy, effective mass, and the spectral function of the fermi polaron
at different impurity frequency $\omega$.
From the expression given above, we can know that the self-energy of fermi polaron is directly related to the coupling constant $g^{-1}$.
From the thrid panel of Fig.4,
we can see that the induced effective mass has a differece compared to the bose polaron one:
For $\omega=2$, the induced effective mass initially has a negative value, which
means that the interaction between impurity and the electron-hole pair initially reduce the effective mass of polaron and thus 
makes it moves faster.
But in large momentum region, the tendency of induced effective mass is the same as the bose polaron one.
The induced effective masses here are calculated through the exact expression $\Delta m^{*}=\frac{\partial^{2}|E(p)|}{\partial p^{2}}$.
Here we only present the results for the bare coupling $g_{b}=-0.5$, but
by changing the value of $g_{b}$, we obtain that the real part of self-energy (no matter how large the $\omega$ is) at large
momentum will always equal to $g_{b}$,
i.e., the attractive feature of the polaron enhanced by the increasing $|g_{b}|$,
while the induced effective masses will decrease rapidly with the increase of $|g_{b}|$ ($g_{b}<0$).
From (e), we can see that the electron spectral function ($p>0$) is much wider than the hole spectral function ($p<0$),
and by changing the value of $g_{b}$,
we find that, for positive $g_{b}$ (repulsive interaction),
the hole spectral function will becomes wider than the electron spectral function.
In Fig.5, we present the density plot of the fermi polaron spectral function.
As can be seen, the electron-hole symmetry is broken due to the different value of electron momentum $k$ and hole momentum $p$,
and the peaks in particle spectral function is much higher than the electron one.
%Note that the spectral function in the hole part ($p<0$) should be negative, which do not showed in the figure.
Our results are in agree with Fig.12-13 of Ref.\cite{Schmidt R}.
 %{Excitation spectra and rf response near the polaron-to-molecule transition from the functional renormalization group}
Besides, for electron part of the spectral function,
we can see the double-peak structure (BCS-type) like the ones
found in Refs.\cite{Hwang E H2,Tsuchiya S}.
  %{Single-particle properties and pseudogap effects in the BCS-BEC crossover regime of an ultracold Fermi gas above Tc}
  %{Plasmon-pole approximation for many-body-effects in extrinsic graphene}

In Fig.6,
the quasiparticle residue is presented.
We can see that, for zero coupling constant $g=0$ ($\Lambda\rightarrow \infty$),
the residue is always equals to one, which means that there does not exists the interacting particle,
and thus the polaron is absent.
For $0<g<1$, a unusual fluctuation of residue emerges in low momentum region.
We suppose such behavior is related to the singular self-energy in this momentum region.
While for $g\ge 1$,
the large fluctuation in residue vanishes, and we can always see a dip structure at momentum $p=2.75$.
We note that the momentum corresponds to such dip structure is independent of the coupling constant $g$,
but it varies with the solution of the self-consistent equation (Eq.()): the value of
this momentum increases with the increase of polaron energy $E(p)$.
  %{Fermi polaron-polaritons in charge-tunable atomically thin semiconductors}
  %{Electron-phonon interactions from first principles}
  %{Repulsive Fermi Polarons in a Resonant Mixture of Ultracold 6Li Atoms}
  %{Field-theoretical study of the Bose polaron}
At large momentum, the residue in all panels tend to one.
For larger coupling $g$, the residue tends to one at larger momentum.

Fig.5 shows the self-energy induced by the interaction between impurity and the electron-hole pair
at strong-coupling regime where the bare coupling $g(\Lambda)\rightarrow \infty$,
 although $g^{-1}$ is in fact weakly depends on the relative
momentum $q$.
We can see that the sign of coupling directly affects the sign of self-energy $\Sigma_{e}$,
and the range of negative self-energy increase with the increaseing impurity frequency.
%We can see that for too large impurity momentum,
%the imaginary part of the self-energy vanishes which means the momentum of the polaron is conserved
%and thus the energy conservation is also confirmed in these areas by the effective mass approximation.
For $\omega\neq 0$,
the imaginary part of self-energy (also the linewidth) has two peaks throughout the whole momentum range,
that is in agree with the experimental result about the Fermi polaron as resported in Ref.\cite{Sidler M}.
  %{Semiclassical transport and phonon scattering of electrons in semiconducting carbon nanotubes}
  %{Optical conductivity of black phosphorus with a tunable electronic structure}
While for the slow impurity in the bose polaron with momentum $p\ll {\rm max}[\mu_{\uparrow},\sqrt{2m_{\downarrow}\Omega_{ph}^{0}(q')}]$,
  %{Highly polarized Fermi gases in two dimensions}
  %{Fr?hlich polaron and bipolaron recent developments}
the polaron energy is purely real and thus the momentum and energy are conserved.
Such stable bound state could also be found in the
polaron-polaron continuum region or the bipolaron-phonon continuum region\cite{Sous J},
where the polarons are unbounded or bounded but through emission and absorption of one phonon.
  %{Phonon-mediated repulsion, sharp transitions and (quasi)self-trapping in the extended Peierls-Hubbard model}
  %{Bipolarons bound by repulsive phonon-mediated interactions}

The above self-energy of impurity electron gives the second-order correction to the intrinsic electron energy,
$E=\frac{\hbar^{2}p^{2}}{2m_{\downarrow}}+\Sigma_{e}(\omega,p)$.
This energy is purely real for a slow impurity,
in which case the polaron is large and continuum,
  %{Fr?hlich polaron and bipolaron: recent developments}
%{https://books.google.co.jp/books?id=wayqcNvBjFgC&pg=PA90&lpg=PA90&dq=half-bandwidth+polaron&source=bl&ots=-gJkJ7tNBx&sig=ACfU3U0gSxds2QQN9z2jGeWAHtqiFSgb4A&hl=en&sa=X&ved=2ahUKEwjyh8mUweHhAhUB6bwKHddYA0cQ6AEwBnoECAkQAQ#v=onepage&q=half-bandwidth%20polaron&f=false}
and the effective mass approximation can be used in this case,
  %{Fr?hlich polaron and bipolaron: recent developments}
  %{Bipolarons in a Bose-Einstein condensate Supplemental}
i.e., $E=\frac{\hbar^{2}p^{2}}{2m^{*}}+E_{0}$
where 
$E_{0}\approx -\frac{g^{2}}{2m_{e-h}\omega^{2}_{e-h}}-N_{ph}\frac{\varepsilon_{e-ph}}{\Omega_{ph}^{0}(q')}<0$
 is the impurity momentum-independent energy of the polaron.
  %{Highly polarized Fermi gases in two dimensions}
  %{Fr?hlich polaron and bipolaron recent developments}
  %{Phonon-mediated repulsion, sharp transitions and (quasi)self-trapping in the extended Peierls-Hubbard model}
%term and can be obtained by measuring the energy of polaron in antiadiabatic limit.
  %{Light Bipolarons Stabilized by Peierls Electron-Phonon Coupling}
The large polaron (compared to the lattice constant)
has a small value of hopping and thus with a positive and large binding energy 
(corresponds to a negative and large polaron energy\cite{Piegari E}).
The distortion as well as the ions would also slow the large polaron down.
%Fig.5(a) shows the second order partial derivatives of the pair propagator of Fermi polaron,
%$\frac{\partial^{2} (-L(p+q,\omega+\nu_{q}))}{\partial p^{2}}\big|_{p=p_{GS}}$,
%for impurity ground state momenta at zero or
%close to the Brillouin zone boundary.

%We can see that the induced effective mass is positive,
%and it descreases at the begining and then increase with the impurity frequency $\omega$.
%In Fig.6,
%we show t

\section{polaron-polaron interaction}

As we state above,
the electron-phonon coupling matrix element $|g_{p+q',p}|^{2}$, 
which describes the phonon-induced electron transitions (interband or intraband),
  %{Electron-phonon interactions for optical-phonon modes in few-layer graphene First-principles calculations}
depends on both the impurity momentum and phonon momentum in the nonlinear case for $q'$ extended to the $K$-point (e.g., $q'=0.665\frac{2\pi}{a}$ for graphene\cite{Yan J A2}),
just like the polaron-polaron interaction vertex as we mention below\cite{Marchand D J J,Sous J}, 
but it depends only on the phonon momentum $q'$ in the
linear case which with adiabatic motion of the impurity.
   %{Electron-phonon interactions for optical-phonon modes in few-layer graphene First-principles calculations}
The absorption and emission of phonons which leads to incoherence, and also possible to gives rise to the polaron-polaron interaction,
for two impurities with opposite spins,
e.g., in the light-enhanced case\cite{Sentef M A}.

The phonon-related polaron-polaron interaction has two kinds.
One of them is the phonon-mediated density-density (direct) interaction,
  %{Bipolarons bound by repulsive phonon-mediated interactions}
  %{Light Bipolarons Stabilized by Peierls Electron-Phonon Coupling}
  %{Phonon-mediated repulsion, sharp transitions and (quasi)self-trapping in the extended Peierls-Hubbard model}
  %{Many-body effects in the normal-state polaron system}
which is related to the dielectric function $\epsilon(q',\nu)$ in the method of GW-random phase approximation (RPA)
(or equivalently, the spectral function $A(q')$ in plasmon-pole approximation (PPA)),
  %{Many-body effects in the normal-state polaron system}
  %{Plasmon-pole approximation for many-body effects in extrinsic graphene}
  %{Fully self-consistent GW self-energy of the electron gas}
and we can write the self-energy of such bipolaron as
\begin{equation} 
\begin{aligned}
\Sigma(p,\omega)=&\int\frac{d\nu}{2\pi}\int\frac{d^{2}k}{(2\pi)^{2}}G_{0}(p-k,\omega-\nu)V_{p-p}(q')
+\\
&\int\frac{d\nu}{2\pi}\int\frac{d^{2}k}{(2\pi)^{2}}G_{0}(p-k,\omega-\nu)V_{p-p}(q')[\epsilon^{-1}(q',\nu)-1],
\end{aligned}
\end{equation}
where the first term is the Hartree-Fock part while the second term is the correlation part.
Here $V_{p-p}(q')$ is the polaron-polaron interaction strength, and
$\epsilon(q',\nu)=1-2V_{p-p}(q')\sum_{p}\frac{N_{F}(p+q')-N_{F}(p)}{\nu+i\eta+\varepsilon_{p+q'}-\varepsilon_{p}}$.
In Landau Fermi-liquid theory, the density-density interaction between polarons must be repulsive\cite{Scazza F},
and the Coulomb interaction is certainly screened.
  %{Breakdown of Fermi liquid theory in topological multi-Weyl semimetals}
This kind of polaron-polaron interaction is dominate for the 2D system on a polar substrate,
and then the effective mass of polaron becomes dependent on both the electron-phonon coupling constant and the phonon occupation factor
\cite{Zhang S,Wong W P D}.
In the presence of effective dielectric function,
  %{Collective modes in multi-Weyl semimetals}
  %{Many-body effects in the normal-state polaron system}
  %{Plasmon-polaron of the topological metallic surface states.pdf}
the residue $Z$ also related to the energy loss function as $Z^{-1}=1-\pi [-{\rm Im}\epsilon^{-1}(q',\nu)]^{-1}$.

The second type of the phonon-related polaron-polaron interaction is the pair-hopping type.
We found the interaction vertex of this type of polaron-polaron interaction is similar to the pair propagator as we mention above,
which is related to the propagation of a pair of coupled impurity-majority, but for here is happen between a pair of impurity with opposite spins,
like the singlet.
  %{Fermi polaron-polaritons in charge-tunable atomically thin semiconductors}
In this case, 
the polaron-polaron propagator, for a pair of polarons with seperartion distance evolute
from 1 to $n$ ($n\ge 1$), can be obtained by the Dyson function as 
$G^{1,n}=G_{0}^{1,n}+g\sum_{n'}G_{0}^{n',n}\sum_{i}\frac{e^{ik(r_{i}+\frac{n'a}{2})(r_{i}+\frac{a}{2})}}{\sqrt{N}}\langle 0|
G(\omega)c_{i}c_{i+1}c^{\dag}_{i}c^{\dag}_{i+n''}b^{\dag}_{i+n'-n''}|0\rangle $,
  %{Phonon-mediated repulsion, sharp transitions and (quasi)self-trapping in the extended Peierls-Hubbard model}
where $G(\omega)=\frac{1}{\omega+i\eta-H_{p-p}}$ is the Green's function of the bipolaron
with the Hamiltonian 
$H_{p-p}=\sum_{p}\varepsilon_{p}c^{\dag}_{p}c_{p}+(\Sigma_{e}+\Pi_{e})\sum_{p}c^{\dag}_{p}c_{p}+(2(\Sigma_{e}+\Pi_{e})+V_{p-p})\sum_{pp'Q}
c^{\dag}_{p'+Q}c^{\dag}_{p-Q}c_{p}c_{p'}$ in momentum space.
  %{Dropping an impurity into a Chern insulator A polaron view on topological matter}
  %{Light Bipolarons Stabilized by Peierls Electron-Phonon Coupling}
The unperturbed state $|0\rangle$ could be treated as the vacuum state at zero-temperature limit.
  %{Fr?hlich polaron and bipolaron recent developments}
Note that here $p$ is no more the momentum of bare impurity but of the polarons.
The Green's function here vanishes in the large distance limit $n\rightarrow\infty$ where $\omega$ outside the polaron-polaron continuum
(unbound) region\cite{Berciu M},
  %{Few-Particle Green’s Functions for Strongly Correlated Systems on Infinite Lattices}
and equation of motion\cite{Sous J2,Sous J3,Berciu M} of this Green's function at ground state (for Brillouin zone boundary) can be obtained as
  %{Phonon-mediated repulsion, sharp transitions and (quasi)self-trapping in the extended Peierls-Hubbard model}
  %{Light Bipolarons Stabilized by Peierls Electron-Phonon Coupling}
  %{Bipolarons bound by repulsive phonon-mediated interactions}
$(\omega+i\eta-[4(\Sigma_{e}+\Pi_{e})+V_{p-p}])G_{1,n}=\delta_{n,1}$.
Note that within the Hamiltonian $H_{p-p}$ the interation vertex $V_{p-p}(p,p',Q)$ ($Q$ is the exchange momentum)
takes an important role in the case of translational invariant which guarantees 
that the total momentum of the bipolaron is a good quantum number.
Thus this expression, which described the pair-hopping process, is valid as long as 
$\frac{\varepsilon_{e-ph}}{3t\Omega^{0}_{ph}}<1$ since the strong electron-phonon coupling will breaks the translational invariance.
  %{Breakdown of the Migdal-Eliashberg theory in the strong-coupling adiabatic regime}
Base on Peierls coupling,
the avoided crossing in the band structure happen for both the single polaron and the bipolaron 
when the electron-phonon interaction strength turns up and reaches $\frac{\varepsilon_{e-ph}}{3t\Omega^{0}_{ph}}>0.5$
\cite{Marchand D J J,Sous J,Sous J2,Sous J3},
which means that the nonadiabaticity appears,
and this is more easy to found in the case of large phonon energy.
  %{Sharp Transition for Single Polarons in the One-Dimensional Su-Schrieffer-Heeger Model}
This avoided crossing phenomenon (similar to the band inversion of the topological materials) is very important
  %{Composite Band Structures and Lattice Dynamics of Bi 2 Te 3 -Based Compounds and Solid Solutions}
because it reveals the topological properties of the polaron band structure,
and we can see that the polaron indeed shares the topological phase with some of the topological insulator or superconductor
\cite{Grusdt F,Qi F,Ezawa M} no matter it's been formed in the solid state or the ultracold gases\cite{Haussmann R}.
  %{Topological phase transition based on the attractive Hubbard model}
  %{Topological polarons, quasiparticle invariants and their detection in 1D symmetry-protected phases}
  %{Valley-Polarized Metals and Quantum Anomalous Hall Effect in Silicene}
The avoided crossing band structure corresponds to the nonadiabatic (or nonlinear) case where the lowest energy of fermionic excitation 
   %{Spectral functions and rf response of ultracold fermionic atoms}
does not located at $p=0$,
and it's possible to appears for both the fermi polaron and boson polaron.
Besides, due to the strong (interatomic) interaction in this case, the superfluid can emerges 
within the cloud, and thus the impurity can moves without lossing energy and 
also without excites other particles.
The superfluid in such strong-coupling regime is usually topological trivial
and with a chemical potential $\mu<0$\cite{Qin F}.
   %{Excitation spectra and rf response near the polaron-to-molecule transition from the functional renormalization group}
  %{polaron in a p+ip fermi topological superfluid}
That's why the could forms the superfluid can formed within the hard-sphere bosons (with interaction effect) 
while the free bosons can not.
  %{Oscillating Superfluidity of Bosons in Optical Lattices}
In such a coherent state, the dissipation of the bath is also fastest\cite{Visuri A M}.
  %{Bose-Einstein Condensate in a Honeycomb Optical Lattice Fingerprint of Superfluidity at the Dirac Point}

\section{Conclusion}

We construct a model where the composite polaron formed in the surface of a 2D Dirac honeycomb lattice.
The impurity is firstly dressed by a cloud of electron-hole pairs (with unequal momenta),
and then excited by the phonons.
The phonons can be viewed in a coherence state while the existence of electron-phonon coupling for small band gap
(absorption and emission of phonons by the impurity) induces the incoherence.
  %{Light-enhanced electron-phonon coupling from nonlinear electron-phonon coupling}
In the presence of electron-phonon coupling, it is possible for impurity to has a quadratic dispersion 
or linear dispersion (due to the effect of Dirac cone which can be treated as a perturbtion origin from the relativistic particles).
Note that we focus on the weak coupling regime throughout the paper,
no matter for the interaction between impurity and the electron-hole pair or that between the impurity and phonons.
For electron-phonon coupling, both the adiabatic and nonadiabatic cases are studied
(we focus on acoustic phonon).
Specially, the avoided crossing of band structure appears for both the single polaron and bipolaron (the pair-hopping one)
in nonadiabatic regime, where the perturbation theory becomes more accurate than Migdal-Eliashberg theory.
  %{Light Bipolarons Stabilized by Peierls Electron-Phonon Coupling}
The avoided crossing for bipolaron can be found in the strong-coupling antiadibatic limit,
  %{Bipolarons bound by repulsive phonon-mediated interactions}
  %{Polaron-phonon interaction in a finite-size lattice A perturbative approach}
  %{Light-enhanced electron-phonon coupling from nonlinear electron-phonon coupling}
where the resonance vanishes and the impurity is surrounded by the virtual phonon emission and absorption.
For the pair-hopping type bipolaron,
which is direactly pairing in the real space unlike the density-density type,
it remains the coherence property with nonadiabatic motion
 (for $\frac{\varepsilon_{e-ph}}{3t\Omega^{0}_{ph}}>0.5$)
and it is possible to forms the superfluid at low temperature.
  %{Bipolarons bound by repulsive phonon-mediated interactions}
This type of bipolaron appears when the next-nearest-neightbor hopping is forbidden by the hard-core statistics\cite{Sous J3,Sous J2}
(for spinless Fermions).
   %{Bipolarons bound by repulsive phonon-mediated interactions}
   %{Phonon-mediated repulsion, sharp transitions and (quasi)self-trapping in the extended Peierls-Hubbard model}
   %{Few-Particle Green’s Functions for Strongly Correlated Systems on Infinite Lattices}
The method of RPA together with the Ward identity
  %{Collective Modes of a Helical Liquid}
 are used in calculating the electron self-energy induced by electron-phonon coupling
  %{Electron-Phonon Interaction and Transport in Semiconducting Carbon Nanotubes}
  %{Many-body effects in the normal-state polaron system}
and the dielectric function which describes the density-density correlation (also the electronic susceptibility)
 between the polarons (in momentum space).
  %{Collective Modes of a Helical Liquid}
  %{Anomalous Acoustic Plasmon Mode from Topologically Protected States}
The RPA here is applicable since the phonon momentum $q'$ is small and the electron-phonon coupling is weak in this paper\cite{Jia X}.

The dimensionless electron-phonon coupling in GW calculation\cite{Attaccalite C} is slightly different
from that in the Peierls (Su-Schrieffer-Heeger) model\cite{Marchand D J J}.
  %{Sharp Transition for Single Polarons in the One-Dimensional Su-Schrieffer-Heeger Model}
  %{Bipolarons bound by repulsive phonon-mediated interactions}
  %{Doped Graphene as Tunable Electron-Phonon Coupling Material}
The most direct evidence is that the electron-phonon coupling matrix element is independent of the electron momentum $p$ in the former one
  %{Phonon-induced giant linear-in-T resistivity in magic angle twisted bilayer graphene Ordinary strangeness and exotic superconductivity}
as applied in Refs.\cite{Zhang S,Wu F},
while the $p$-dependence may emerges for the later one after the Fourier transform of interaction vertex.
  %{http: electron-phonon interactions}
%repulsive-attractive
  %{Bipolarons bound by repulsive phonon-mediated interactions}
  %{Phonon-induced giant linear-in-T resistivity in magic angle twisted bilayer graphene Ordinary strangeness and exotic superconductivity}

Through the density plot of spectral function.
we found that the polaronic effect induced modes are damped in the regions of $p-\omega$ space
where the imaginary part of self-energy is large.
We also found that, for small $q'$,
the effect electron-phonon couping leads to a linear dispersion in the large-impurity momentum (stable) region (see Fig.3).
In the absence of umklapp scattering,
  %{Breakdown of the Landau-Fermi liquid in two dimensions due to umklapp scattering}
this dampling dynamics can be well described by the Landau-Fermi liquid theory (for free impurity).
This inspires us that, for topological insulator systems,
the topologically protected umklapp-free plasmon-polaron mode\cite{Shvonski A,Jia X} can emerges and then leads to the band dispersion 
which is gapless 
in the middle point of BZ edge---the $M$ point.
We can certainly know that such situation corresponds to the antiadiabatic limit which with strong (short-ranged) interaction in the $K$-point,
  %{Bose-Einstein Condensate in a Honeycomb Optical Lattice: Fingerprint of Superfluidity at the Dirac Point}
and can also be observed in the optical honeycomb lattice in the presence of strong interatomic interaction\cite{Chen Z}.
Furthermore,
GW calculation (with dynamically screened Coulomb potential) shows that the band gap in $M$-point
varies with the doping level:
when the doping level is made away from zero through the gate voltage (no matter $n$-type or $p$-type),
accompanied by the increasing short-range (Hubbard-type) 
  %{Equations-of-motion method for triplet excitation operators in graphene}
  %{Breakdown of Fermi liquid theory in topological multi-Weyl semimetals}
  %{Electromagnetic modes from Stoner enhancement Graphene as a case study}
electron-electron interaction and decreasing deformation potential\cite{Attaccalite C}
(due to the competitive effect between electron-phonon coupling and electron-electron Coulomb
repulsion),
the gap in $M$-point decreases.
In the other hand, the above mentioned umklapp-free mode provides another way to realize such nonadiabatic dynamics,
even when the fermi wave vector is much smaller than phonon one ${\bf q}'$.
It relys on the low energy helical character of the topological state,
  %{Collective Modes of a Helical Liquid}
  %{Plasmon-polaron of the topological metallic surface states}
that is,
within the first Brillouin zone,
  %{Fr?hlich polaron and bipolaron recent developments}
for strong enough electron-phonon coupling with $p\ll q'$, we have ${\bf p}+{\bf q}'\approx {\bf q}'$,
and the overlap factor $\langle f_{{\bf q}'-{\bf G}}|f_{{\bf q}'}\rangle\rightarrow 0$.
 This is indeed an nonadiabatic limit since $E_{M} \ll \Omega_{ph}^{0}(q')=\beta q'$ 
where $E_{M}$ is the absolute value of energy of the lowest conduction band at $M$-point
  %{Enhanced Superconducting State in FeSe=SrTiO 3 by a Dynamic Interfacial Polaron Mechanism}
  %{Breakdown of the Migdal-Eliashberg theory in the strong-coupling adiabatic regime}
(note that here $\Omega_{ph}^{0}(q')$
of the $\alpha$-mode\cite{Jia X,Zhang S}
  %{nonadiabatic limit: Breakdown of the Migdal-Eliashberg theory in the strong-coupling adiabatic regime}
keeps increasing linearly even when it passes the $M$-point or $K$-point\cite{Jia X} which is compeletely different to the normal 
monoatomic 2D honeycomb lattices\cite{Ge X J}).
Also, $p\ll q'$ shows that the impurity moves slowly and the phonons can respond instantaneously to it\cite{Zhang S}.
   %{Enhanced Superconducting State in FeSe=SrTiO 3 by a Dynamic Interfacial Polaron Mechanism}
The nonadiabaticity can also be seen from the band structure of the helical edge states\cite{Michiardi M}.
  %{Bulk band structure of Bi2Te3}

In weak coupling regime, both the effective masses induced by the electron-phonon interaction
and the interaction between impurity and the electron-hole pair
are studied through the approximated and exact expressions.
We found that in stable region ($p>1$),
the effective masses increase as $q^{a}$ with $a>1$.
For impurity with smaller electron energy, the induced effective mass is larger.
We also found that, the effective mass derived from approximated expression is larger than that derived from exact
expression.
The reason for this phenomenon is related to the impurity dispersion (although may be variable in the polaronic dynamics)
as esplained in the main tetx.

The polaronic effect investigated here is meaningful to the study of pairing mechanism as well as the 
 phonon-mediated high-temperature superconductivity.
  %{Light Bipolarons Stabilized by Peierls Electron-Phonon Coupling}
  %{Phonon-induced giant linear-in-T resistivity in magic angle twisted bilayer graphene Ordinary strangeness and exotic superconductivity}
For example,
the twisted bilayer graphene near the magic angle could induces the
superconducting instability\cite{Cao Y,Wu F}
with the attractive electron-acoustic phonon interaction which even overcomes the Coulomb repulsion
through the retardation effect
\cite{Wu F,Wellein G}.
  %{Electron–phonon interaction in dielectric bilayer systems. effect of the electronic polarizability}
The phonon velocity is been estimated as $s=0.02v_{F}$ in Ref.\cite{Wu F}, 
and the electron-acoustic phonon coupling will flatten the pristine electron dispersion,
and thus lowers the group velocity of electron (narrower bandwidth due to the incoherence).
We can also infer that the nonadiabaticity is increased
since the strong electron-phonon coupling changes the cone-like dispersion
into the avoided crossing-like dispersion\cite{Wu F}
and lengthen the relaxation time of impurity-momentum $p$.
 In the other hand, due to the lowered slope of bands near Dirac cone,
the gap in $M$ point should be reduced,
which is also consistent with our above conclusions.
It is recently found that at a tilt Weyl semimetal in the presence of electric field and magnetic field,
the motion of occupied state along the axis connecting to Weyl nodes with opposite chirality
can excites the electron-hole pairs in the momentum space\cite{Zubkov M A} which can be viewed as a chiral anomaly.
In the mean time, the deformation-induced vibration, although with the sound velocity much slower that the fermi velocity, can excites the fermi quasiparticle\cite{Zubkov M A}.
Then, in the presence of multi-phonon (classical), the polaronic effect as well as the interesting modes 
mentioned in this paper would be produced in such model.
  %{the type ii weyl semimetals at low temperatures chiral anomaly, elastic deformations, zero sound}

%The strong many-body interaction effects may drive the
%2D metallic Fermi liquid ground state into gapped insulator phase with vanishing Drude weight,
%as the strong correlation goes beyond the Fermi-liquid paradigm\cite{Wu F,Sagi Y}.
%  %{Breakdown of the fermi liquid description for strongly interacting fermions}
%Such a largely increased resistivity with the breakdown of previous Fermi-liquid picture
%has been observed in the twisted bilayer graphene near the magic angle\cite{Wu F} (at metallic phase) and the cuprates.
%Such strange metallic behavior is been ascribed to the strong electron-acoustic phonon interaction at higher temperature
%accompanied with the flattening of conduction band and valence band.
%  %{Phonon-induced giant linear-in-T resistivity in magic angle twisted bilayer graphene: Ordinary strangeness and exotic superconductivity}
%With the increace of electron-phonon interaction, the slope of band is lowered and thus the Fermi velocity of electron also
%decreased.
%That leads to the nonadiabatic effects\cite{Wums21}, like the decrease of quasiparticle residue and increase of effective mass.
%While at low-temperature limit,
%similar effect could also be observed (even more obvious) in the Weyl system with high-order dispersion.
%It is because the dispersion at Dirac or Weyl cone has a decisive effect on the long-range Coulomb interaction.
%  %{Breakdown of Fermi liquid theory in topological multi-Weyl semimetals}

\clearpage

\section{Appendix.A: Peierls model in real space representation}

We rewrite the local phonon and the e-ph coupling terms in Eq.(\ref{871}) in space representation as,
\begin{equation} 
	\begin{aligned}
		\mathcal{H}=H_{ph}+H_{e-ph}=
		\omega \sum_{i}b^{\dag}_{i}b_{i}+
		g\omega \sum_{ij}(c_{i}^{\dag}c_{j}+c_{j}^{\dag}c_{i})
		(b^{\dag}_{i}+b_{i}-b_{j}^{\dag}-b_{j}).
	\end{aligned}
\end{equation}
Consider the Peierls phonons quasiclassically
as the harmonic oscillators,
like the hydrogen bond vibration\cite{Araki,Chang} whose electron trapping effect can be observed by far-infrared spectrum,
\begin{equation} 
	\begin{aligned}
		H_{ph}=
		\sum_{i}(\frac{q_{i}^2}{2m_{ph}}+\frac{1}{2}m_{ph}\omega_{0}^{2}x_{i}^{2})
		=\omega_{p}
		\sum_{i}(\mathcal{P}_{i}^{2}+\mathcal{X}_{i}^{2}),
	\end{aligned}
\end{equation}
where we define
\begin{equation} 
	\begin{aligned}
		&
		\mathcal{P}_{i}=(b^{\dag}_{i}+b_{i}-X_{i})\cos (\omega_{p}t),\\
		&
		\mathcal{X}_{i}=(b^{\dag}_{i}+b_{i}-X_{i})\sin (\omega_{p}t),\\
		&X_{i}=
		b^{\dag}_{i}+b_{i}-\sqrt{b_{i}b_{i}^{\dag}}.
	\end{aligned}
\end{equation}
In terms of a canonical  momentum-related function
$f(\hat{\mathcal{P}}_{i}):=b^{\dag}_{i}+b_{i}-X_{i}=\sqrt{b^{\dag}_{i}b_{i}}$
and the canonical coordinate
$\hat{\mathcal{X}}_{i}:=\omega_{p}t$,
we can apply the Poisson bracket,
$\frac{\partial \mathcal{P}_{i}}{\partial 
	\hat{\mathcal{P}}_{i}}
\frac{\partial \mathcal{X}_{i}}{\partial \hat{\mathcal{X}}_{i}}
-\frac{\partial \mathcal{P}_{i}}{\partial 
	\hat{\mathcal{X}}_{i}}
\frac{\partial \mathcal{X}_{i}}{\partial \hat{\mathcal{P}}_{i}}=1$,
and obtain $f(\hat{\mathcal{P}}_{i})=\sqrt{2
	\hat{\mathcal{P}}_{i}}$.

In terms of the Holstein phonon-assisted electron relaxation,
the current in time domain reads\cite{Fetherolf}
\begin{equation} 
	\begin{aligned}
		J(t)=-ig_{p}\omega_{p}\sum_{ij}
		(c^{\dag}_{i}c_{j}-c^{\dag}_{j}c_{i})
		([b^{\dag}_{i}+b_{i}](t)-[b^{\dag}_{j}+b_{j}](t)),
	\end{aligned}
\end{equation}
which can be cast into the following simple form
\begin{equation} 
	\begin{aligned}
		J(t)=\sum_{ij}J_{ij}(t)c^{\dag}_{i}c_{j},
	\end{aligned}
\end{equation}
with $J_{ij}$ a Peierls phonons-related strength constant which could be Gaussian distributed.
%From the general form of the Hubbard Hamiltonian (including the tight-binding Hubbard part and the potential part)
%{Effect of resonance impurities on Drude peak broadening in uniaxially strained graphene}
%{Bose-Einstein condensate in a honeycomb optical lattice: Fingerprint of superfluidity at the Dirac point}
%For weakly bound many-body system,

Most importantly, the effect of classical time evolution on the current-current correlation can be estimated by the following equality
\begin{equation} 
	\begin{aligned}
		&
		\mathcal{C}_{JJ}(t)
		=\frac{{\rm Tr}[J(t)J(0)e^{-\beta H}]}
		{{\rm Tr}e^{-\beta H}}\\
		&=\sum_{\mathcal{P},\mathcal{X}}
		F(\mathcal{P},\mathcal{X})
		\frac{{\rm Tr}[e^{i\int^{t}_{0}d\tau H(\tau)}
			J(t)e^{-i\int^{t}_{0}d\tau H(\tau)}J(0)e^{-\beta H}}
		{{\rm Tr}e^{-\beta H}},
	\end{aligned}
\end{equation}
where $F(\mathcal{P},\mathcal{X})$ is the phase-space distribution of the classical Peierls variables.

In low-carrier-density limit,
the dc conductivity in the presence of weak-binding-energy-supported polaron 
provides the well-defined mobility (or ideal conductivity; due to the absence of ergodicitys)
$\mu_{m}=\sigma_{dc}/n$.
While at high electron concentration,
the long-range charge order or the superconductivity is possible to form.
When the long-range Coulomb effect enhance the mobility,
the Peierls polaron becomes unstable,
which is in agree with the
semilocal generalized gradient approximation (GGA)
as well as the experimentally observed higher conductivity in
anatase compares to rutile\cite{Elmaslmane}.
As higher mobility will delocalize the charge density and suppress the formation of polaron.
The method of DFT+U (with the Hubbrd interaction estimated from RPA)
as well as the range-hybrid approaches (which gives rise to piecewise linear),
aiming at the localization correction,
but sometimes they still underestimate the band gap and Coulomb screening,
i.e., overestimate the mobility,
and thus leads to an unstable Peierls-type polaron.
Similar reason results in the different results by IR spectroscopy
and STM or optical absorption\cite{Elmaslmane},
where the IR spectroscopy usually closes to the result predicted by
large mobility as well as the Holstein model.
Thus the comparasion between IR spectral and the optical absorption one
can help to estimate the competition between 
localization (small Holstein polaron) and delocalization
(large Peierls polaron) effect.
Compares to WTe$_{2}$ which has a good (low-resistance) Ohmic contact as can be verified by its nearly linear current-voltage (I-V) curve. In contrast,
the WSe$_{2}$ usually has a higher resistance for Ohmic contact due to the existence of
Schottky barrier.
Since WSe$_{2}$ is a semiconductor with high mobility, and it
is more easy to realizing the n-type doping
(using potassium or lithium), a low work function contact metal is often used as the electrode\cite{Chuang}.
Thus our study on the polaronic dynamic concerning the surface charge trapping
and its effect to the system mobility is also helpful in solve the problem of
high-resistance Ohmic contact in both the chemically or electrostatically ways.

\section{Appendix.B: ZA phonon mode with higher order dispersion: $\Omega_{ph}^{0}(q')=q'^{\alpha}(\alpha>2)$}

From the Fig.6, we can see that, for $p<2.5$, the effective mass $\Delta m^{*}$ diverges dramatically at certain momenta $p$
and impurity frequencies $\omega$.
Such divergence of effective mass origins from the instability of the fermi polaron (as a metastable excitation
which is the origin of the small imaginary part in the right-hand-side of Eq.(27)),
   %{Highly polarized Fermi gases in two dimensions}
and it has
also been found in the strong coupling region of Bose polaron in a condensate
\cite{Ardila L A P,Ichmoukhamedov T,Van Loon S}.
  %{Impurity in a Bose-Einstein condensate Study of the attractive and repulsive branch using quantum Monte Carlo methods}
  %{Feynman path-integral treatment of the Bose polaron beyond the Fr¨ohlich model}
  %{Analytical theory of the dressed bound state in highly polarized Fermi gases}
We note that the effective masses obtained here have the negative part,
furthermore, the region where $m_{0}+\Delta m^{*} <0$
can be viewed as a signal of the emergence of avoided crossing dispersion (nonadiabatic),
 %{Light Bipolarons Stabilized by Peierls Electron-Phonon Coupling}
which implies that the minimum of fermi polaron dispersion is not at $p=0$\cite{Parish M M}
  %{Highly polarized Fermi gases in two dimensions}
(because in effective mass approximation the negative kinetic term leads to 
larger momentum-independent polaron energy than the total one)
  %{DIVERGENCE OF THE EFFECTIVE MASS OF A POLARON IN THE STRONG COUPLING LIMIT}
and thus also signals the instability\cite{Combescot R3}.
We will see in the follwing that this is in contrast to the effective mass induced by electron-phonon coupling,
which is completely positive over the whole momentum range (for small finite $\beta$).
We also found that, for $p\ge 2.5$, the effective mass is always positive.
%where $Z=[1-\partial_{\omega}{\rm Re}\Sigma(\omega,p)]^{-1}\big|_{\omega=E(p)}$ is the spectral weight.
In Fig.7, we plot the results of the induced effective masses within such a fermi polaron through the exact expression 
(upper panels) approximated expression (lower panels)
of the effective mass ($\Delta m^{*}=-[\frac{\partial ^{2}E(p)}{\partial p^{2}}]^{-1}=[\frac{\partial ^{2}|E(p)|}{\partial p^{2}}]^{-1}$ and 
$\Delta m^{*}=p/\frac{\partial E(p)}{\partial p}$, respectively).
We can see that in large impurity momentum ($p>1.5$) the effective mass increase powerly with $p$ and decrease with $\omega$.

In Fig.8-9, we show both the exact result $\Delta m^{*}=-[\frac{\partial ^{2}E(p)}{\partial p^{2}}]^{-1}$
and the approximated result of the effective mass $\Delta m^{*}$ induced by the electron-phonon coupling,
for free impurity (quadratic dispersion) and the one near Dirac cone 
(the Dirac cone could be viewed as a defect or perturbation but we do not consider its topological properties 
here, i.e., without the spin/pseudospin-momentum coupling),
respectively.
We can see from the figures that the effective mass obtained through the approximated expression
is larger than the exact one.
From the expressions of the induced effective mass,
we can know that the approximated effective mass only takes the second-order dispersion of momentum $p$ 
into account,
which makes it more accurate for the impurity with quadratic dispersion.
That is why the approximated results (the full-line) in Fig.9 is farther away from the exact results (the dashed-line) than Fig.8.
Here for the Dirac-type (initial) impurity electron, we suppose the linear dispersion $\varepsilon_{p}=\alpha p$
where we set $\alpha\approx 30 \beta$ (for LA) according to the first-principle calculations\cite{Ge X J,Yao Y},
i.e., the group velocity of the Dirac Fermi quasiparticle $\partial E(p)/\partial p=\alpha$
 is much larger than the sound velocity.
  %{Spectral functions and rf response of ultracold fermionic atoms}
That allows the quasiparticle decaying to happen through the emission of phonons,
and also gives rise to the finite spectral width base on the BCS-type particle-quasiparticle (or quasiparticle-quasiparticle)
interaction (like the third term of Eq.(20) with a finite exchange momentum $q$).
  %{Spectral functions and rf response of ultracold fermionic atoms}
For attractive polaron ($a_{\psi\phi}<0$), the broadening of the spectral function of polaron vanishes at
the minimum of particle-branch or the maximum of hole-branch when the gap between these two branches is zero,
because there has no enough phase space for quasiparticles to decay.
  %{Spectral functions and rf response of ultracold fermionic atoms}
Note that the polaron branches are parabolic here, so that the group velocity $\alpha$ could
becomes smaller than the sound velocity $s\beta$ (when consider the interaction between 
impurity and the phonon (gapless boson quasiparticle)) in the level closest to the chemical potential (e.g., $\mu_{\uparrow}$).
  %{Spectral functions and rf response of ultracold fermionic atoms}
This phenomenon is also been verified by our previous work\cite{polaron2} and other Refs.\cite{Haussmann R,Rath S P}.
Also, we can clearly find that the induced effective mass for impurity which is initially a Dirac-like electron (Fig.9) is much larger than 
the one which is initially a conventional electron (Fig.8).
That is reasonable because it requires stronger interaction strength
to excites a initially linear Dirac electron to a quadratic electron (e.g., enters into the surface 2D electron gas),
  %{Plasmon-polaron of the topological metallic surface states}
and then through the expression of effective mass given above we can easily know that the impurity
has larger effective mass.
  %{Quasiparticle Properties of a Mobile Impurity in a Bose-Einstein Condensate}

Here for effective masses induced by both the impurity-electron-hole pair interaction and the electron-phonon coupling,
we have $Z\rightarrow 1$ when $\Delta m^{*}\rightarrow 0$, i.e., the noninteracting particles is totally dominating.
In this case, for fermi polaron with $\omega\neq 0$ (see Fig.5(a)),
in the region where the imaginary part of self-energy (impurity damping rate)
acts as $\sim \omega^{a}$ ($a>1$),
   %{Breakdown of Fermi liquid theory in topological multi-Weyl semimetals}
the residue $Z$ tends to a small but nonzero value in the limit of vanishing fermi energy,
which means that the fermi-liquid theory as well as the perturbation theory are still applicable in this scenario\cite{Christensen R S,Wang J R}
   %{Quasiparticle Properties of a Mobile Impurity in a Bose-Einstein Condensate}
(while for bose polaron, an additional requirement is the nonadiabatic limit).
   %{Polaron-phonon interaction in a finite-size lattice A perturbative approach}
We note that the consideration of multiple particle-hole excitations become important in the strong coupling regime where 
$Z\rightarrow 0$.
  %{Quasiparticle properties of an impurity in a Fermi gas}
For impurity with quadratic dispersion, the residue will remains finite even in the zero-energy limit,
   %{Breakdown of Fermi liquid theory in topological multi-Weyl semimetals}
and it has smaller value in nonadiabatic case compared to the adiabatic case due to the coherent effect.
  %{Bose-Einstein condensate in a honeycomb optical lattice: Fingerprint of superfluidity at the Dirac point}

In Fig.10,
we show the quasiparticle residue, self-energy of the composite polaron which containing the effects
of both the fermi polaron and electron-phonon coupling.
According to the self-energy computed above (Fig.2 and Fig.5),
we can see that the composite polaron is stable (well-defined) only when the momentum larger or comparable with 1,
above where the imaginary part of the self-energy tends to zero.
This means that it is only meaningful to observe the related physical quantities in the region $p\ge 1$ (the exact value of momentum 
varies with impurity frequency),
while in the low momentum region, the nonadiabatic effect may induce the instabilities.
From Fig.10,
we can see that the quasiparticle residue is very close to 1 ($Z\gtrsim 0.9$) for $p\ge 1$,
and the self-energy is also always negative in this momentum region,
which implies that the composite polaron still has an attractive overall effect.
As we stated above, here the self-energy of composite polaron is obtained by the relation 
${\rm Re}\Sigma(E(p),p)={\rm Re}\Sigma_{e}(\omega,p)+{\rm Re}\Pi_{e}(p)$.
The large residue $Z$ is also agrees with the restriction of weak interaction throughout this paper.

\begin{figure}
 \centering
   \begin{center}
     \includegraphics*[width=0.8\linewidth]{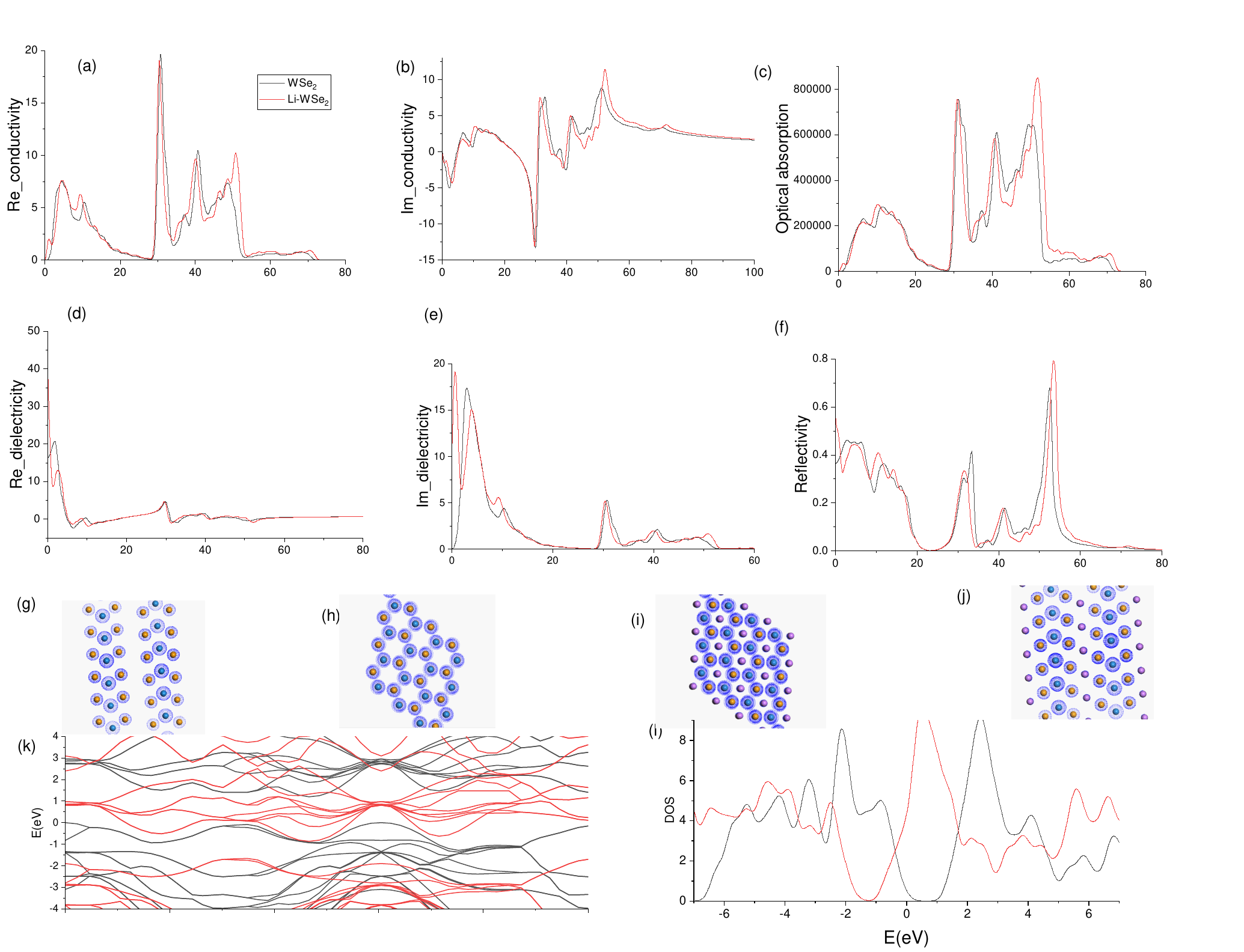}
\caption{DFT study for the pristine and Li-doped Wse$_{2}$.
In Li-doped Wse$_{2}$ as a n-type semiconductor, the Li provided extra electrons and slightly localized the electrons around it, and thus lower the charge mobility
as can be seem from the larger charge distance between the sphere of the charge densities.
(a)-(f) show the optical conductivity, absorption, dielectrivity, reflectivity,
where x-axis is teh frequency in unit of eV.
(g)-(j) show the charge density of WSe$_{2}$ before and after Li doping.
The optical conductivity and dielectricity also in agree with this statement:
The doped Wse$_{2}$ has a lower (in average) conductivity and higher dielectricity
(in zero frequency limit),
higher absorption, higher reflectivity, compares to the pristine one,
which indicates its lower mobility and stronger Coulomb screening (which cause the charge localization).
}
   \end{center}
\end{figure}

\begin{tiny}

\renewcommand\refname{References}

\end{tiny}
	
\clearpage

\begin{figure}[!ht]
   \centering
 \centering
   \begin{center}
     \includegraphics*[width=0.8\linewidth]{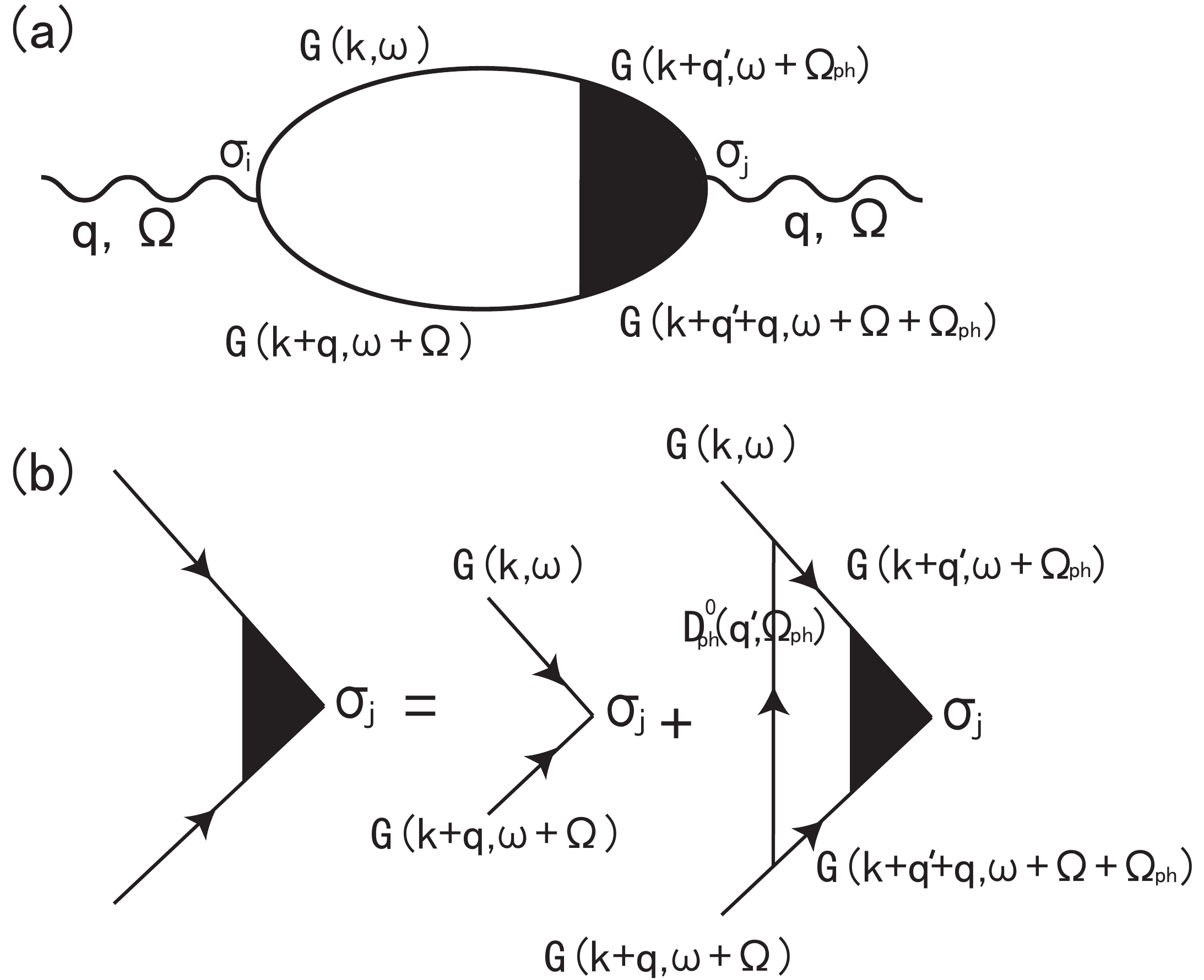}
\caption{Real part (left) and imaginary part (right) of the pair propagator at non-chiral case
as a function of the impurity momentum $p$ and majority momentum $q$.
The rows from top to bottom correspond to the Bosonic frequency (impurity) $\omega=-1,\ 0,\ 1,\ 2$, respectively.
The momentum cutoff $\Lambda$ is setted as 1 and the chemical potential is zero.
The vertical axis is in unit of $\frac{1}{2\pi}$.
}
   \end{center}
\end{figure}
\clearpage

\begin{figure}[!ht]
   \centering
 \centering
   \begin{center}
     \includegraphics*[width=0.8\linewidth]{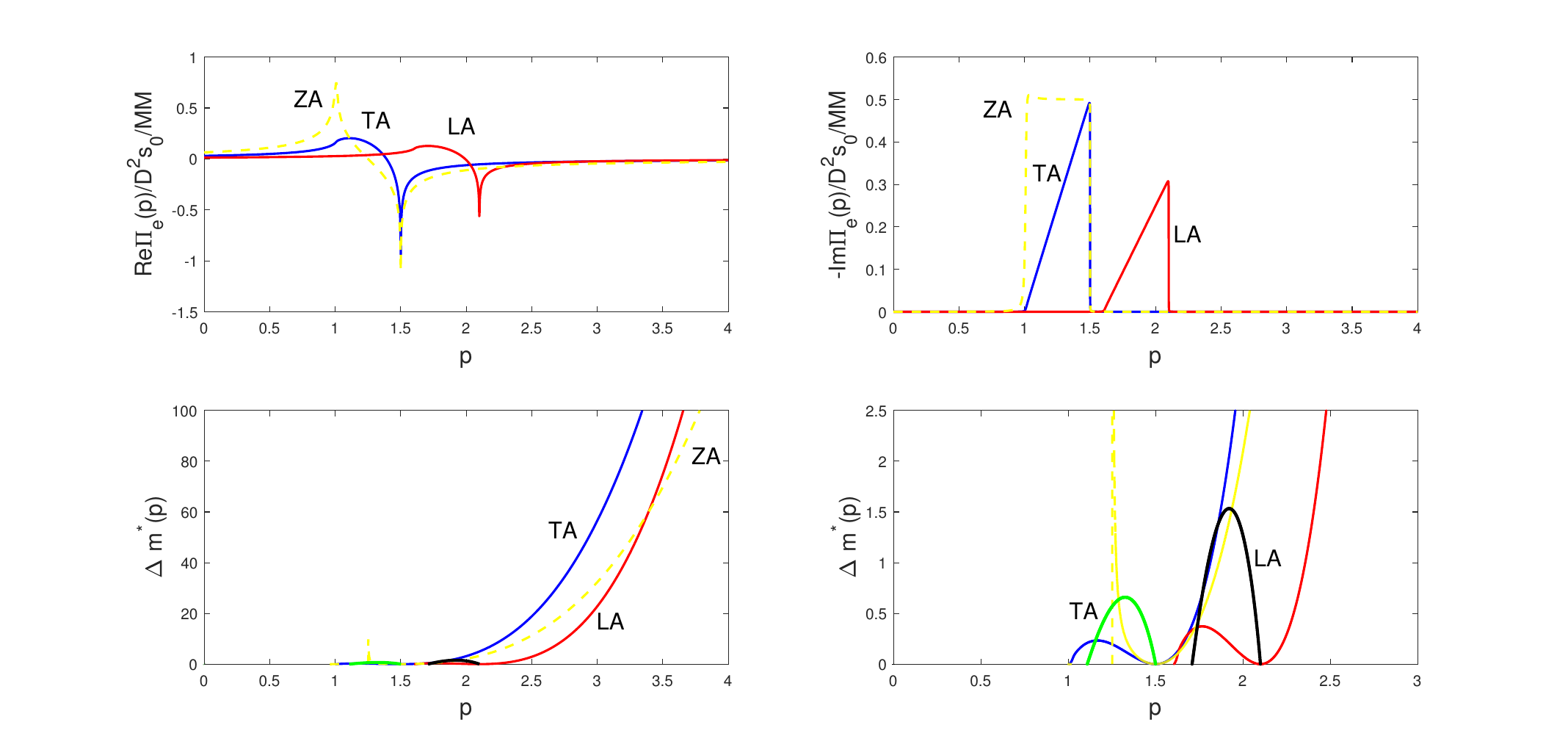}
\caption{The upper panels show the real part and imaginary part of the self-energy induced by the electron-phonon coupling.
For TA, LA, and ZA phonon modes,
we apply the dispersions $q'$, $1.6q'$, and $q'^{2}$ respectively.
The imaginary part of the self-energy here can also be regarded as the impurity scattering rate (inversed lifetime)
according to the RPA.
  %{Electron-Phonon Interaction and Transport in Semiconducting Carbon Nanotubes}
  %{Intrinsic electrical transport properties of monolayer silicene and MoS2 from first principles}
The thrid panel shows the induced effective masses by the electron-phonon coupling,
and the green line and black line correspond to the induced effective masses calculated by the approximated expression 
for the TA and LA phonon, respectively.
The chemical potential is setted as zero and the Debye redius is setted as $q_{D}=1$.
The impurity mass is also setted as 1.
The last panel shows the enlarged view of the thrid panel.
%and the effective masses (not the one induced by the polaronic effect) of the impurity 
}
   \end{center}
\end{figure}
\clearpage

\begin{figure}
\centering
\begin{subfigure}
  \centering
  \includegraphics[width=0.8\linewidth]{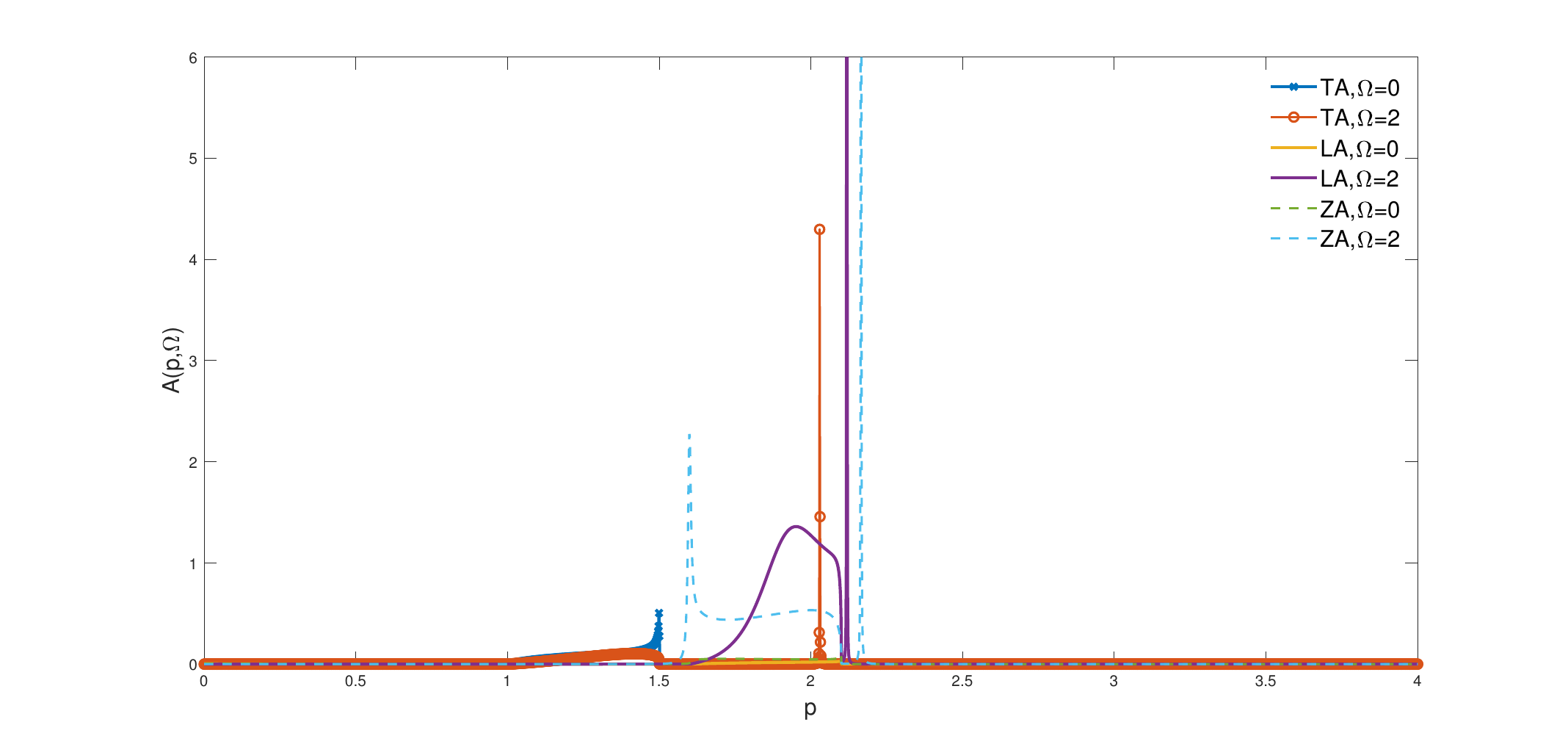}
  %\caption{A subfigure}
\end{subfigure}\\
\begin{subfigure}
  \centering
  \includegraphics[width=0.8\linewidth]{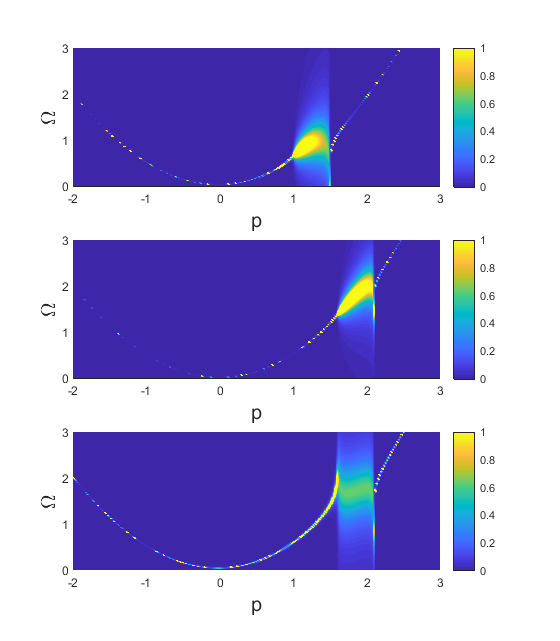}
 % \caption{A subfigure}
\end{subfigure}
\caption{Polaron spectral function $A(p,\Omega)$ due to the electron-phonon coupling.
The three intensity plots from top to the bottom correspond to the TA, LA, and ZA phonons.}
\end{figure}
\clearpage

\begin{figure}[!ht]
   \centering
 \centering
   \begin{center}
     \includegraphics*[width=1\linewidth]{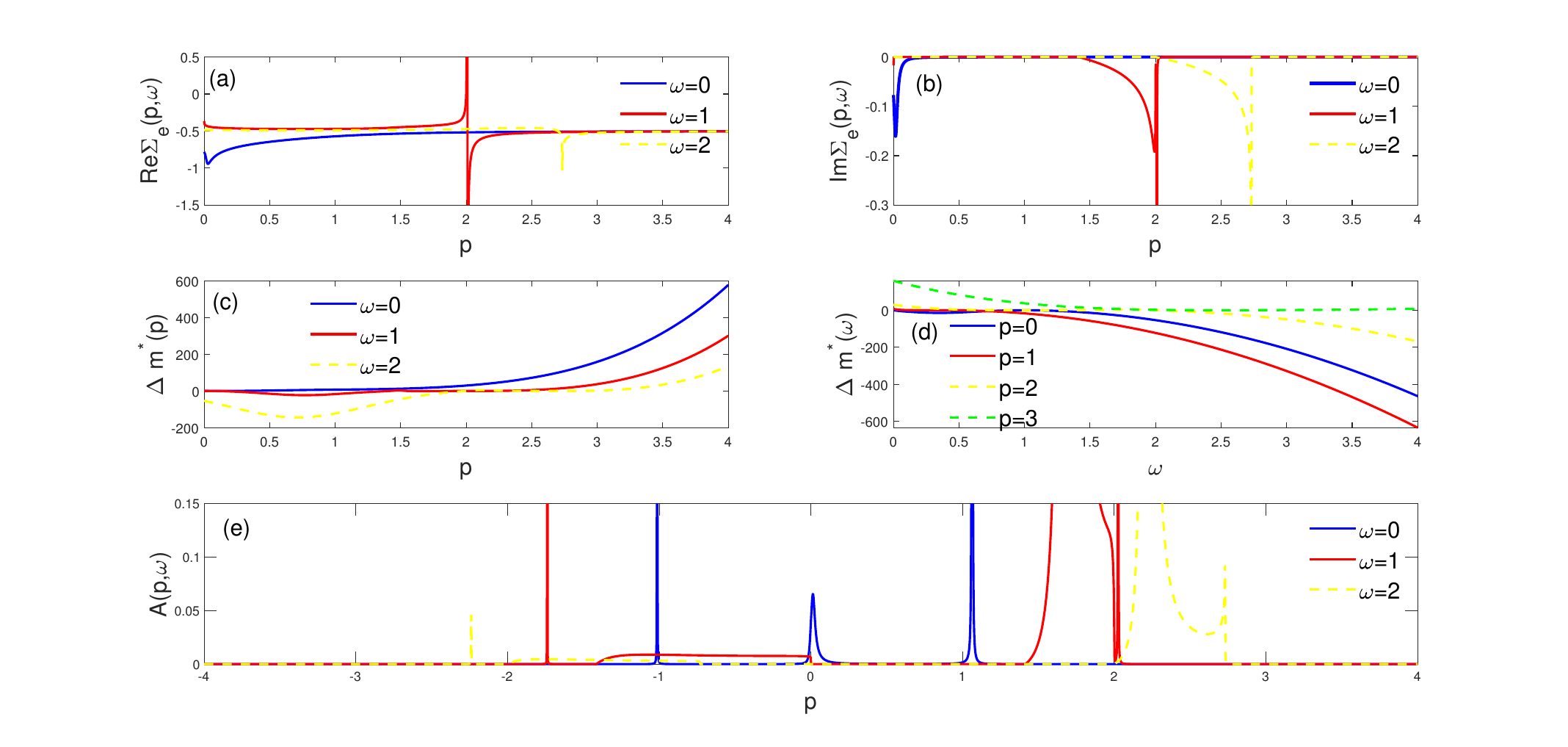}
\caption{Real part (a) and imaginary part (b) of the self-energy induced by the interaction between 
impurity and the electron-hole pair (in coherence case).
We set the chemical potential as zero, and bare coupling as $g_{b}=-0.5$.
(c) and (d) shows the induced effective masses as a function of impurity momentum and frequency respectively.
(e) shows the calculated spectral function as function of momentum.
%For simplicity, we set the momentum of hole and the chemical potential as zero.
% ($q=0$, $\mu=0$).
Here we only present the results for the bare coupling $g_{b}=-0.5$, but
by changing the value of $g_{b}$, we obtain that the real part of self-energy (no matter how large the $\omega$ is) at large
momentum will always equal to $g_{b}$,
i.e., the attractive feature of the polaron enhanced by the increasing $|g_{b}|$.
while the induced effective masses will decrease with the increase of $|g_{b}|$ ($g_{b}<0$).
}
   \end{center}
\end{figure}

\begin{figure}[!ht]
   \centering
 \centering
   \begin{center}
     \includegraphics*[width=1.2\linewidth]{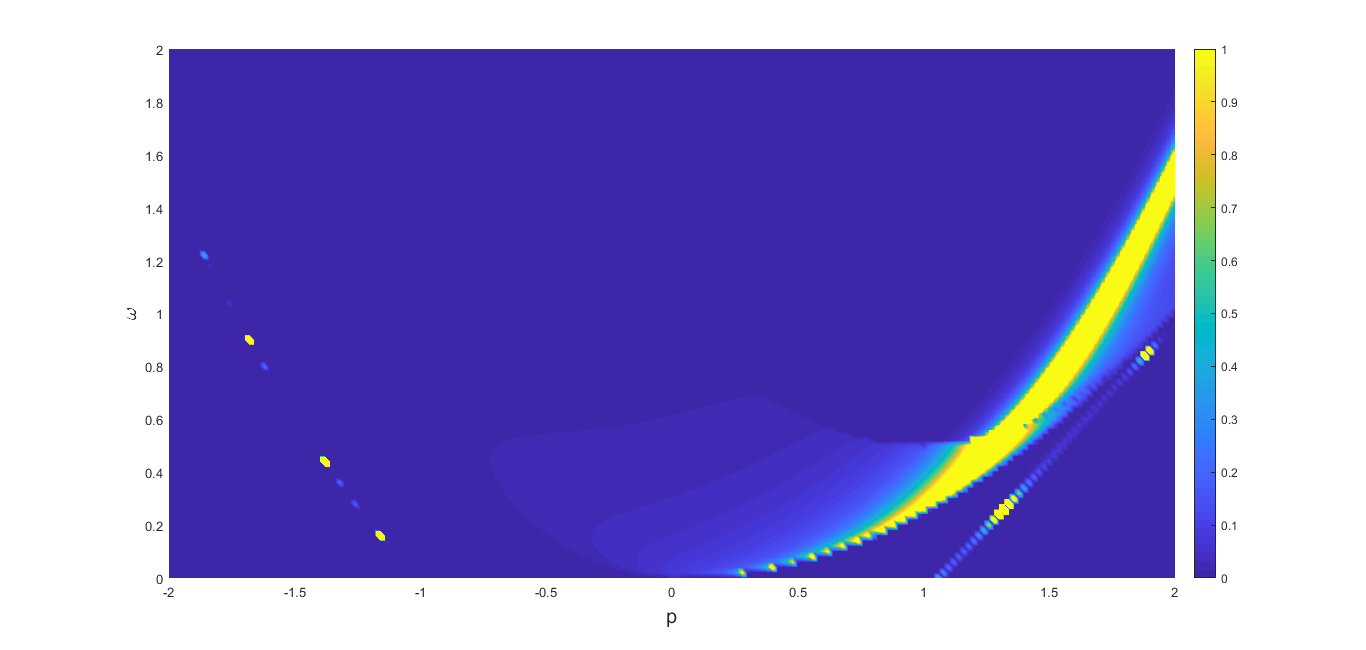}
\caption{Intensity of the spectral function $A(p,\omega)$  due to the interaction between impurity and the electron-hole pair.
}
   \end{center}
\end{figure}

\begin{figure}[!ht]
   \centering
 \centering
   \begin{center}
     \includegraphics*[width=1\linewidth]{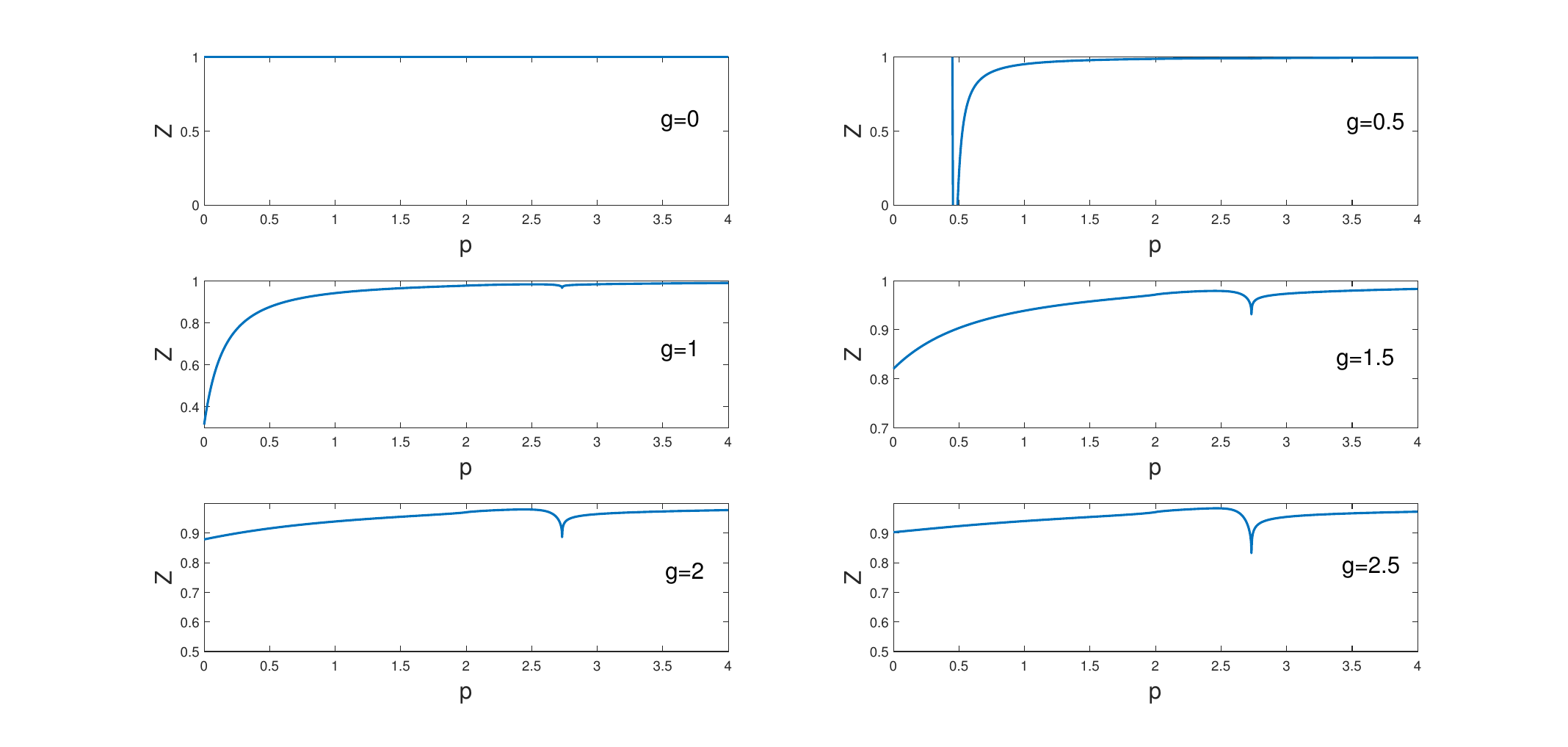}
\caption{Quasiparticle residue.
}
   \end{center}
\end{figure}

\begin{figure}[!ht]
   \centering
 \centering
   \begin{center}
  \includegraphics*[width=1\linewidth]{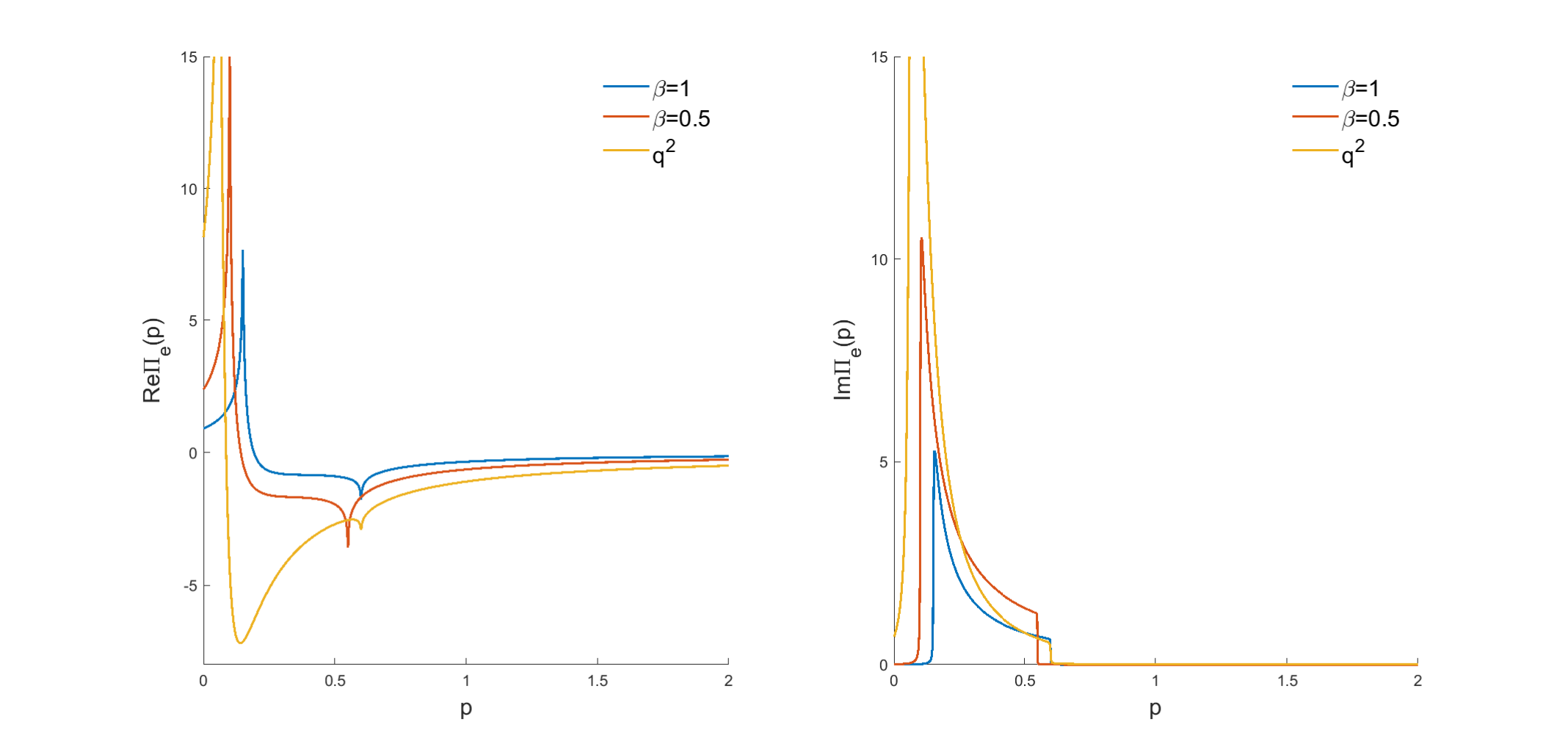}
\caption{Electron self-energy induced by the electron-phonon interaction.
The blue and red curves correspond to the linear phonon dispersion $\Omega_{ph}^{0}(q')=\beta q'$ 
(as found in the LA, TA branchs in 2D hexagonal material) with $\beta=1$ and $\beta=0.5$, respectively.
The yellow curve corresponds to $\Omega_{ph}^{0}(q')= q'^{2}$. 
  %{PHONONICS IN LOW-DIMENSIONS ENGINEERING PHONONS IN NANOSTRUCTURES AND GRAPHENE}
  %{Plasmon-polaron of the topological metallic surface states}
The vertical axis is in unit of $(\frac{3\sqrt{3}t}{a\sqrt{4MN}})^{2}$.
Here we apply the quadratic dispersion of the impurity, and we can see that the resulting impurity self-energy
is momentum $p$-dependent.
While for the Dirac dispersion, where the $p$-dependence vanishes due to the linear relation,
and the resulting self-energy can be obtained as
$-0.0793994839+i(1.0702\times 10^{-5})$, $-0.1561074601+i(2.0684\times 10^{-5})$, $-0.3732+i(2.0348\times 10^{-5})$
for the three cases of $\Omega_{ph}^{0}(q')=q'$, $\Omega_{ph}^{0}(q')=0.5q'$, $\Omega_{ph}^{0}(q')=q'^{2}$,
respectively.
It is obvious that around a Dirac cone, the polaron formed by the coupling between electron and the lattice vibration mode
  %{Ultrafast THz Probe of Photoinduced Polarons in Lead-Halide Perovskites}
is an attractive one and with much smaller imaginary part compared to the free electron.
  %{Bose polarons near quantum criticality}
}
   \end{center}
\end{figure}

\begin{figure}[!ht]
   \centering
 \centering
   \begin{center}
     \includegraphics*[width=1\linewidth]{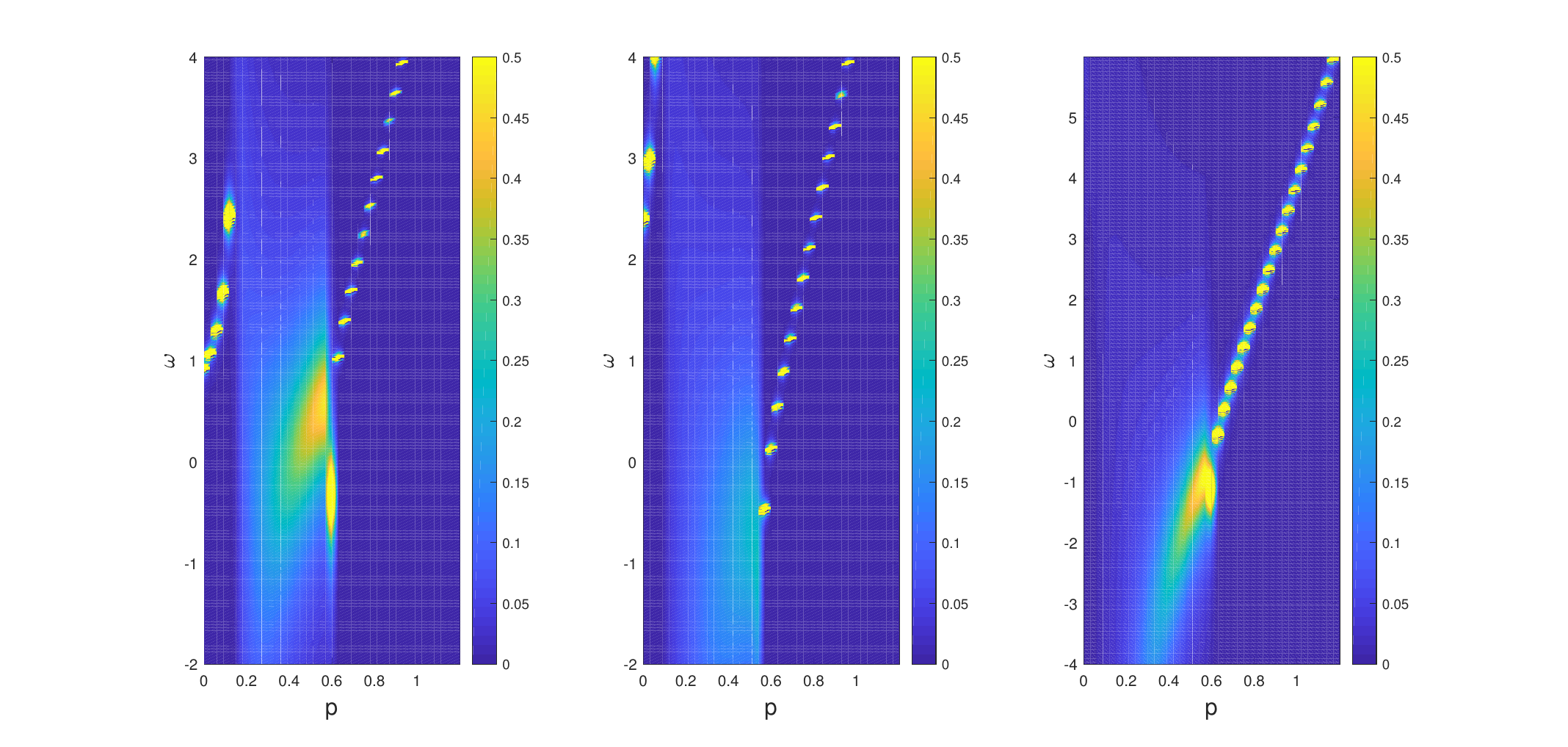}
\caption{Spectral function according to the electron-phonon coupling induced self-energy.
The left, middle, right panels are corresponding to the phonons with dispersion $\Omega_{ph}^{0}=q',\ 0.5q',\ q'^{2}$,
respectively.
The bright regions are the polaronic modes induced by the electron-phonon coupling.
In the region where the imaginary part of self-energy is large,
the polaronic mode is been damped, resembling to the damping of other collective excitations in the single particle 
continuum regime.
  %{Plasmon-polaron of the topological metallic surface states}
}
   \end{center}
\end{figure}

\begin{figure}[!ht]
   \centering
 \centering
   \begin{center}
     \includegraphics*[width=1\linewidth]{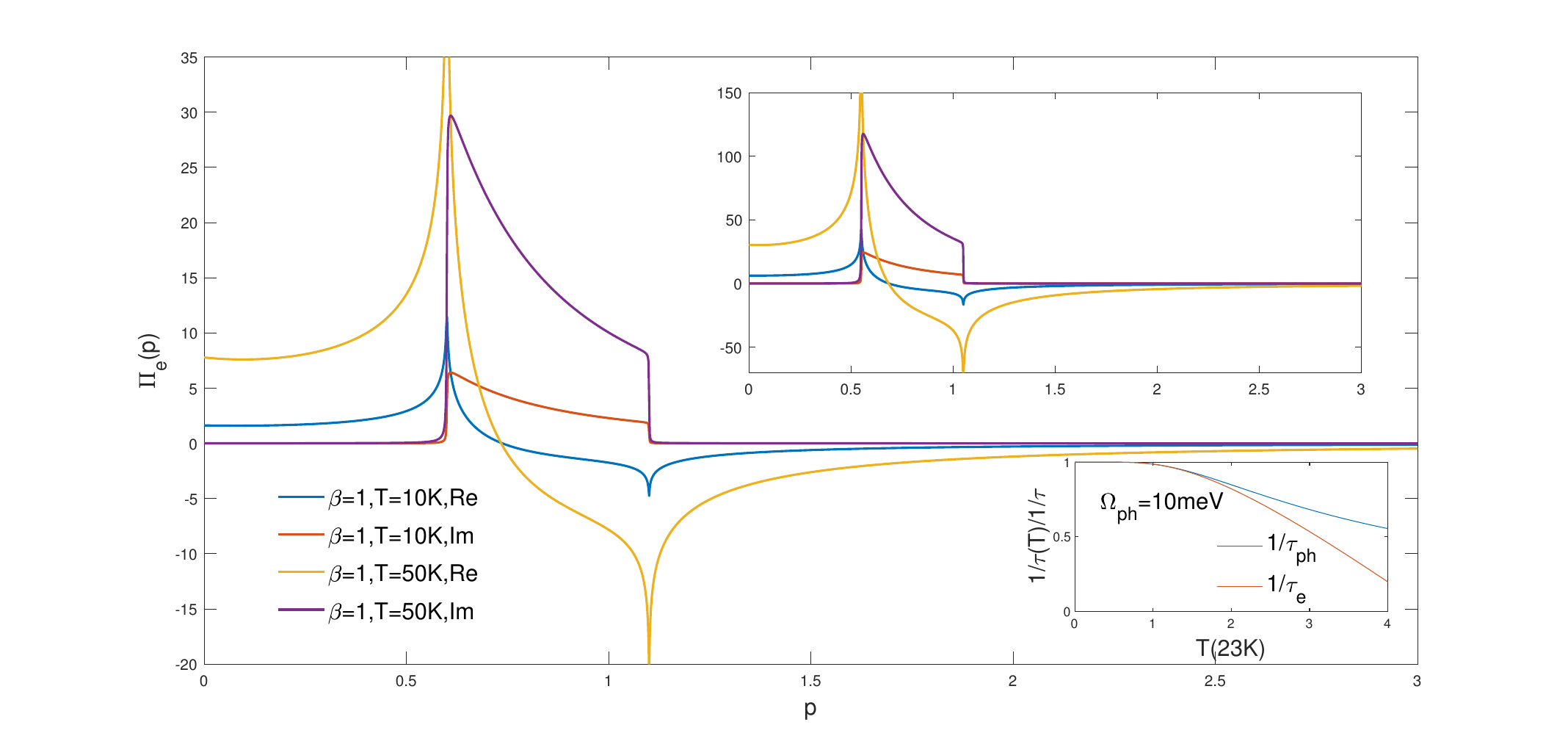}
\caption{Electron self-energy induced by the electron-phonon interaction with $\beta=1$ at finite temperature.
The vertical axis is in unit of $(\frac{3\sqrt{3}t}{a\sqrt{4MN}})^{2}$.
The upper inset shows the case with $\beta=0.5$.
Similar to the Fig.2, the peaks are higher in the case of small $\beta$,
and such difference is more obvious at finite-temperature compared to the zero-temperature.
The lower inset shows the temperature dependence of the electron linewidth ($1/\tau_{e}$) and the phonon linewidth ($1/\tau_{ph}$).
  %{Phonon Anharmonicities in Graphite and Graphene}
Note that, for the calculation of the finite-temperature case, we use the approximation of the series expansion near zero phonon frequency
$\Omega_{ph}\rightarrow 0$,
   %{Exp(\Omega_{ph}/T)\approx 1+\Omega_{ph}/T}
thus the selection of $q'$ is restricted within the small values.
}
   \end{center}
\end{figure}

\begin{figure}[!ht]
   \centering
 \centering
   \begin{center}
     \includegraphics*[width=0.8\linewidth]{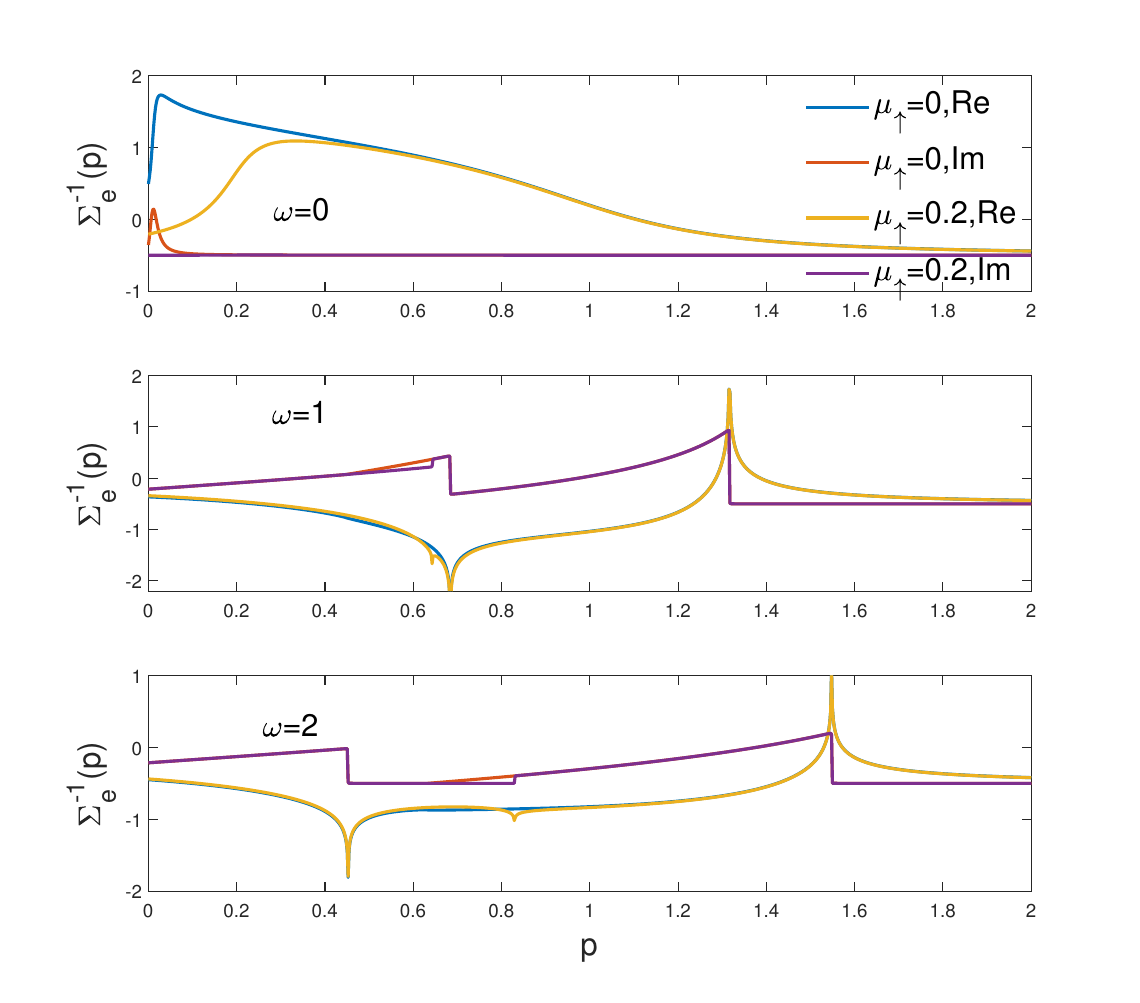}
\caption{Electron self-energy induced by the interaction between impurity and the eletron-hole pair.
 %Here we set the momentum of the majority-electron $k$ near one,
 %$k\sim 1$, as be labeled in each panel.
The curves for $\mu_{\uparrow}=0$ and $\mu_{\uparrow}=0.2$ are presented.
Here $\mu_{\uparrow}\approx E_{F}$ in weak coupling BCS limit.
  %{BCS to Bose crossover: Broken-symmetry state}
  %{Spectral functions and rf response of ultracold fermionic atoms}
We note that here the case of $\mu_{\uparrow}=0$ corresponds to the vacuum limit\cite{Devreese J T} or the two-body limit,
  %{Highly polarized Fermi gases in two dimensions}
  %{Fr?hlich polaron and bipolaron recent developments}
i.e., we only take one particle-hole excitation into account,
which is applicable due to the destructive interference in a system with the superposition between states with different number of
particle-hole pairs.
  %{Normal State of Highly Polarized Fermi Gases Full Many-Body Treatment}
We also set $q=0$ here for the validity of one particle-hole part ansanz (see Ref.\cite{Punk M,Combescot R}).
  %{Polaron-to-molecule transition in a strongly imbalanced Fermi gas}
The imaginary part of the self-energy also equal to the electron scattering rate $\hbar/2\pi\tau$.
}
   \end{center}
\end{figure}

\begin{figure}[!ht]
   \centering
 \centering
   \begin{center}
     \includegraphics*[width=0.8\linewidth]{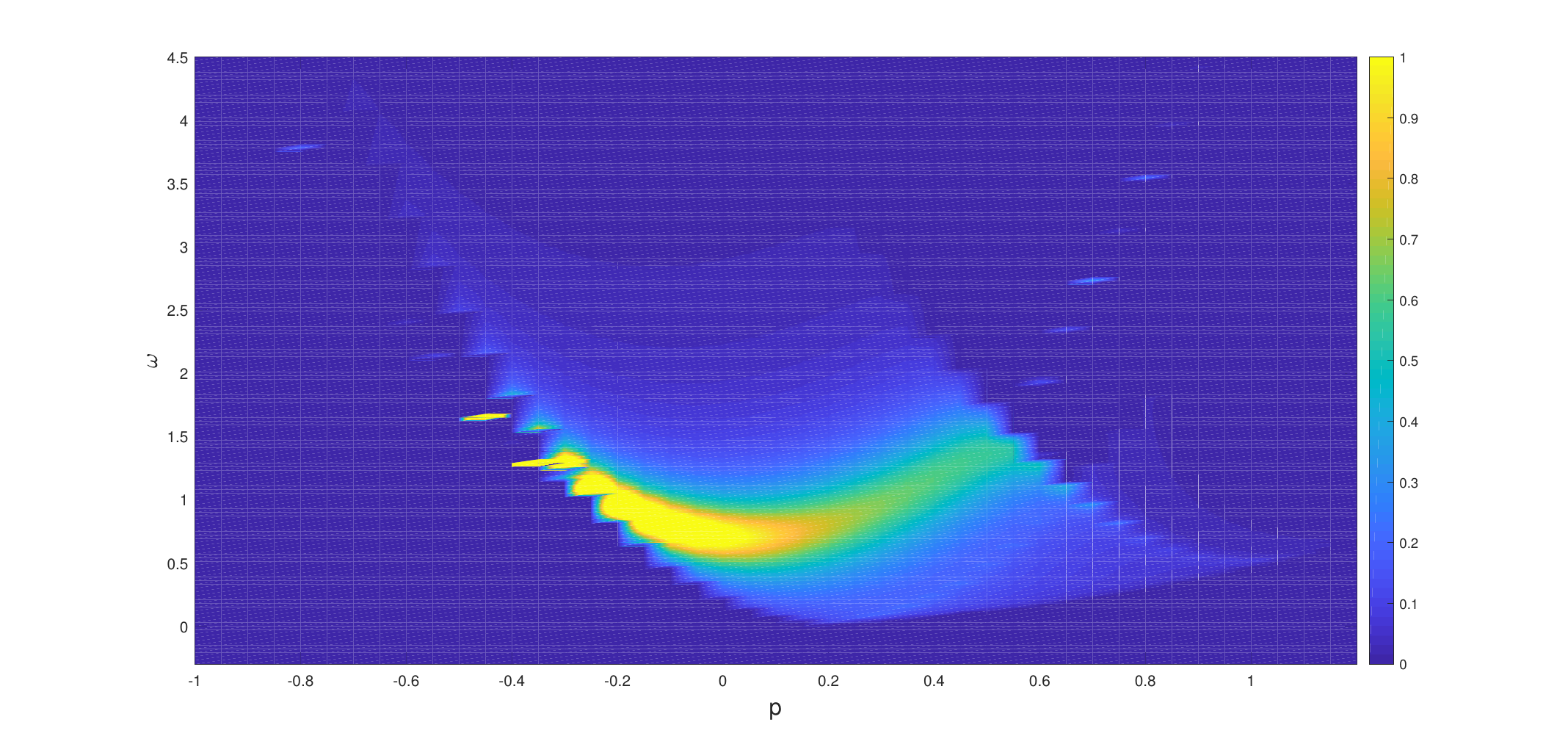}
\caption{Spectral function of the fermi polaron self-energy.
}
   \end{center}
\end{figure}

\begin{figure}[!ht]
   \centering
 \centering
   \begin{center}
     \includegraphics*[width=1\linewidth]{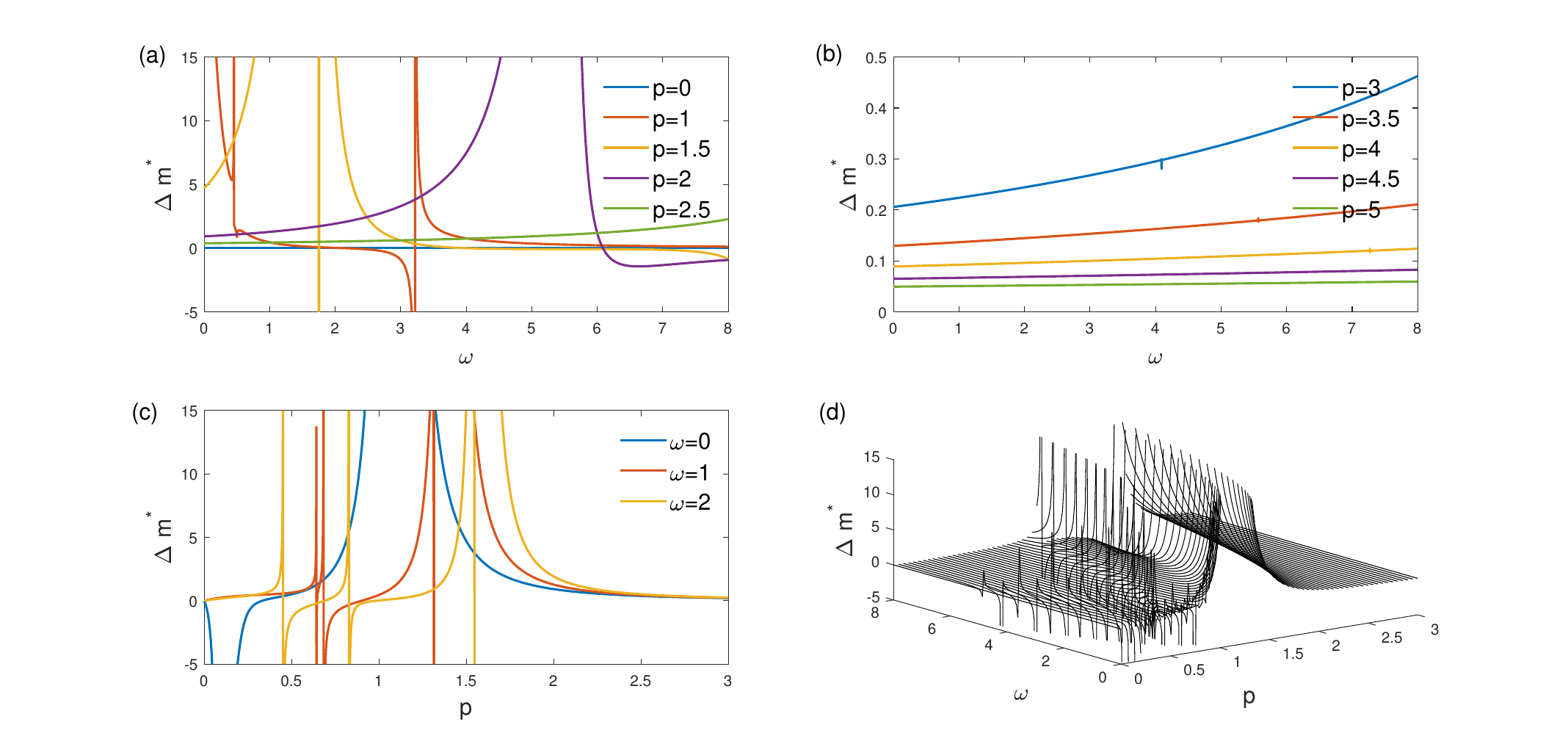}
\caption{Effective mass induced by the interaction between impurity and the eletron-hole pair in the strong coupling regime
($g^{-1}(\Lambda)\rightarrow 0$).
Here we use the approximated expression of $\Delta m^{*}(p,\omega)=\frac{p}{\partial_{p}\Sigma_{e}(p,\omega)}$\cite{Rath S P,Scazza F}.
  %{Field-theoretical study of the Bose polaron}
We set $\mu_{\uparrow}=k_{F}=0.2$ here.
}
   \end{center}
\end{figure}

\begin{figure}[!ht]
   \centering
 \centering
   \begin{center}
     \includegraphics*[width=1\linewidth]{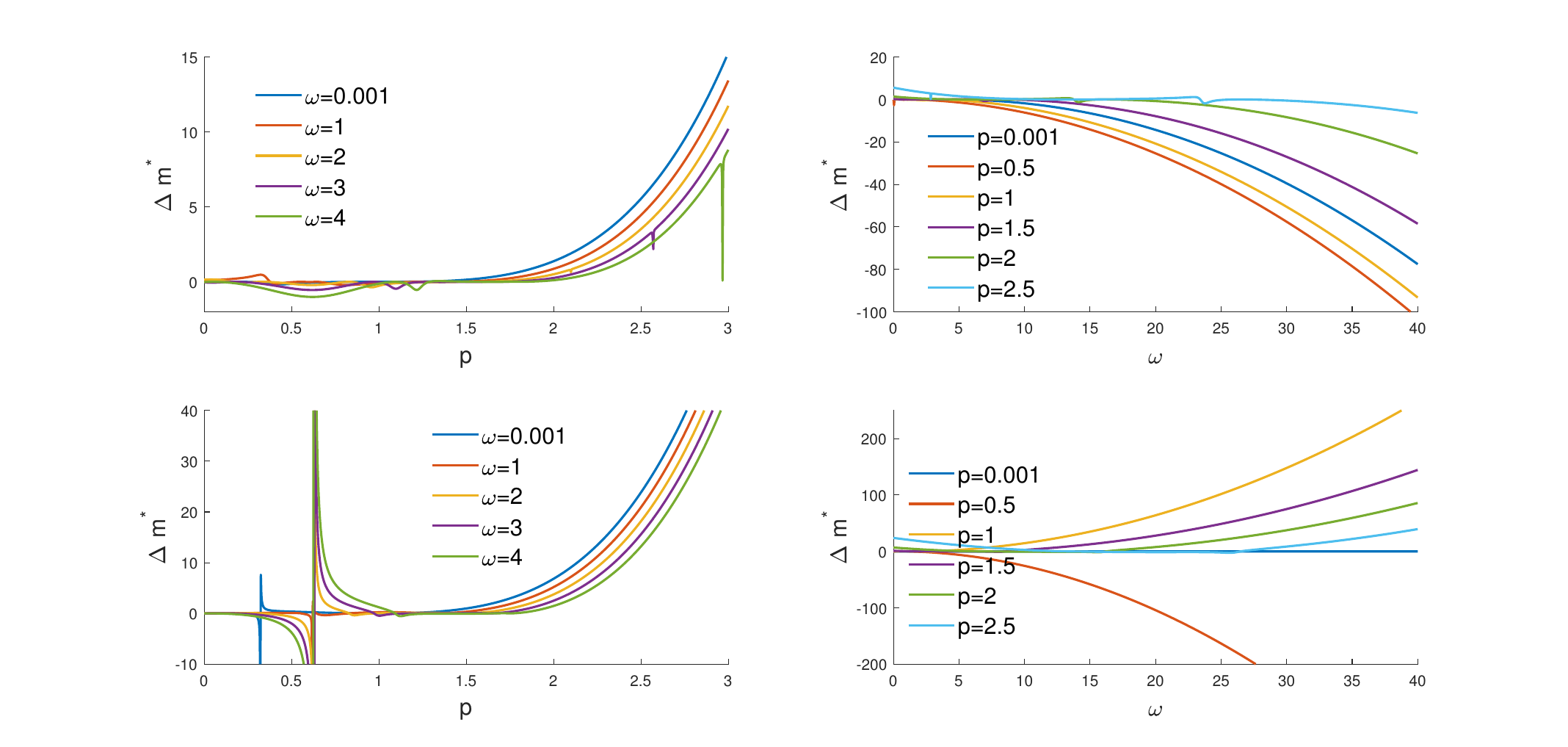}
\caption{Effective mass induced by the interaction between impurity and the eletron-hole pair
with the bare coupling setted as $(g^{b})^{-1}=-0.5$.
 Exact expression (upper panels) approximated
expression (lower panels) are used (see main text).
We set $\mu_{\uparrow}=k_{F}=0.2$ here.
}
   \end{center}
\end{figure}

\begin{figure}[!ht]
   \centering
 \centering
   \begin{center}
     \includegraphics*[width=1\linewidth]{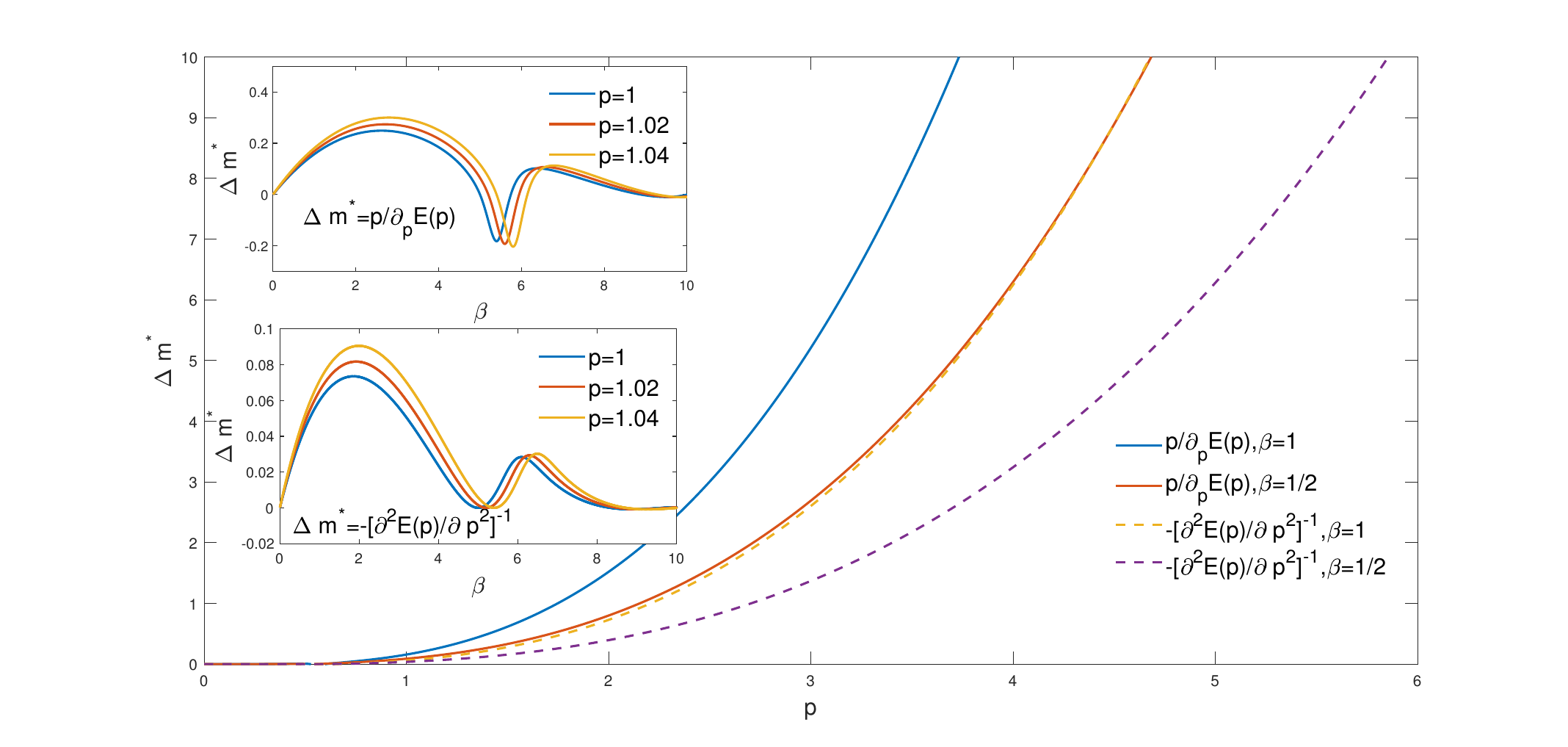}
\caption{Effective mass induced by the electron-phonon coupling.
We present both the approximation result of $m^{*}=\frac{p}{\partial_{p}E(p)}$
and the exact result $m^{*}=-[\frac{\partial ^{2}E(p)}{\partial p^{2}}]^{-1}$.
We can clearly see that the expression of $m^{*}=\frac{p}{\partial_{p}E(p)}$ is applicable in the low-impurity-momentum region.
We also present the effective mass as a function of slope rate of phonon spectrum base on these two expressions
in the upper inset and lower inset, respectively.
The difference is that the negative effective mass does not emerges in the lower inset (the exact result)
which means that the electron-phonon coupling only increases the effective mass of the polaron.
While for the upper inset for momentum $p\sim 1$,
the negative effective mass emerges near the point $\beta=5.5$.
}
   \end{center}
\end{figure}

\begin{figure}[!ht]
   \centering
 \centering
   \begin{center}
     \includegraphics*[width=1\linewidth]{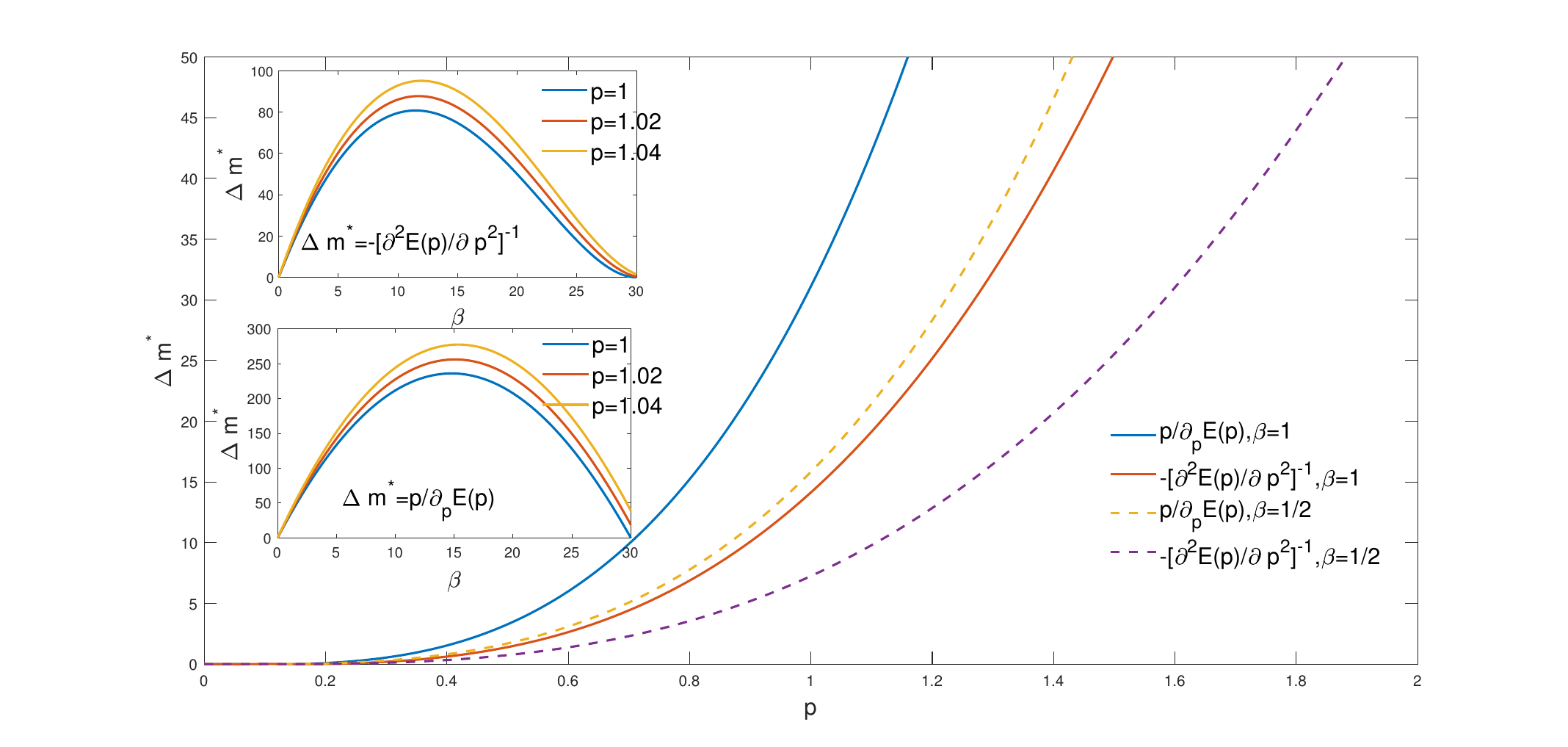}
\caption{The same with Fig.7 but for initial unperturbated impurity located near the Dirac cone.
Here we set $\alpha\approx 30 \beta$ (for LA) according to the first-principle calculations\cite{Ge X J,Yao Y}.  %1 THz=0.00413 eV=4 meV
}
   \end{center}
\end{figure}

\begin{figure}[!ht]
   \centering
 \centering
   \begin{center}
     \includegraphics*[width=1\linewidth]{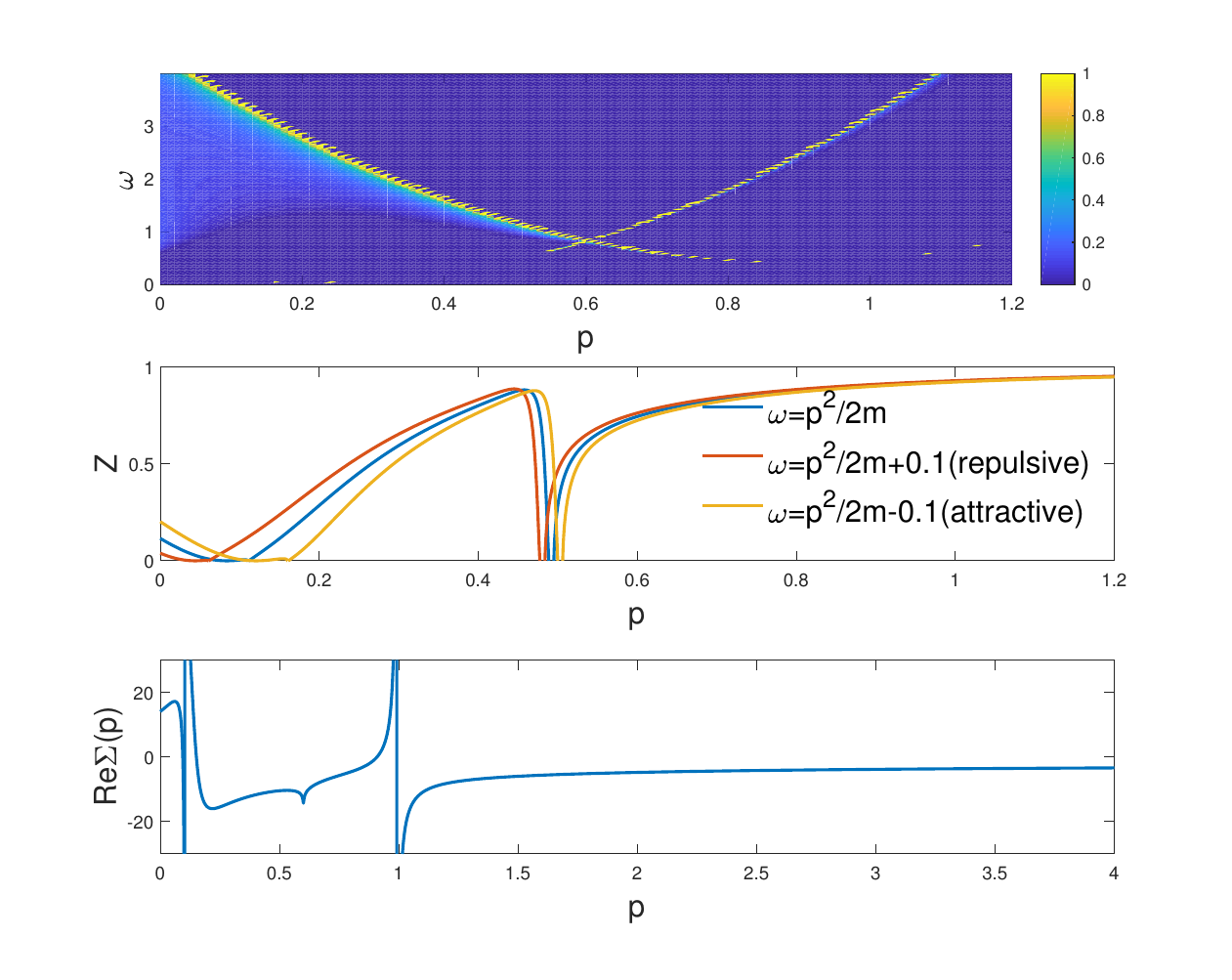}
\caption{(upper panel) Two-dimensional map of 
the quasiparticle residue $Z$ of the composite polaron.
(middle panel) the approximated result of residue $Z$.
(lower panel) the self-energy of the composite polaron.
Note that here the residue in middle panel,
which has two minimal singular points in $p=0.1$ and $p=0.5$,
is not the exact result since 
the choosed impurity frequency $\omega$ is not exactly the pole of impurity Green's function.
Through the lower panel, we can see that the imaginary part of self-energy induced instabilities
are mainly located on the $p=0.1$ and $p=1$.
Thus we can reasonably think that the two minimal singular points of exact residue should also located on the $p=0.1$ and $p=1$,
rather than the $p=0.1$ and $p=0.5$.
}
   \end{center}
\end{figure}
\clearpage
\end{small}


\begin{thebibliography}{99}
\bibitem{Scazza F}Scazza F, Valtolina G, Massignan P, et al. Repulsive Fermi polarons in a resonant mixture of ultracold Li 6 atoms[J]. Physical review letters, 2017, 118(8): 083602.
\bibitem{Li W}Li W, Sarma S D. Variational study of polarons in Bose-Einstein condensates[J]. Physical Review A, 2014, 90(1): 013618.
\bibitem{Fratini E}Fratini E, Pieri P. Mass imbalance effect in resonant Bose-Fermi mixtures[J]. Physical Review A, 2012, 85(6): 063618.
  %\bibitem{Haussmann R}Haussmann R, Punk M, Zwerger W. Spectral functions and rf response of ultracold fermionic atoms[J]. Physical Review A, 2009, 80(6): 063612.
\bibitem{Sidler M}Sidler M, Back P, Cotlet O, et al. Fermi polaron-polaritons in charge-tunable atomically thin semiconductors[J]. Nature Physics, 2017, 13(3): 255.
\bibitem{polaron2}Wu C H. Attractive fermi polaron in a semi-Dirac system within ladder approximation[J]. arXiv preprint arXiv:1901.07881, 2019.
\bibitem{polaron1}Wu C H. Attractive polaron in a Dirac system within ladder approximation[J]. arXiv preprint arXiv:1812.04833, 2018.
\bibitem{Qin F}Qin F, Cui X, Yi W. Polaron in a $ p+ ip $ Fermi topological superfluid[J]. arXiv preprint arXiv:1901.02766, 2019.
\bibitem{Caldas H}Caldas H, Mota A L, Farias R L S, et al. Superfluidity in two-dimensional imbalanced Fermi gases[J]. Journal of Statistical Mechanics: Theory and Experiment, 2012, 2012(10): P10019.
\bibitem{Wang D W}Wang D W, Lukin M D, Demler E. Engineering superfluidity in Bose-Fermi mixtures of ultracold atoms[J]. Physical Review A, 2005, 72(5): 051604.
\bibitem{Kinnunen J J}Kinnunen J J, Wu Z, Bruun G M. Induced p-wave pairing in Bose-Fermi mixtures[J]. Physical review letters, 2018, 121(25): 253402.
\bibitem{Fukuhara T}Fukuhara T, Kantian A, Endres M, et al. Quantum dynamics of a mobile spin impurity[J]. Nature Physics, 2013, 9(4): 235.
\bibitem{Koschorreck M}Koschorreck M, Pertot D, Vogt E, et al. Attractive and repulsive Fermi polarons in two dimensions[J]. Nature, 2012, 485(7400): 619.
\bibitem{Kasprzak J}Kasprzak J, Richard M, Kundermann S, et al. Bose-Einstein condensation of exciton polaritons[J]. Nature, 2006, 443(7110): 409.
\bibitem{Camacho-Guardian A}Camacho-Guardian A, Goldman N, Massignan P, et al. Dropping an impurity into a Chern insulator: a polaron view on topological matter[J]. Physical Review B, 2019, 99(8): 081105.
%\bibitem{Jia X}Jia X, Zhang S, Sankar R, et al. Anomalous acoustic plasmon mode from topologically protected states[J]. Physical review letters, 2017, 119(13): 136805.
\bibitem{Polini M}Polini M, Asgari R, Borghi G, et al. Plasmons and the spectral function of graphene[J]. Physical Review B, 2008, 77(8): 081411.
\bibitem{Wang Z}Wang Z, Zhao J, Frank B, et al. Plasmon–polaron coupling in conjugated polymer on infrared nanoantennas[J]. Nano letters, 2015, 15(8): 5382-5387.
\bibitem{Blume D}Blume D. Efimov physics and the three-body parameter for shallow van der waals potentials[J]. Few-Body Systems, 2015, 56(11-12): 859-867.
\bibitem{Wang Y}Wang Y, Julienne P S. Universal van der Waals physics for three cold atoms near Feshbach resonances[J]. Nature Physics, 2014, 10(10): 768.
\bibitem{Rath S P}Rath S P, Schmidt R. Field-theoretical study of the Bose polaron[J]. Physical Review A, 2013, 88(5): 053632.
\bibitem{Narayanaswamy A}Narayanaswamy A, Zheng Y. van der Waals energy and pressure in dissipative media: Fluctuational electrodynamics and mode summation[J]. Physical Review A, 2013, 88(1): 012502.
\bibitem{Chen C}Chen C, Avila J, Wang S, et al. Emergence of Interfacial Polarons from Electron–Phonon Coupling in Graphene/h-BN van der Waals Heterostructures[J]. Nano letters, 2018, 18(2): 1082-1087.
\bibitem{Massignan P}Massignan P. Polarons and dressed molecules near narrow Feshbach resonances[J]. EPL (Europhysics Letters), 2012, 98(1): 10012.
\bibitem{Bruun G M}Bruun G M, Jackson A D, Kolomeitsev E E. Multichannel scattering and Feshbach resonances: Effective theory, phenomenology, and many-body effects[J]. Physical Review A, 2005, 71(5): 052713.
\bibitem{Ge X J}Ge X J, Yao K L, Lü J T. Comparative study of phonon spectrum and thermal expansion of graphene, silicene, germanene, and blue phosphorene[J]. Physical Review B, 2016, 94(16): 165433.
%\bibitem{Bruun G M}Bruun G M, Jackson A D, Kolomeitsev E E. Multichannel scattering and %Feshbach resonances: Effective theory, phenomenology, and many-body effects[J]. Physical %Review A, 2005, 71(5): 052713.
\bibitem{Combescot R}Combescot R, Giraud S. Normal state of highly polarized Fermi gases: full many-body treatment[J]. Physical review letters, 2008, 101(5): 050404.
\bibitem{Levine M}Levine M, Tjon J, Wright J. Nonsingular Bethe-Salpeter Equation[J]. Physical Review Letters, 1966, 16(21): 962.
\bibitem{Karmanov V A}Karmanov V A, Carbonell J. Solving Bethe-Salpeter equation in Minkowski space[J]. The European Physical Journal A-Hadrons and Nuclei, 2006, 27(1): 1-9.
\bibitem{Camacho-Guardian A2}Camacho-Guardian A, Ardila L A P, Pohl T, et al. Bipolarons in a Bose-Einstein condensate[J]. Physical review letters, 2018, 121(1): 013401.
%\bibitem{Massignan P}Massignan P. Polarons and dressed molecules near narrow Feshbach resonances[J]. EPL (Europhysics Letters), 2012, 98(1): 10012.
\bibitem{Diestler D J}Diestler D J, Zewail A H. Vibronic dephasing of anharmonic molecules. I. Theory and its application to the separability of intra‐and intermolecular processes[J]. The Journal of Chemical Physics, 1979, 71(7): 3103-3112.
\bibitem{Andrews M R}Andrews M R, Kurn D M, Miesner H J, et al. Propagation of sound in a Bose-Einstein condensate[J]. Physical review letters, 1997, 79(4): 553.
\bibitem{Nielsen K K}Nielsen K K, Ardila L A P, Bruun G M, et al. Critical slowdown of non-equilibrium polaron dynamics[J]. New Journal of Physics, 2019.
\bibitem{Visuri A M}Visuri A M, Kinnunen J J, Baarsma J E, et al. Decoherence of an impurity in a one-dimensional fermionic bath with mass imbalance[J]. Physical Review A, 2016, 94(1): 013619.
\bibitem{Hadzibabic Z}Hadzibabic Z, Krüger P, Cheneau M, et al. Berezinskii–Kosterlitz–Thouless crossover in a trapped atomic gas[J]. Nature, 2006, 441(7097): 1118.
\bibitem{Merkl M}Merkl M, Jacob A, Zimmer F E, et al. Chiral confinement in quasirelativistic Bose-Einstein condensates[J]. Physical review letters, 2010, 104(7): 073603.
\bibitem{Haddad L H}Haddad L H, Carr L D. The nonlinear Dirac equation in Bose–Einstein condensates: Foundation and symmetries[J]. Physica D: Nonlinear Phenomena, 2009, 238(15): 1413-1421.
\bibitem{Kim N Y}Kim N Y, Kusudo K, L{\"o}ffler A, et al. Exciton–polariton condensates near the Dirac point in a triangular lattice[J]. New Journal of Physics, 2013, 15(3): 035032.
\bibitem{Mahan G D}Mahan G D. Many-particle physics[M]. Springer Science $\&$ Business Media, 2013.
\bibitem{Li X}Li X, Mullen J T, Jin Z, et al. Intrinsic electrical transport properties of monolayer silicene and MoS 2 from first principles[J]. Physical Review B, 2013, 87(11): 115418.
\bibitem{Matthes L}Matthes L, Gori P, Pulci O, et al. Universal infrared absorbance of two-dimensional honeycomb group-IV crystals[J]. Physical Review B, 2013, 87(3): 035438.
\bibitem{Borysenko K M}Borysenko K M, Mullen J T, Barry E A, et al. First-principles analysis of electron-phonon interactions in graphene[J]. Physical Review B, 2010, 81(12): 121412.
\bibitem{Yan J A}Yan J A, Stein R, Schaefer D M, et al. Electron-phonon coupling in two-dimensional silicene and germanene[J]. Physical Review B, 2013, 88(12): 121403.
\bibitem{Yan J A2}Yan J A, Ruan W Y, Chou M Y. Electron-phonon interactions for optical-phonon modes in few-layer graphene: First-principles calculations[J]. Physical Review B, 2009, 79(11): 115443.
\bibitem{Alexandrov A S}Alexandrov A S. Breakdown of the Migdal-Eliashberg theory in the strong-coupling adiabatic regime[J]. EPL (Europhysics Letters), 2001, 56(1): 92.
\bibitem{Gusynin V P}Gusynin V P, Sharapov S G. Transport of Dirac quasiparticles in graphene: Hall and optical conductivities[J]. Physical Review B, 2006, 73(24): 245411.
\bibitem{Pyatkovskiy P K}Pyatkovskiy P K. Dynamical polarization, screening, and plasmons in gapped graphene[J]. Journal of Physics: Condensed Matter, 2008, 21(2): 025506.
\bibitem{Chen Z}Chen Z, Wu B. Bose-Einstein condensate in a honeycomb optical lattice: Fingerprint of superfluidity at the Dirac point[J]. Physical review letters, 2011, 107(6): 065301.
\bibitem{Liu J}Liu J, Wu B, Niu Q. Nonlinear evolution of quantum states in the adiabatic regime[J]. Physical review letters, 2003, 90(17): 170404.
\bibitem{Wu B}Wu B, Niu Q. Nonlinear landau-zener tunneling[J]. Physical Review A, 2000, 61(2): 023402.
\bibitem{Liu J2}Liu J, Fu L, Ou B Y, et al. Theory of nonlinear Landau-Zener tunneling[J]. Physical Review A, 2002, 66(2): 023404.
\bibitem{Liu F}Liu F, Liu C C, Wu K, et al. d+id Chiral Superconductivity in Bilayer Silicene[J]. Physical review letters, 2013, 111(6): 066804.
\bibitem{Yi W}Yi W, Zhang W. Molecule and polaron in a highly polarized two-dimensional Fermi gas with spin-orbit coupling[J]. Physical review letters, 2012, 109(14): 140402.
\bibitem{Chen J G}Chen J G, Deng T S, Yi W, et al. Polarons and molecules in a Fermi gas with orbital Feshbach resonance[J]. Physical Review A, 2016, 94(5): 053627.
\bibitem{Band Y B}Band Y B, Trippenbach M. Bose-Einstein condensates in time-dependent light potentials: Adiabatic and nonadiabatic behavior of nonlinear wave equations[J]. Physical Review A, 2002, 65(5): 053602.
\bibitem{Berman O L}Berman O L, Lozovik Y E, Gumbs G. Bose-Einstein condensation and superfluidity of magnetoexcitons in bilayer graphene[J]. Physical Review B, 2008, 77(15): 155433.
\bibitem{Min H}Min H, Bistritzer R, Su J J, et al. Room-temperature superfluidity in graphene bilayers[J]. Physical Review B, 2008, 78(12): 121401.
\bibitem{Blakie P B}Blakie P B, Clark C W. Wannier states and Bose–Hubbard parameters for 2D optical lattices[J]. Journal of Physics B: Atomic, Molecular and Optical Physics, 2004, 37(7): 1391.
\bibitem{Wu B2}Wu B, Niu Q. Superfluidity of Bose–Einstein condensate in an optical lattice: Landau–Zener tunnelling and dynamical instability[J]. New journal of Physics, 2003, 5(1): 104.
\bibitem{Wellein G}Wellein G, R{\"o}der H, Fehske H. Polarons and bipolarons in strongly interacting electron-phonon systems[J]. Physical Review B, 1996, 53(15): 9666.
%\bibitem{}Camacho-Guardian A, Ardila L A P, Pohl T, et al. Bipolarons in a Bose-Einstein condensate[J]. Physical review letters, 2018, 121(1): 013401.
\bibitem{Alexandrov A S2}Alexandrov A S, Mott N F. Bipolarons[J]. Reports on Progress in Physics, 1994, 57(12): 1197.
\bibitem{Jotzu G}Jotzu G, Messer M, G?rg F, et al. Creating state-dependent lattices for ultracold fermions by magnetic gradient modulation[J]. Physical review letters, 2015, 115(7): 073002.
\bibitem{Kohstall C}Kohstall C, Zaccanti M, Jag M, et al. Metastability and coherence of repulsive polarons in a strongly interacting Fermi mixture[J]. Nature, 2012, 485(7400): 615.
\bibitem{Carneiro L M}Carneiro L M, Cushing S K, Liu C, et al. Excitation-wavelength-dependent small polaron trapping of photoexcited carriers in $\alpha$-Fe2O3[J]. Nature materials, 2017, 16(8): 819.
\bibitem{Vlietinck J}Vlietinck J, Ryckebusch J, Van Houcke K. Diagrammatic Monte Carlo study of the Fermi polaron in two dimensions[J]. Physical Review B, 2014, 89(8): 085119.
\bibitem{Panochko G}Panochko G, Pastukhov V, Vakarchuk I. Behavior of the impurity atom in a weakly-interacting Bose gas[J]. arXiv preprint arXiv:1703.10390, 2017.
\bibitem{Giustino F}Giustino F, Cohen M L, Louie S G. Electron-phonon interaction using Wannier functions[J]. Physical Review B, 2007, 76(16): 165108.
\bibitem{Perebeinos V}Perebeinos V, Tersoff J, Avouris P. Electron-phonon interaction and transport in semiconducting carbon nanotubes[J]. Physical review letters, 2005, 94(8): 086802.
\bibitem{Muntwiler M}Muntwiler M, Yang Q, Tisdale W A, et al. Coulomb barrier for charge separation at an organic semiconductor interface[J]. Physical review letters, 2008, 101(19): 196403.
\bibitem{Verissimo-Alves M}Verissimo-Alves M, Capaz R B, Koiller B, et al. Polarons in carbon nanotubes[J]. Physical review letters, 2001, 86(15): 3372.
\bibitem{Sous J}Sous J, Chakraborty M, Krems R V, et al. Light bipolarons stabilized by Peierls electron-phonon coupling[J]. Physical review letters, 2018, 121(24): 247001.
\bibitem{Allen P B}Allen P B. Neutron spectroscopy of superconductors[J]. Physical Review B, 1972, 6(7): 2577.
\bibitem{Zijlstra E S}Zijlstra E S, Tatarinova L L, Garcia M E. Laser-induced phonon-phonon interactions in bismuth[J]. Physical Review B, 2006, 74(22): 220301.
\bibitem{Marchand D J J}Marchand D J J, De Filippis G, Cataudella V, et al. Sharp transition for single polarons in the one-dimensional Su-Schrieffer-Heeger model[J]. Physical review letters, 2010, 105(26): 266605.
\bibitem{Piegari E}Piegari E, Cataudella V, Ramaglia V M, et al. Comment on “polarons in carbon nanotubes”[J]. Physical review letters, 2002, 89(4): 049701.
\bibitem{Knupfer M}Knupfer M, Pichler T, Golden M S, et al. Size of electron-hole pairs in n-conjugated systems[J]. Physical review letters, 1999, 83(7): 1443.
\bibitem{Combescot R2}Combescot R, Recati A, Lobo C, et al. Normal state of highly polarized Fermi gases: simple many-body approaches[J]. Physical review letters, 2007, 98(18): 180402.
\bibitem{Devreese J T}Devreese J T, Alexandrov A S. Fr?hlich polaron and bipolaron: recent developments[J]. Reports on Progress in Physics, 2009, 72(6): 066501.
\bibitem{Sentef M A}Sentef M A. Light-enhanced electron-phonon coupling from nonlinear electron-phonon coupling[J]. Physical Review B, 2017, 95(20): 205111.
\bibitem{Berciu M}Berciu M. Few-particle Green’s functions for strongly correlated systems on infinite lattices[J]. Physical review letters, 2011, 107(24): 246403.
\bibitem{Sous J2}Sous J, Berciu M, Krems R V. Bipolarons bound by repulsive phonon-mediated interactions[J]. Physical Review A, 2017, 96(6): 063619.
\bibitem{Sous J3}Sous J, Chakraborty M, Adolphs C P J, et al. Phonon-mediated repulsion, sharp transitions and (quasi) self-trapping in the extended Peierls-Hubbard model[J]. Scientific reports, 2017, 7(1): 1169.
\bibitem{Yao Y}Yao Y, Ye F, Qi X L, et al. Spin-orbit gap of graphene: First-principles calculations[J]. Physical Review B, 2007, 75(4): 041401.
\bibitem{Haussmann R}Haussmann R, Punk M, Zwerger W. Spectral functions and rf response of ultracold fermionic atoms[J]. Physical Review A, 2009, 80(6): 063612.

\bibitem{Grusdt F}Grusdt F, Yao N Y, Demler E. Topological polarons, quasiparticle invariants and their detection in 1D symmetry-protected phases[J]. arXiv preprint arXiv:1904.00220, 2019.
\bibitem{Qi F}Qi F, Cao J, Cao J, et al. Topological phase transition based on the attractive Hubbard model[J]. arXiv preprint arXiv:1904.00171, 2019.
\bibitem{Ezawa M}Ezawa M. Valley-polarized metals and quantum anomalous Hall effect in silicene[J]. Physical review letters, 2012, 109(5): 055502.

\bibitem{Ardila L A P}Ardila L A P, Giorgini S. Impurity in a Bose-Einstein condensate: Study of the attractive and repulsive branch using quantum Monte Carlo methods[J]. Physical Review A, 2015, 92(3): 033612.
\bibitem{Ichmoukhamedov T}Ichmoukhamedov T, Tempere J. Feynman path-integral treatment of the Bose polaron beyond the Fr\" ohlich model[J]. arXiv preprint arXiv:1905.07368, 2019.
\bibitem{Van Loon S}Van Loon S, Casteels W, Tempere J. Ground-state properties of interacting Bose polarons[J]. Physical Review A, 2018, 98(6): 063631.
\bibitem{Combescot R3}Combescot R, Giraud S, Leyronas X. Analytical theory of the dressed bound state in highly polarized Fermi gases[J]. EPL (Europhysics Letters), 2010, 88(6): 60007.
\bibitem{Parish M M}Parish M M, Levinsen J. Highly polarized Fermi gases in two dimensions[J]. Physical Review A, 2013, 87(3): 033616.
\bibitem{Christensen R S}Christensen R S, Levinsen J, Bruun G M. Quasiparticle properties of a mobile impurity in a Bose-Einstein condensate[J]. Physical review letters, 2015, 115(16): 160401.
\bibitem{Wang J R}Wang J R, Liu G Z, Zhang C J. Breakdown of Fermi liquid theory in topological multi-Weyl semimetals[J]. Physical Review B, 2018, 98(20): 205113.
\bibitem{Cinquanta E}Cinquanta E, Meggiolaro D, Motti S G, et al. Ultrafast THz Probe of Photoinduced Polarons in Lead-Halide Perovskites[J]. Physical review letters, 2019, 122(16): 166601.
\bibitem{Kopnin N B}Kopnin N B, Sonin E B. BCS superconductivity of Dirac electrons in graphene layers[J]. Physical review letters, 2008, 100(24): 246808.
\bibitem{Punk M}Punk M, Dumitrescu P T, Zwerger W. Polaron-to-molecule transition in a strongly imbalanced Fermi gas[J]. Physical Review A, 2009, 80(5): 053605.
\bibitem{Wong W P D}Wong W P D, Yin J, Chaudhary B, et al. Large Polaron Generation and Dynamics in 3D Metal-Halide Perovskites[J]. arXiv preprint arXiv:1905.11704, 2019.
\bibitem{Dong X Y}Dong X Y, Li R Z, Deng J P, et al. Interlayer exciton-polaron effect in transition metal dichalcogenides van der Waals heterostructures[J]. Journal of Physics and Chemistry of Solids, 2019.
\bibitem{Shvonski A}Shvonski A, Kong J, Kempa K. Plasmon-polaron of the topological metallic surface states[J]. Physical Review B, 2019, 99(12): 125148.
\bibitem{ZZ Yan}ZZ Yan, Y Ni, C Robens, MW Zwierlein. Bose polarons near quantum criticality. arXiv preprint arXiv:1904.02685
%\bibitem{Wang J R}Wang J R, Liu G Z, Zhang C J. Breakdown of Fermi liquid theory in topological multi-Weyl semimetals[J]. Physical Review B, 2018, 98(20): 205113.
\bibitem{Burkov A A}Burkov A A, Hook M D, Balents L. Topological nodal semimetals[J]. Physical Review B, 2011, 84(23): 235126.
\bibitem{Jafari S A}Jafari S A, Baskaran G. Equations-of-motion method for triplet excitation operators in graphene[J]. Journal of Physics: Condensed Matter, 2012, 24(9): 095601.
\bibitem{Zhang S}Zhang S, Wei T, Guan J, et al. Enhanced Superconducting State in FeSe/SrTiO 3 by a Dynamic Interfacial Polaron Mechanism[J]. Physical review letters, 2019, 122(6): 066802.

\bibitem{Wong W P D}Wong W P D, Yin J, Chaudhary B, et al. Large Polaron Generation and Dynamics in 3D Metal-Halide Perovskites[J]. arXiv preprint arXiv:1905.11704, 2019.

%Haussmann R, Punk M, Zwerger W. Spectral functions and rf response of ultracold fermionic atoms[J]. Physical Review A, 2009, 80(6): 063612.
\bibitem{Vukmirovi? N}Vukmirovi{\'c} N, Bruder C, Stojanovi{\'c} V M. Electron-phonon coupling in crystalline organic semiconductors: microscopic evidence for nonpolaronic charge carriers[J]. Physical review letters, 2012, 109(12): 126407.
%\bibitem{Vukmirovi? N}Vukmirovi? N, Bruder C, Stojanovi? V M. Electron-phonon coupling in crystalline organic semiconductors: microscopic evidence for nonpolaronic charge carriers[J].         Physical review letters, 2012, 109(12): 126407.

\bibitem{Attaccalite C}Attaccalite C, Wirtz L, Lazzeri M, et al. Doped graphene as tunable electron? phonon coupling material[J]. Nano letters, 2010, 10(4): 1172-1176.
\bibitem{Wu F}Wu F, Hwang E, Sarma S D. Phonon-induced giant linear-in-T resistivity in magic angle twisted bilayer graphene: Ordinary strangeness and exotic superconductivity[J]. Physical Review B, 2019, 99(16): 165112.
\bibitem{Jia X}Jia X, Zhang S, Sankar R, et al. Anomalous acoustic plasmon mode from topologically protected states[J]. Physical review letters, 2017, 119(13): 136805.

\bibitem{Cao Y}Cao Y, Fatemi V, Fang S, et al. Unconventional superconductivity in magic-angle graphene superlattices[J]. Nature, 2018, 556(7699): 43.
\bibitem{Wellein G}Wellein G, R{\"o}der H, Fehske H. Polarons and bipolarons in strongly interacting electron-phonon systems[J]. Physical Review B, 1996, 53(15): 9666.
\bibitem{Attaccalite C}Attaccalite C, Wirtz L, Lazzeri M, et al. Doped graphene as tunable electron? phonon coupling material[J]. Nano letters, 2010, 10(4): 1172-1176.
\bibitem{Michiardi M}Michiardi M, Aguilera I, Bianchi M, et al. Bulk band structure of Bi 2 Te 3[J]. Physical Review B, 2014, 90(7): 075105.
\bibitem{Altman E}Altman E, Auerbach A. Oscillating superfluidity of bosons in optical lattices[J]. Physical review letters, 2002, 89(25): 250404.

\bibitem{Qin F}Qin F, Cui X, Yi W. Polaron in a p+ip Fermi topological superfluid[J]. Physical Review A, 2019, 99(3): 033613.
\bibitem{Zubkov M A}Zubkov M A, Lewkowicz M. The type II Weyl semimetals at low temperatures: Chiral anomaly, elastic deformations, zero sound[J]. Annals of Physics, 2018, 399: 26-52.

\bibitem{Shubnyi V O}Shubnyi V O, Skrypnyk Y V, Sharapov S G, et al. Effect of resonant impurity scattering of carriers on the Drude-peak broadening in uniaxially strained graphene[J]. Physical Review B, 2019, 99(23): 235421.
\bibitem{Rosch A}Rosch A, Kopp T. Heavy particle in a d-dimensional fermionic bath: A strong coupling approach[J]. Physical review letters, 1995, 75(10): 1988.
\bibitem{Kantian A}Kantian A, Schollw?ck U, Giamarchi T. Competing regimes of motion of 1D mobile impurities[J]. Physical review letters, 2014, 113(7): 070601.
\bibitem{Meden V}Meden V, Schmitteckert P, Shannon N. Orthogonality catastrophe in a one-dimensional system of correlated electrons[J]. Physical Review B, 1998, 57(15): 8878.
\bibitem{ms18}Wu C H. Electronic properties of the Dirac and Weyl systems with first-and higher-order dispersion in non-Fermi-liquid picture[J]. arXiv preprint arXiv:1811.08809, 2018.
\bibitem{Bhandari}Bhandari S, Cheung M S, Geva E, et al. Fundamental gaps of condensed-phase organic semiconductors from single-molecule calculations using polarization-consistent optimally tuned screened range-separated hybrid functionals[J]. Journal of Chemical Theory and Computation, 2018, 14(12): 6287-6294.

\bibitem{Ong Z Y}Ong Z Y, Pop E. Effect of substrate modes on thermal transport in supported graphene[J]. Physical Review B, 2011, 84(7): 075471.
\bibitem{Giustino F2}Giustino F. Electron-phonon interactions from first principles[J]. Reviews of Modern Physics, 2017, 89(1): 015003.
\bibitem{Hwang E H}Hwang E H, Sarma S D. Acoustic phonon scattering limited carrier mobility in two-dimensional extrinsic graphene[J]. Physical Review B, 2008, 77(11): 115449.
\bibitem{Bistritzer R}Bistritzer R, MacDonald A H. Electronic cooling in graphene[J]. Physical Review Letters, 2009, 102(20): 206410.
\bibitem{Schmidt R}Schmidt R, Enss T. Excitation spectra and rf response near the polaron-to-molecule transition from the functional renormalization group[J]. Physical Review A, 2011, 83(6): 063620.

\bibitem{Kim S}Kim S, Woo S, Min H. Vertex corrections to the dc conductivity in anisotropic multiband systems[J]. Physical Review B, 2019, 99(16): 165107.
\bibitem{Pyatkovskiy P K}Pyatkovskiy P K. Dynamical polarization, screening, and plasmons in gapped graphene[J]. Journal of Physics: Condensed Matter, 2008, 21(2): 025506.

\bibitem{Hwang E H2}Hwang E H, Throckmorton R E, Sarma S D. Plasmon-pole approximation for many-body effects in extrinsic graphene[J]. Physical Review B, 2018, 98(19): 195140.
\bibitem{Tsuchiya S}Tsuchiya S, Watanabe R, Ohashi Y. Single-particle properties and pseudogap effects in the BCS-BEC crossover regime of an ultracold Fermi gas above T c[J]. Physical Review A, 2009, 80(3): 033613.
\bibitem{Wu F}Wu F, Qu F, MacDonald A H. Exciton band structure of monolayer MoS 2[J]. Physical Review B, 2015, 91(7): 075310.

\bibitem{Rana F}Rana F, George P A, Strait J H, et al. Carrier recombination and generation rates for intravalley and intervalley phonon scattering in graphene[J]. Physical Review B, 2009, 79(11): 115447.
\bibitem{Borysenko K M}Borysenko K M, Mullen J T, Barry E A, et al. First-principles analysis of electron-phonon interactions in graphene[J]. Physical Review B, 2010, 81(12): 121412.
\bibitem{Alexandrov A}Alexandrov A, Capellmann H. Phonons in a strongly coupled electron-phonon system[J]. Physical Review B, 1991, 43(3): 2042.
\bibitem{Sio W H}Sio W H, Verdi C, Poncé S, et al. Ab initio theory of polarons: Formalism and applications[J]. Physical Review B, 2019, 99(23): 235139.
\bibitem{Nocera}Nocera A, Sous J, Feiguin A E, et al. Bipolaron liquids at strong Peierls electron-phonon couplings[J]. Physical Review B, 2021, 104(20): L201109.
\bibitem{Parish}Parish M M. Polaron-molecule transitions in a two-dimensional Fermi gas[J]. Physical Review A, 2011, 83(5): 051603.
\bibitem{Jiang}Jiang Y, Fan Y, Li S, et al. Photocatalytic Methane Conversion: Insight into the Mechanism of C (sp3)–H Bond Activation[J]. CCS Chemistry, 2023, 5(1): 30-54.
\bibitem{Dias}Dias R P, Yoo C S, Kim M, et al. Insulator-metal transition of highly compressed carbon disulfide[J]. Physical Review B, 2011, 84(14): 144104.
\bibitem{Kalosakas}Kalosakas G, Aubry S, Tsironis G P. Possibility of observation of polaron normal modes at the far-infrared spectrum of acetanilide and related organics[J]. Physics Letters A, 1998, 247(6): 413-416.
\bibitem{Kick}Kick M, Scheurer C, Oberhofer H. Polaron-Assisted Charge Transport in Li-Ion Battery Anode Materials[J]. ACS Applied Energy Materials, 2021, 4(8): 8583-8591.
\bibitem{Peng}Peng C W, Liao W B, Chen T Y, et al. Efficient spin-orbit torque generation in semiconducting WTe2 with hopping transport[J]. ACS Applied Materials $\&$ Interfaces, 2021, 13(13): 15950-15957.
\bibitem{Fetherolf}Fetherolf, Jonathan H., Denis Golež, and Timothy C. Berkelbach. "A unification of the Holstein polaron and dynamic disorder pictures of charge transport in organic crystals." Physical Review X 10.2 (2020): 021062.


\bibitem{Elmaslmane}Elmaslmane A R, Watkins M B, McKenna K P. First-principles modeling of polaron formation in TiO2 polymorphs[J]. Journal of chemical theory and computation, 2018, 14(7): 3740-3751.
\bibitem{Chuang}Chuang H J, Tan X, Ghimire N J, et al. High mobility WSe2 p-and n-type field-effect transistors contacted by highly doped graphene for low-resistance contacts[J]. Nano letters, 2014, 14(6): 3594-3601.
\bibitem{Araki}Araki, Gako, et al. "Polaronlike vibrational bands of molecular crystals with one-dimensional hydrogen-bond chains: N-methylacetamide." Physical Review B 43.15 (1991): 12662.
\bibitem{Chang}Chang, Chun-Min, AH Castro Neto, and A. R. Bishop. "Long-range charge transfer in periodic DNA through polaron diffusion." Chemical physics 303.1-2 (2004): 189-196.



\end{thebibliography}
\end{document}